\documentclass[journal=jceaax,manuscript=article]{achemso}

\usepackage[version=3]{mhchem} 
\usepackage[T1]{fontenc}       
\SectionNumbersOn

\author{Karin I. \"Oberg}
\email{koberg@cfa.harvard.edu}
\affiliation{Harvard-Smithsonian Center for Astrophysics, 60 Garden St, Cambridge, MA 02138, U.S.A.}

\title{Photochemistry and astrochemistry: photochemical pathways to interstellar complex organic molecules}

\abbreviations{IR,UV,TPD,ISM}

\begin{document}


\begin{abstract}
The interstellar medium is characterized by a rich and diverse chemistry. Many of its complex organic molecules are proposed to form through radical chemistry in icy grain mantles. Radicals form readily when interstellar ices (composed of water and other volatiles) are exposed to UV photons and other sources of dissociative radiation, and if sufficiently mobile the radicals can react to form larger, more complex molecules.  The resulting complex organic molecules (COMs) accompany star and planet formation, and may eventually seed the origins of life on nascent planets. Experiments of increasing sophistication have demonstrated that known interstellar COMs as well as the prebiotically interesting amino acids can form through ice photochemistry. We review these experiments and discuss the qualitative and quantitative kinetic and mechanistic constraints they have provided. We finally compare the effects of UV radiation with those of three other potential sources of radical production and chemistry in interstellar ices: electrons, ions and X-rays. 
\end{abstract}

\tableofcontents

\section{Introduction}

The interstellar medium (ISM), the medium between stars, is host to a rich, organic chemistry. The presence of large molecules may be a surprise considering that the ISM is permeated by high-energy radiation fields including UV radiation. The interstellar UV field has been known or inferred to be present since the dawn of modern astronomy. The potentially destructive power of UV radiation on molecules was a major argument against the presence of an interstellar chemistry. Today we know that some parts of the interstellar medium, so called molecular clouds, are shielded from the onslaught of external UV irradiation by interstellar dust particles that efficiently absorb UV and visible radiation. These cold ($\sim$10~K) clouds are rich in molecules, most of which can be explained by gas-phase chemical reactions between ions and molecules. This gas-phase chemistry cannot explain the presence of some of the most complex organic molecules found in the ISM, however, i.e. molecules such as ethylene glycol and ethanol. These highly saturated organics are instead proposed to originate from radical chemistry in icy mantles embedding the interstellar grains that reside in clouds. 

Radicals form in ices through either successive atom-addition reactions or through dissociation of pre-existing molecules. Radical reactions generally lack energy barriers and can thus proceed at the low temperatures characteristic of interstellar ices (10--100~K), as long as the radical reactants either form on neighboring sites or can diffuse through the ice. Despite the absence of external UV photons, there are several potential sources of dissociative radiation in clouds. Cosmic ray interactions with molecular hydrogen and grain mantles cause secondary UV photons and electrons. Close to young stars there may also be substantial X-ray radiation. The same kind of radiation that precludes molecular survival exterior to clouds is thus in small doses a source of chemical complexity in cloud interiors.

The relative importance of different radiation sources for ice chemistry is contested and may vary between different interstellar environments. The primary aim of this review is to describe and evaluate the constraints that laboratory experiments have provided on ice photochemistry in astrophysical environments. In the final section these results are then compared with ice experiments that use electrons, ions and X-rays as an energy source.

\subsection{Astrochemistry in the Context of Star and Planet Formation}

\begin{figure}[htbp]
\begin{center}
\vspace{-5mm}
\includegraphics[width=5in]{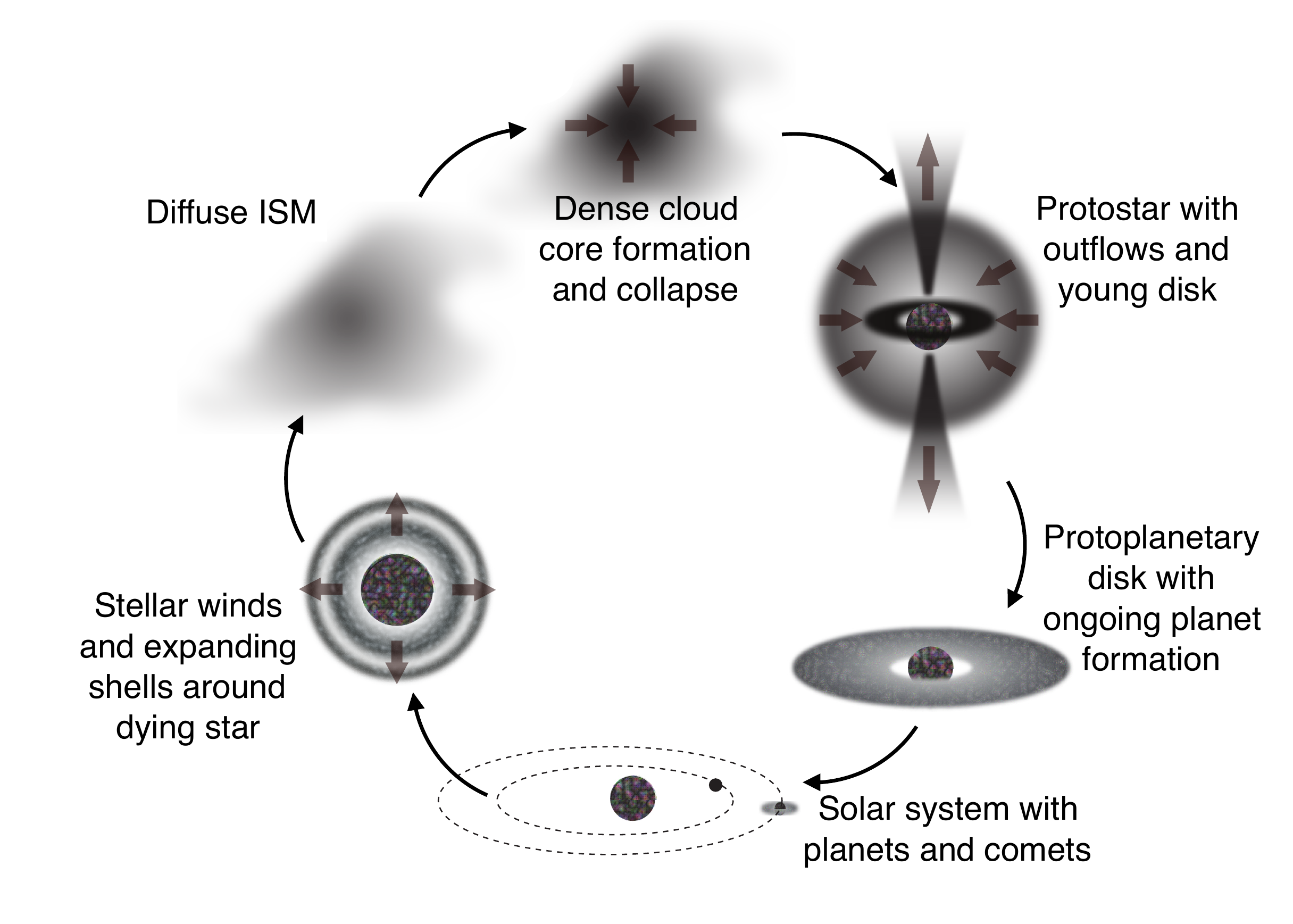}
\vspace{-5mm}
\caption{Solar-type star formation begins with collapse of a core in a dense molecular cloud\cite{Shu77}. During collapse a disk and outflows form to mediate transport of angular momentum. Once collapse has proceeded far enough for the core to heat up a protostar is formed. The remaining cloud envelope accretes onto the star or disperses leaving a pre-main sequence star with a disk. The disk is the formation site of planets, which must form on time scales of 1-10~Myrs, the life time of such disks. }
\label{fig:sf}
\end{center}
\end{figure}

The context of this review is the interstellar and circumstellar media from which stars and planets form. This medium consist of $\sim$99\% gas and 1\% grains, by mass. The gas is mainly atomic or molecular hydrogen with trace amounts of heavier elements. Figure \ref{fig:sf} shows the different phases of star and planet formation for an isolated Solar-type star. In the Galaxy, most of the space between stars is characterized by a diffuse mixture of gas (mainly atomic and molecular hydrogen) and small, sub-micron-sized dust grains. The gas and grains are almost completely exposed to the interstellar radiation field (ISRF), which has a substantial far-ultra-violet (FUV) or vacuum-ultra-violet (VUV) component that readily dissociates most molecules (both FUV and VUV emission is defined as UV emission with shorter wavelengths than 200 nm). The diffuse ISM is therefore characterized by atoms and ions, though low abundances of molecules survive\cite{Snow06}. Over-densities in the ISM are referred to as clouds and the beginning of star formation is the assembly of a dense molecular cloud with even denser cloud cores. The edges of the cloud are still exposed to the ISRF as well as to radiation of any nearby stars, which results in so called photon dominated regions (PDRs) where the combination of high UV fields and dense, relatively cold gas and dust result in a characteristic chemistry\cite{Hollenbach97}. The very edges of such PDRs present, similarly to the diffuse ISM, mainly atoms, ions and radicals, but gas-phase chemistry quickly reforms simple molecules behind the UV front. Some PDRs display a rich molecular chemistry, which may be the result of UV-mediated destruction of macromolecules and volatile grain mantles\cite{Guzman14,Guzman15}.

Molecular cloud interiors are protected from external UV radiation, but not from cosmic rays. Cosmic rays interact with hydrogen to produce secondary UV radiation and electrons, and are responsible for partially ionizing the cloud gas. The energy input from the cosmic rays is relatively low, which in combination with efficient molecular cooling results in low ($\sim$10~K) temperatures. Densities are high compared to the diffuse interstellar medium (the hydrogen density n$_{\rm H}\sim$10$^4$ cm$^{-3}$), but very low compared to terrestrial conditions. Under these conditions, ion-molecule reactions in the gas-phase drive an efficient chemistry resulting in e.g. formation of large carbon chains, as well as many smaller molecules\cite{Bergin07}. Most gas-phase molecules observed in clouds can be explained by this ion-molecule chemistry\cite{Herbst73}.

Dense molecular clouds are also characterized by large volatile depletions from the gas-phase due condensation or freeze-out of molecules onto interstellar dust grains. At grain temperatures of $\sim$10~K all molecules but H$_2$ and He readily stick onto grains upon collision and the freeze-out time scales are set by collision time scales. Freeze-out combined with atom reactions on the grain surfaces produce large quantities of small hydrogenated species such as water, which remain on the grains forming an icy grain mantle. 

If a core within the dense cloud becomes sufficiently massive it begins to collapse due to self-gravity\cite{Shu77}. Initially the core is kept cool by molecules radiating away the heat produced by the collapse. As the collapse proceeds the core becomes optically thick, however, and begins to warm up. Eventually it becomes warm enough for deuterium fusion and a protostar has formed. Protostars remain enveloped in molecular material for some time and are also characterized by the presence of a circumstellar disk and/or outflows. The outflows help transport away angular momenta. The disk serves a similar purpose, but also aids in accretion of additional material onto the star. Chemically this stage is characterized by sublimation of ice mantles, and ensuing gas-phase chemistry, as grains flow toward the protostar\cite{Herbst09}. Some of the cloud chemistry products probably survives incorporation into the accretion disk\cite{Visser09}.

Within a few hundred thousand years, the initial envelope of cloud material either accretes or disperses and what remains is a pre-main sequence star and a circumstellar disk. The disk is the formation site of planets and is generally referred to as a protoplanetary disk. The disk structure and composition regulates the formation of planets and both are topics of very active research (see Williams \& Cieza (2011)\cite{Williams11} for a review). The chemical structure of protoplanetary disks is less well known compared to protostellar envelopes and clouds because of small angular sizes and intrinsically weak emission. The remnant of our own protoplanetary disk, comets, display a rich organic chemistry\cite{Mumma11,LeRoy15,Wright15,Goesmann15} and recent observations suggest that protoplanetary disks can host a complex organic chemistry as well\cite{Oberg15}. This is important when considering the connection between astrochemistry and prebiotic chemistry, since it is the disk material that eventually becomes incorporated into planets and comets, and thus set the prebiotic potential of nascent planets. In following sub-sections we review in some further depth the aspects of astrochemistry that directly relates to the photochemical formation of complex organics.

\subsection{Ice Formation and Demographics in the ISM}

Icy grain mantles are the starting point of complex organic molecule formation in the interstellar medium and therefore worth considering in some detail.  In clouds, protostellar envelopes and protoplanetary disks, such icy mantles constitute the main reservoir of volatiles.
Ices form by a condensation of atoms and molecules from the gas-phase and subsequent grain surface chemistry\cite{Tielens82}.. The latter mostly involves hydrogen atoms, because at grain temperatures of 10~K (typical of molecular clouds), H atoms are much more mobile than any other ice constituent. While the dense cloud is forming out of the diffuse media, atoms dominate the gas-phase. The most abundant atoms are H, He (which does not partake in chemistry), O, C and N. O, C, N and H accrete onto grains and H$_2$O, CH$_4$ and NH$_3$ form by hydrogenation. Some C and O are converted to CO in the gas-phase relatively quickly and then condense out. As long as O condensation is also taking place, OH will be abundant on the grain surface and condensed CO can react with OH to form CO$_2$\cite{Ioppolo11}. As the gas-phase becomes dominated by molecules, CO becomes the most abundant species to accrete onto grains. Some CO will still react to form CO$_2$, while other CO molecules will react with H to form H$_2$CO and CH$_3$OH. 

\begin{figure}[htbp]
\begin{center}
\includegraphics[width=5in]{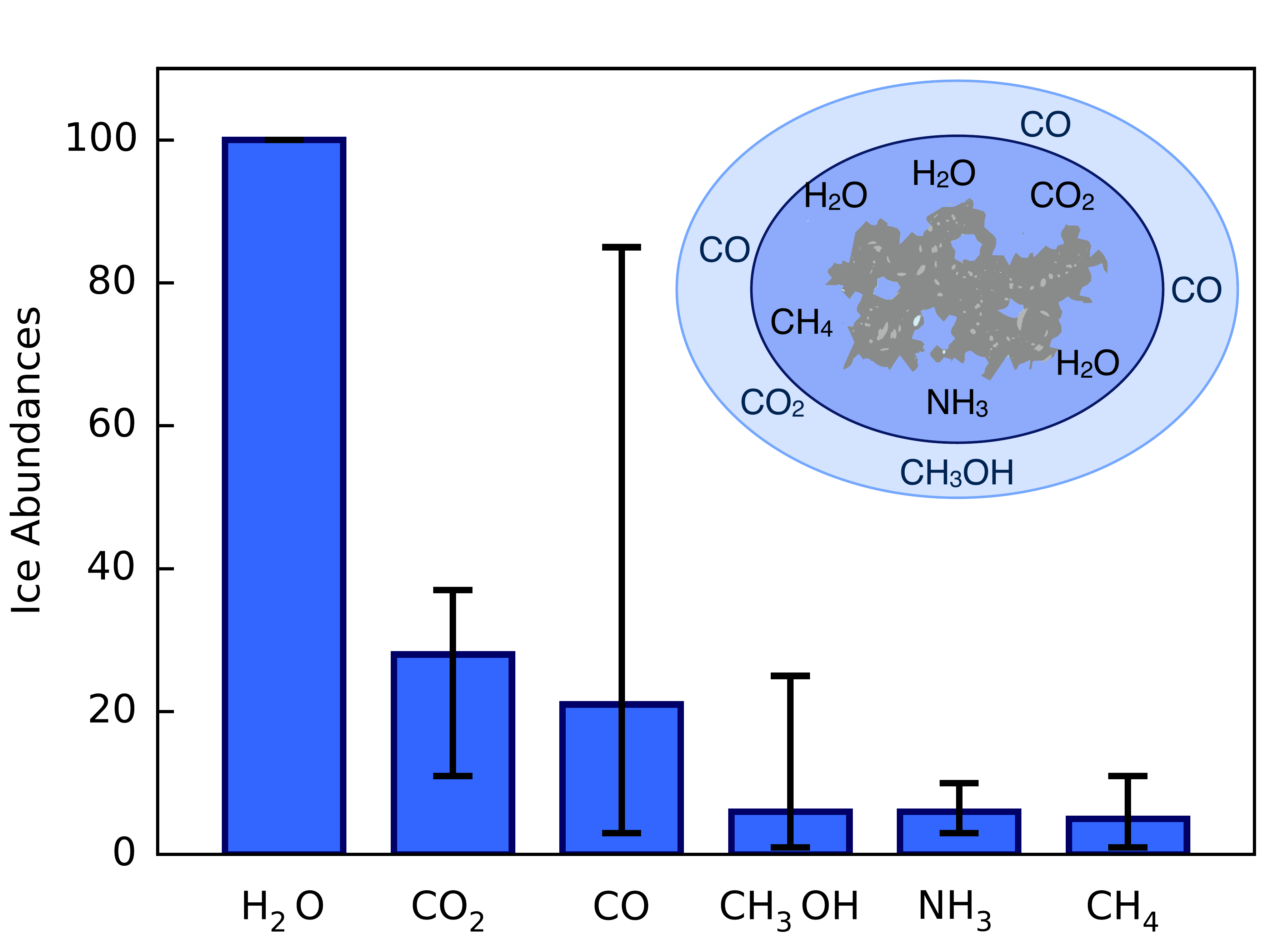}
\caption{The median composition of interstellar icy grain mantles (bars) normalized to the most abundant ice species, water\cite{Boogert15}. The superimposed error bars indicate the minimum and maximum abundance detected of different species, relative to water ice in that line of sight.}
\label{fig:ice}
\end{center}
\end{figure}

Interstellar ice compositions are primarily studied through absorption infrared spectroscopy toward a background source such as a protostar\cite{Gibb04,Oberg11c,Boogert15}. Based on astronomical observations, ice mantle compositions can vary substantially between different clouds, and also between different protostellar envelopes within the same dense cloud. There are some general trends, however. First, water ice is the most abundant ice constituent, followed by CO and CO$_2$ at $\sim$20--30\% each with respect to water. CH$_4$, NH$_3$ and CH$_3$OH are also detected in many lines of sight at typical abundances of $\sim$5\% with respect to water ice. Figure \ref{fig:ice} shows the median ice abundances toward Solar-type protostars, as well as the minimum and maximum abundance with respect to water ice observed for each of the other five species.

Analysis of observed  ice spectral profile reveal that interstellar ices are not perfectly mixed, but rather present in at least two different phases, a water-rich phase and water-poor-phase\cite{Pontoppidan03,Pontoppidan04}. The water-rich phase is observed to form first and contains most of the water and CO$_2$, and probably most NH$_3$ and CH$_4$ ice as well. A second CO-rich ice contains the remaining CO$_2$ ice and probably most of the CH$_3$OH (Fig. \ref{fig:ice})\cite{Penteado15}. Theoretically these two separate phases form because ice formation is dominated by hydrogenation of atoms at early times, and by reactions involving CO at late times\cite{Garrod11}. 

It is important to note that none of the ice experiments reviewed below mimic the observed average ice composition and morphology perfectly. The reasons for differences between experiment and observations vary. Early experiments were conducted before the large surveys of ices toward Solar-type protostars in the 2000s, and thus relied on more anecdotal ice observations. Most of the differences between astrophysical and laboratory ice compositions are instead motivated by experimental concerns however. Simple ice mixtures or even pure ices are often used to isolate the kinetics and mechanisms of particular ice formation pathways. In other experiments the concentrations of the most reactive ice species are increased to increase the yield of prebiotically interesting products.

In astrophysical environments it is difficult to directly detect organics larger than CH$_3$OH in ices due to overlapping ice bands from more complex organic ices. A mixture of moderately complex organics, including CH$_3$CH$_2$OH and HCOOH, have been proposed to contribute to observed ice bands between 5 and 7$\mu$m toward protostars\cite{Gibb04,Oberg11c,Boogert15}, but unique identifications have been controversial. Similarly, families of very large COMs may be possible to detect with the next generation of infrared telescopes using features at $\sim$3.4$\mu$m, but only if the original ice mantles has first sublimated\cite{MunozCaro09,MunozCaro13}. Detections of specific molecules in both ice and ice residues will likely remain elusive even with large increases in sensitivity, however, due to the intrinsic  overlap between infrared ice features of related molecules. 

\subsection{Interstellar UV Fields}

\begin{figure}[htbp]
\begin{center}
\includegraphics[width=5in]{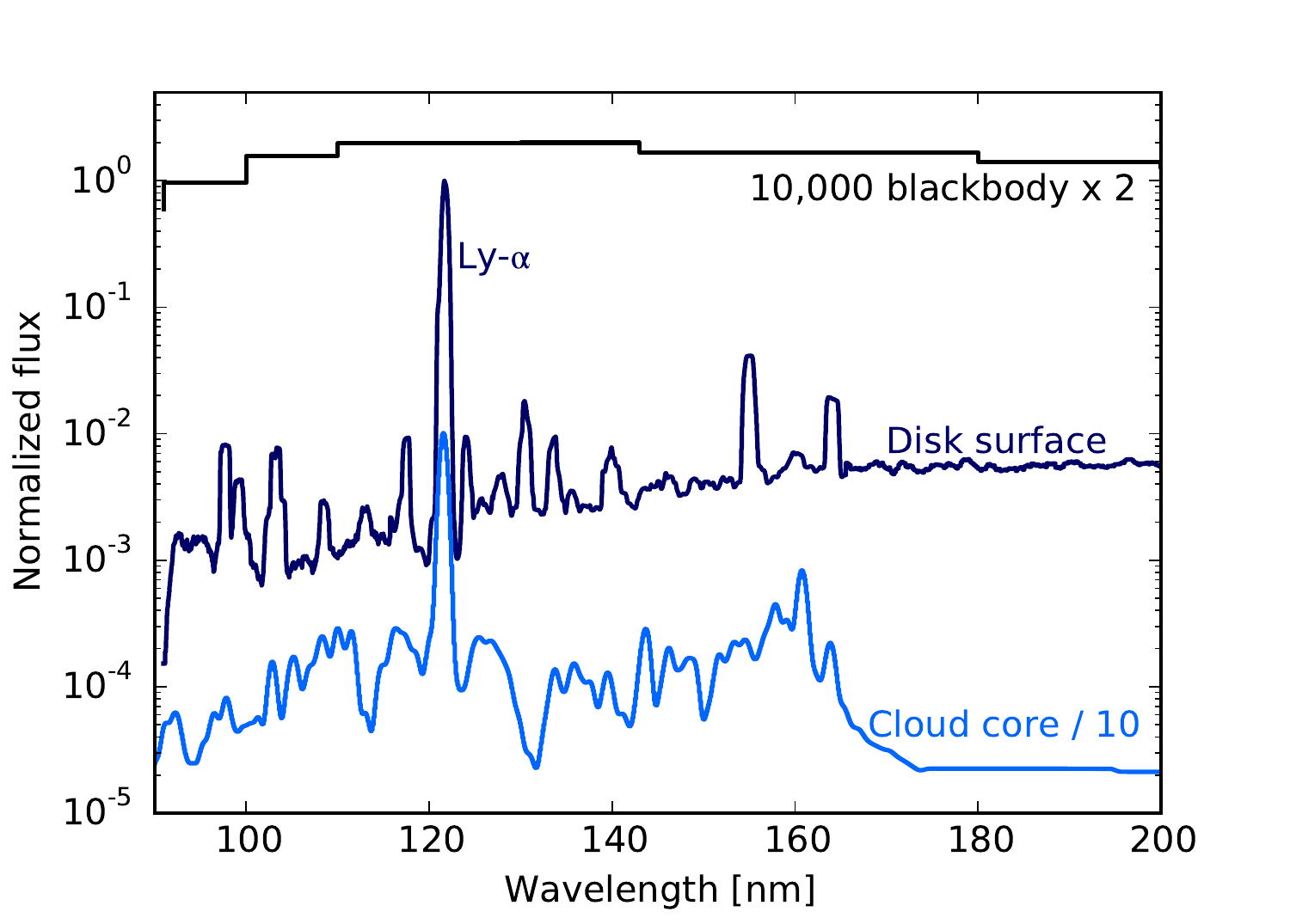}
\caption{Ices can be exposed to a range of UV spectra in the ISM, including black-body radiation from nearby massive stars\cite{Mathis83} (top), line emission from a young Solar-type star\cite{Herczeg02,Valenti03,Johns-Krull07} (middle), and line emission from cosmic ray interactions with molecular hydrogen\cite{Gredel87} (bottom; for a high resolution spectrum see the original publication). All spectra have been normalized to have a peak flux of unity to enable easy comparison of the different spectral energy distributions.}
\label{fig:uv}
\end{center}
\end{figure}

In the context of this review it is important to consider what kind of UV fields icy grain mantles are exposed to from formation to sublimation (Fig. \ref{fig:uv}), and how the energy deposition from UV radiation compares to the energy deposition from other dissociative radiation and particles. Close to the edges of clouds, UV fields are dominated by radiation from massive stars, which can be approximated as a black body with a temperature of 10,000--30,000~K. If there is a massive star nearby, this field will be very intense. If no such star is present close to the clouds, the grain will still be exposed to the interstellar radiation field (ISRF), which is similar in shape, but lower in flux. Models of the interstellar radiation field predict integrated VUV (912--2050 \AA) fluxes of $2.67\times10^{-3}$ erg cm$^{-2}$ s$^{-1}$\cite{Draine78,vanDishoeck06}, corresponding to $\sim$10$^8$ photons cm$^{-2}$ s$^{-1}$. No photons with shorter wavelengths than 912 {\AA}  are present in the diffuse ISM due to absorptions by atomic hydrogen.

In diffuse environments the timescales of ice photodesorption are short. Build-up of icy grain mantles therefore take place only in dense molecular clouds, where the grains are shielded by dust absorption from any external sources of UV radiation. They are instead exposed to a low-intensity UV field, which is due to cosmic ray interactions with H$_2$. The resulting VUV flux is $\sim$10$^{4}$ photons cm$^{-2}$\cite{Shen04}. The VUV spectra that emerges from this process has been calculated (Fig. \ref{fig:uv})\cite{Gredel87}, and the calculation demonstrates that a substantial portion of the VUV flux comes in the form of Lyman-$\alpha$. 

At the later stages of star formation, icy mantles can be directly exposed to UV radiation from the central protostar or pre-main sequence star. Solar-type pre-main sequence stars can have large VUV excesses in the form of Lyman-$\alpha$ radiation\cite{Herczeg02,Valenti03,Johns-Krull07,Bergin03}, which may induce additional UV processing in ices residing in protoplanetary disks.

In addition to UV radiation, interstellar ices are exposed to higher energy photons (especially X-rays), cosmic rays (i.e. high energy atomic nuclei) and electrons. Which kind of radiation that is the most important for ice chemistry in molecular clouds, protostellar envelopes and in protoplanetary disks is debated. Calculations relevant to molecular clouds suggest that an order of magnitude more energy is deposited into grain mantles from secondary UV photons compared to direct interactions with cosmic rays\cite{Shen04}. These estimates may underestimate, however, the importance of secondary electrons created by a cosmic ray as it is passing through the grain and ice mantle\cite{Bennett05}. That is, cosmic rays passing through an icy mantle can generate millions of secondary electrons that can induce dissociations similarly to UV radiation. The interactions of electrons with ices was recently reviewed\cite{Arumainayagam10}, and there are many studies on the generation of secondary electrons by cosmic ray interactions with ice, and their effects on ice chemistry\cite{Moore07,Kaiser13}. In the protoplanetary disk stage, cosmic rays can be efficiently excluded from the disk itself due to shielding by strong T Tauri (pre-main sequence Solar-type stars) stellar winds\cite{Cleeves15}. In this environment, X-rays may be the most important source of molecular dissociation in ices; young stars have X-ray excesses, and X-rays can penetrate deeper into the disk than VUV photons.

\subsection{Complex Organic Molecules during Star Formation}

\begin{figure}[htbp]
\begin{center}
\includegraphics[width=6in]{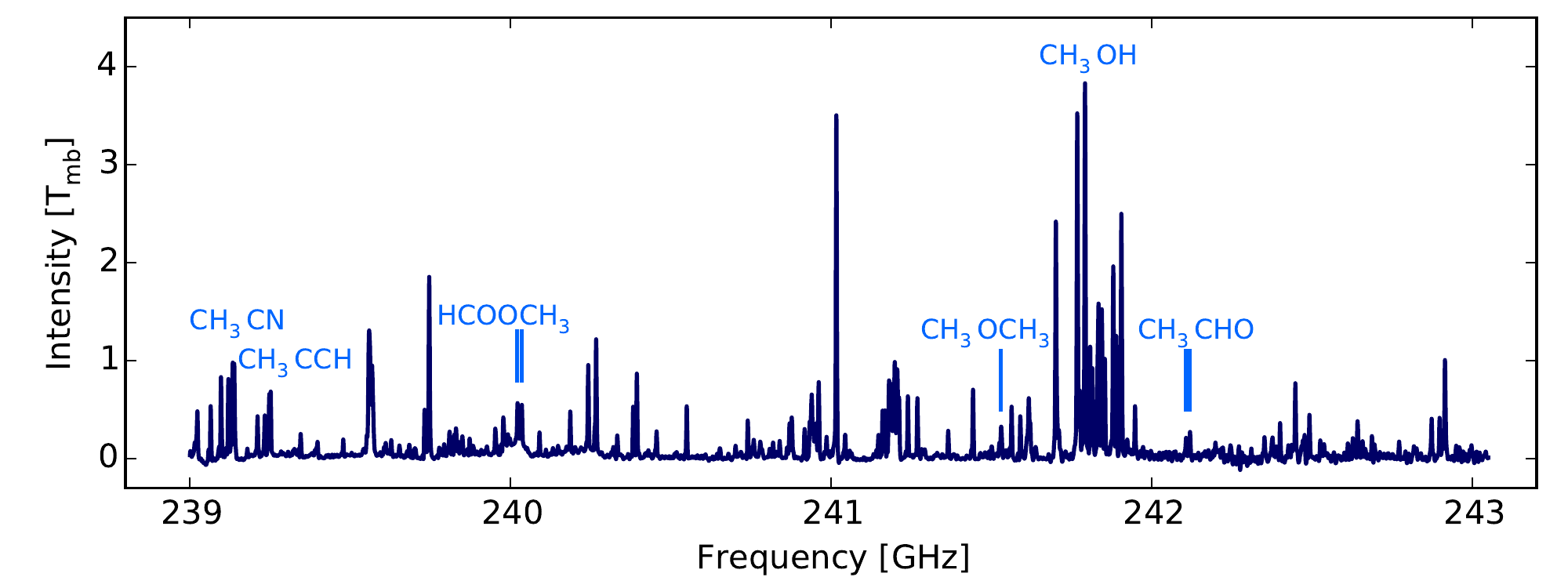}
\caption{Spectra between 239 and 243 GHz toward the massive protostar NGC 7538 IRS1\cite{Oberg14}.}
\label{fig:irs1}
\end{center}
\end{figure}

COMs were first discovered in massive-star forming regions, where clusters of stars, some of which are many times more massive than the Sun, are currently forming and heating up their surrounding dense media and sublimating icy grain mantles\cite{Ball70,Solomon71}. In these, and other interstellar environments, COMs are observed using rotational spectroscopy. Surveys toward large numbers of high-mass star forming regions have revealed that cm and mm emission from COMs are common attributes of high-mass star formation (Fig. \ref{fig:irs1} shows one example) and the presence of lines from acetonitrile and other COMs toward a star forming region is now used as a sign post for ongoing high-mass formation \cite{Bisschop07,Rosero13}. The emitted molecules generally present high excitation temperatures (100--200~K). The high excitation temperatures as well as spatial resolved observations show that the molecular emission originates in a hot core close to the protostar, where ices have sublimated.

In early models, COMs  were proposed to form in the hot gas close to the protostar where they are observed. Sublimation of CH$_3$OH ice and other volatiles close to the protostar were supposed to initiate an efficient gas-phase ion-molecule chemistry resulting in the rapid production of HCOOCH$_3$, CH$_3$OCH$_3$ and other complex organic molecules\cite{Charnley95}. A key step in this ion-molecule chemistry is the formation of protonated form of the final product (e.g. protonated HCOOCH$_3$) followed by a recombination with an electron to produce the neutral molecule. The viability of this scenario was challenged, however, by experimental measurements revealing that the recombination step had a very low yield for most molecules and instead preferentially resulted in dissociation into smaller fragments\cite{Geppert06}. Gas-phase chemistry following ice sublimation is therefore no longer considered a plausible scenario for the formation of the majority of the observed COMs toward protostars, but can still contribute to the formation of a subset of COMs\cite{Garrod08}.

\begin{figure}[htbp]
\begin{center}
\includegraphics[width=5in]{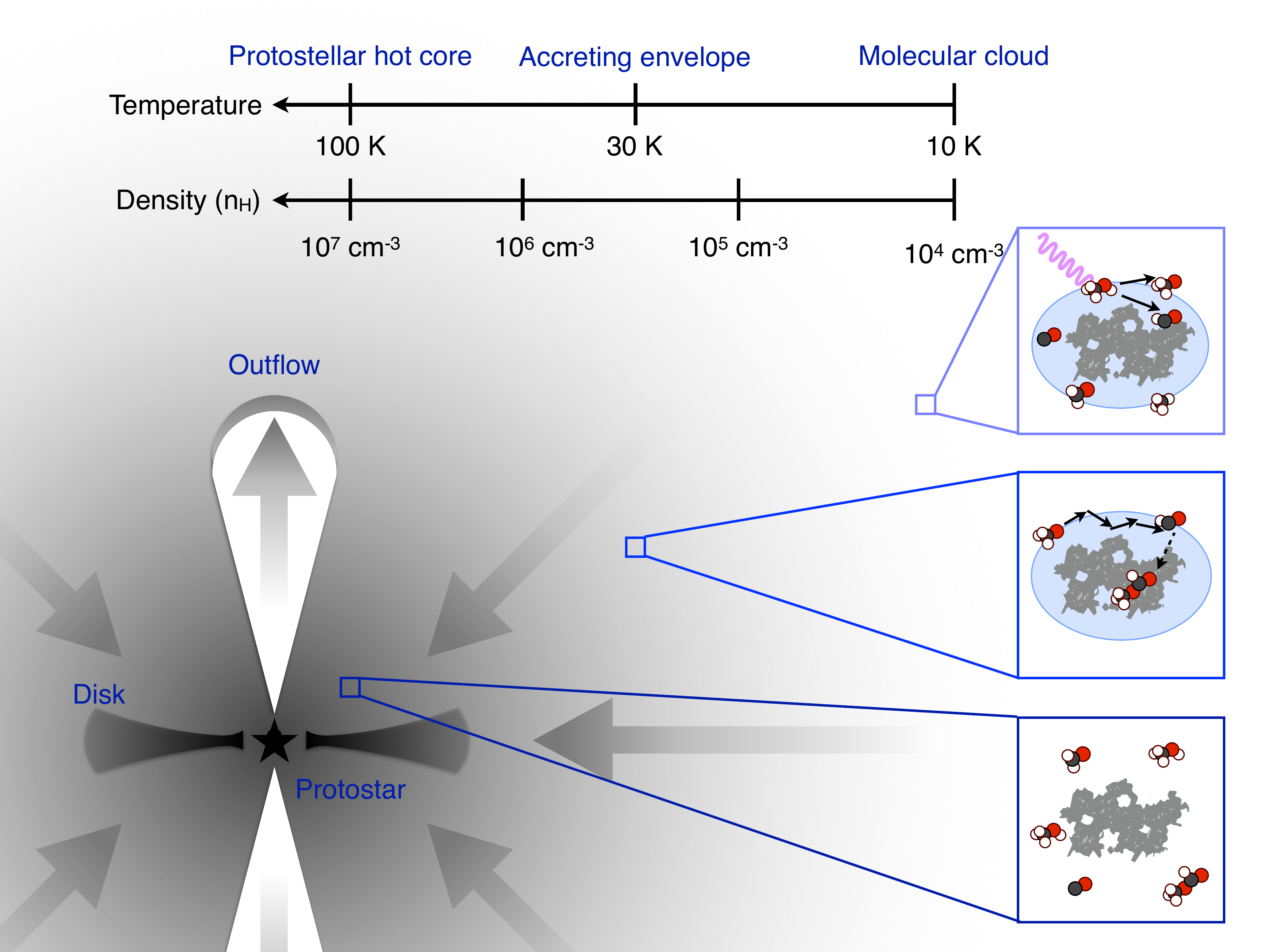}
\caption{Cartoon of complex organic molecule formation through ice photochemistry during star formation. Simple ices form from atom addition reactions at low temperatures in dense molecular clouds. When these ices are exposed to UV radiation, some portion of the original ice is dissociated into radicals. As the icy grain heats up during star formation, the radicals become mobile and can combine to form more complex species. Some radical chemistry may also be possible in the coldest regions dependent on radical concentration and the efficiency of non-thermal diffusion.}
\label{fig:com_form}
\end{center}
\end{figure}

The current understanding is that COMs mainly form through ice chemistry on grains. Figure \ref{fig:com_form} illustrates how COMs are proposed to form during star formation through ice photochemistry. As outlined above and below this model has several caveats, including few constraints on the dominant radiation source, and on the relative importance of thermal and non-thermal diffusion of reactants. There are also alternative COM formation pathways that have been proposed\cite{Fedoseev15,Balucani15}. Nevertheless, this is the formation scheme employed in most contemporary astrochemistry models, and it thus provides a useful starting point\cite{Herbst09}. In this scheme, the simple ices formed during the molecular cloud phase partially dissociate into radicals when exposed to UV radiation from e.g. cosmic ray interactions with H$_2$. For example CH$_3$OH can be dissociated into CH$_3$+OH, CH$_2$OH+H and other species.
When a protostar forms and begins to accrete nebular material it accretes dust grains together with gas. As the dust grain approaches the protostar, the icy mantle becomes warm enough for radicals to become mobile in the ice. The radicals combine to form new, often more complex species, such as CH$_3$CH$_2$OH from CH$_3$ and CH$_2$OH\cite{Garrod06,Garrod08,Garrod13}. The icy grain continues to flow toward the central protostar and eventually becomes warm enough for the ices to sublimate (T$\sim$100--150~K) populating the gas-phase with a mixture of simple molecules from the original ice mantle and the newly formed more complex organics. This model explains observed COM abundances toward massive protostars very well\cite{Herbst09}.

During the past couple of decades it has become clear that interstellar COMs are not limited to hot cores in high-mass star forming regions. Large numbers of complex organics have been detected toward the low-mass Solar-type protostar IRAS 16293-2422 \cite{Cazaux03,Jorgensen11}. In this protostar the excitation pattern and spatial profiles \cite{Bisschop08} of the detected molecules indicate that the molecules are emitting from a hot and compact region that is associated with ice sublimation close to the protostar, analogously to the hot cores found around massive protostars. The chemical composition is also similar to what is found toward more massive protostars, and the chemistry has been successfully modeled using adaptations of the scenario presented in Fig. \ref{fig:com_form} \cite{Herbst09}. 

More recently several COMs have been observed in a number of sources that are cold compared to the hot cores associated with the innermost regions of low-mass and high-mass protostars. These observations cannot be explained by thermal sublimation of complex ices, but may still be consistent with an ice photochemistry origin. In fact some such observations, including one of COMs in a protostellar outflow, provide some of the strongest support of an icy origin of gas-phase interstellar COMs\cite{Arce08}. Protostellar outflows are energetic outflows of gas and dust that appear during the early stages of star formation. The outflows cause shocks, which induce ice mantle sublimation. Compared to protostars, the timescales in these outflows are short, too short for any significant gas-phase formation of COMs. Outflows thus provide a good opportunity to observe  ice chemistry products in isolation from gas-phase COM chemistry. The presence of COMs in such an outflow is thus consistent with an ice formation scenario, but not with gas-phase formation of COMs. 

COMs have also been detected in the lukewarm envelopes of several low-mass and high-mass protostars. The COM excitation temperatures in these sources are well below the expected ice sublimation temperature of $\sim$100~K\cite{Bottinelli07,Oberg11b,Oberg13,Oberg14,Fayolle15}. COMs are thus present in protostellar gas prior to the onset of thermal ice sublimation.  The protostellar envelopes where these COMs are detected are generally warm enough for diffusion-limited COM formation to be efficient in ices. The detections do not then pose a fundamental challenge to the ice photochemistry model, as long as non-thermal desorption, e.g. UV photodesorption, chemical desorption and sputtering from cosmic rays and energetic electrons, is sufficiently efficient\cite{Oberg08,Garrod07,Dartois15}. These observations do, however, complicate the scenario presented in Fig. \ref{fig:com_form}. On a more positive note, they also provide a potential window into COM formation at different ice temperatures.

Finally, COMs have been detected in few very cold, molecular cloud core environments\cite{Oberg10a,Bacmann12,Cernicharo12}. These observations are more challenging to explain by an ice radical combination chemistry, since radical diffusion is expected to be inhibited in ice at such low temperatures\cite{Garrod08}. Non-thermal or hot-atom diffusion of radicals following dissociation of the parent molecule may explain these observations, since it would allow radicals forming in the ice at some distance from one another to meet and react, but there is a lack of data or theory on the efficiency of non-thermal diffusion in ices. Another possible explanation is cosmic ray driven chemistry, since a single cosmic ray interaction with the ice will result in a large number of secondary electrons, which may locally heat up the ice and thus enable radical diffusion for a short period of time\cite{Bennett05}.

\subsection{Review Outline}

In summary, complex organic molecules are observed in a range of interstellar sources and ice photochemistry provides an attractive explanation for the their existence. The idea that UV irradiation of interstellar ices is a source of chemical complexity has its roots in a series of laboratory experiments carried out in the 1970s and 1980s\cite{Greenberg76}, that found that UV irradiation of interstellar ice analogs resulted in the production of large abundances of large and complex organics. This kind of experiment revealed that ice photochemistry is possible pathway to chemical complexity in space. To determine that ice photochemistry is the dominant pathway requires additional information, however, on the kinetics and mechanisms of ice photochemistry reactions, and on competing formation pathways that are regulated by electrons, cosmic rays and other sources of energy. Some of this information has been acquired over the past decades following dedicated efforts to unravel ice photochemical systems and their electron and cosmic ray induced counterparts. The information is far from complete, however. The aim of this review is therefore not to give a final answer to how much ice photochemistry contributes to chemical complexity in space, but rather to present and evaluate the experiments that have taken us to the current level of understanding of how UV photons interact with interstellar ice analogs to form new organic molecules.

Prior to the experimental work, we briefly review the theoretical models that have been developed to interpret experiments and astronomical observations of ice photochemistry (\S2). \S3 reviews experimental techniques employed in laboratory studies of ice photochemistry. In \S4--7 we then review ice photochemistry experiments and their contribution to our understanding of ice photochemistry. \S4 evaluates existing data on the initiation of photochemistry in ices, i.e. ice photodissociation, and the constraints this data provides on the photodissociation process in different ice set-ups. \S5 reviews photochemistry experiments on pure ices, with a focus on the constraints they have provided on photodissociation branching ratios (i.e. what radicals ices dissociate into and how this differ from the gas-phase) and radical diffusion. Photochemistry in simple ice mixtures is treated in \S6. In \S7 we then present and discuss photochemistry experiments that employ more complex ice mixtures as the starting material. The experiments are typically motivated by the question of what prebiotically interesting molecules could form in analogs to interstellar ices, but sometime provide mechanistic constraints as well. \S8 reviews how UV photochemistry compares with the chemistry induced by other types of dissociative radiation, most notably X-rays, energetic protons, and electrons, before a summary and concluding remarks in \S9. 

\section{Theoretical Methods}

\begin{figure}[htbp]
\begin{center}
\includegraphics[width=5in]{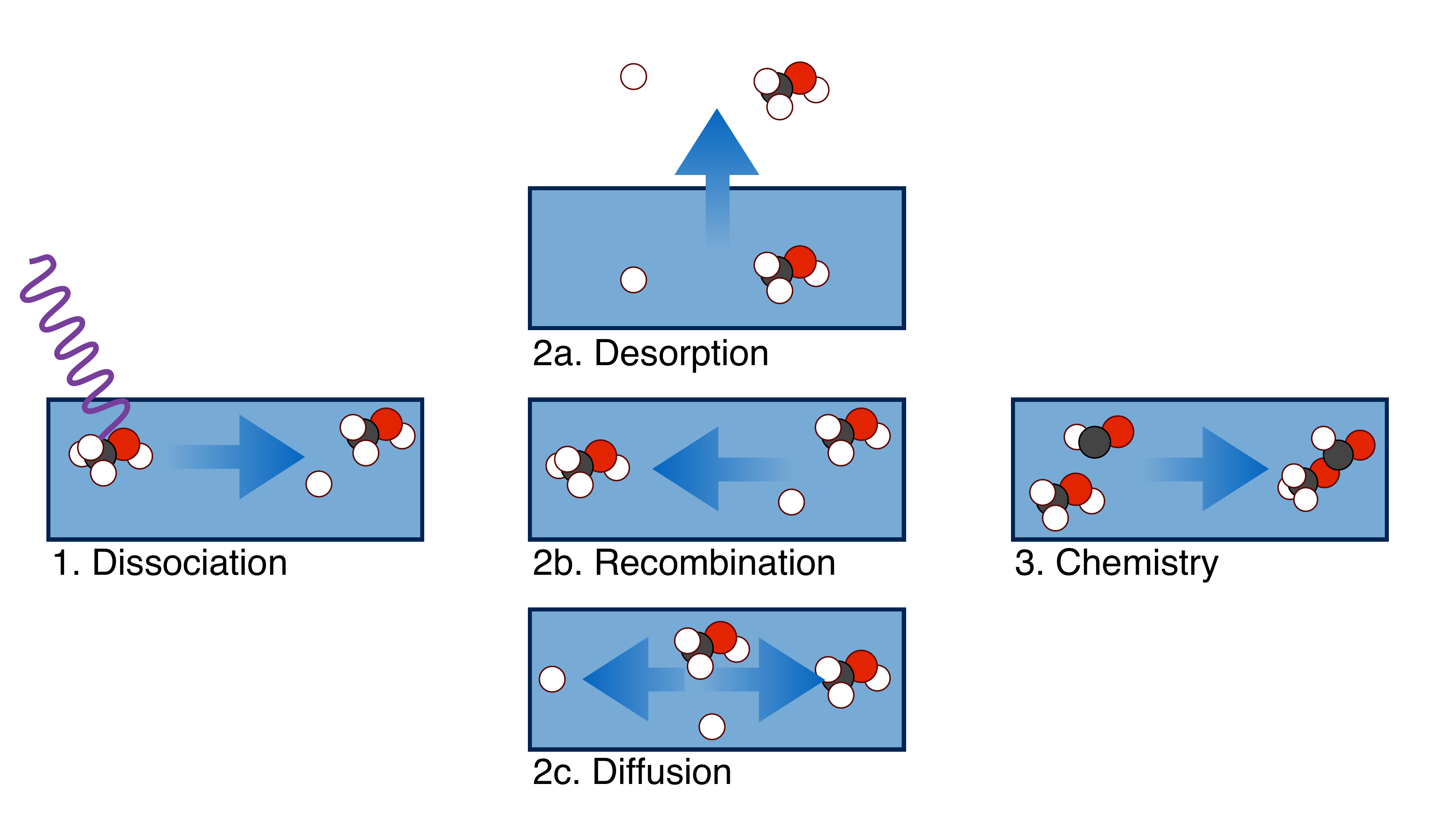}
\caption{The fundamental steps involved in ice photochemistry. Radicals formed through photodissociation can desorb, recombine into the parent molecule or diffuse away from the dissociation site. These processes all take place in competition with one another and with  diffusion+combination with radicals from other photodissociation events to form new molecules.}
\label{fig:ice_chem}
\end{center}
\end{figure}

Photochemistry in ices is initiated by the absorption of a UV photon into a dissociative electronic level of an ice molecule followed by its dissociation into radicals or ions. For the vast majority of experiments the products are (almost) exclusively radicals, since most small molecules have either ionization thresholds higher than 11 eV (the typical UV cutoff in experiments) or simply very low ionization yields below 11~eV\cite{vanDishoeck06}. The most notable exceptions are experiments that include polycyclic aromatics hydrocarbons (PAHs) embedded in the ice \cite{Bouwman09}. Following dissociation, the photo-produced  radicals can either recombine into the original molecule, react with other radicals or molecules in the ice, become trapped in the ice matrix or desorb. Which path dominates will depend on a combination of radical concentration, diffusion and desorption barriers, the importance of non-thermal ('hot atom') diffusion and desorption, the ice temperature, and whether the radicals form close to the surface or deep into the ice (Fig. \ref{fig:ice_chem}).

In experiments it is often difficult to observe the individual reaction steps and theoretical models are therefore key, both to interpret the experimental outcomes, and to translate experimental results to astrophysical environments where time scales are longer and many other environmental factors may differ as well. Different levels of theory, ranging from quantum mechanical calculations and molecular dynamics to macroscopic rate equation models, are used to address different aspects of the ice photochemistry process. Molecular dynamics enables simulations of the fast, non-thermal processes that follows immediately on UV photodissociation and have been key to elucidate photodesorption mechanisms and the competition between desorption and diffusion. Microscopic Monte Carlo models are used to investigate the diffusion, desorption and reaction steps that occur on longer time scales due to thermal processes. This method is too computationally expensive to treat chemically complicated ices, and to predict the outcome of photochemistry for realistic interstellar conditions. Macroscopic stochastic or rate equation models are therefore generally employed in comprehensive models of interstellar photochemistry. The latter has the advantage of a straightforward coupling with concurrent gas-phase chemistry. Two recent reviews\cite{Garrod13b,Cuppen13} covered most of the relevant theory and only a brief summary is presented here.

\subsection{Molecular Dynamics Simulations}

Molecular dynamics (MD) modeling has been applied to astrochemical ice photochemistry primarily to study the influence of UV radiation on water ice\cite{Andersson06,Andersson08,Arasa11}, but this could be extended to more systems. Most ice observed in the ISM is amorphous and to mimic these conditions, astrochemically relevant MD modeling is typically applied to an amorphous ice slab\cite{Al-Halabi04}. A single (water) molecule in the ice slab is then selected to be photodissociated. This molecule is allowed to be completely flexible and its interactions with the surrounding ice is governed by an analytic potential energy surface for an electronically excited state that leads to molecule dissociation. This process can be repeated for molecules at different depths into the ice to determine how the competition between recombination, desorption and entrapment depend on how close to the surface the molecule is excited. Excitation energies can be chosen to match experimental conditions, with most experiments using UV photons around 8--10~eV.

Following excitation of the molecule into a dissociative state, the trajectory of the dissociation is followed for 10s of ps, which is sufficient to reach the final outcome if  slow processes such as thermalized hopping/diffusion are ignored. Upon dissociation the intermolecular interactions switch from an interaction between the original molecule and the water ice slab, to two separate interactions between dissociation products (e.g. H and OH) with the ice, using the relevant potentials. If there is a recombination, the interaction switches back to the molecule-ice potential.

This kind of simulation has been used to demonstrate a decreasing photodesorption efficiency and increasing radical entrapment efficiency for molecules dissociated deeper into the ice. It has also been used to demonstrate the importance of energy transfer between 'hot atoms' and molecules in the ice, i.e. photodesorption of whole water molecules mainly proceed through a 'kick-out' by the energetic H produced in the photodissociation step\cite{Andersson08}. This suggests that non-thermal or 'hot atom' processes may generally be important in ice photochemistry experiments. The main limitation of MD simulations is that the simulations can only follow reactions for very short (ps) timescales. This is sufficient to explore the immediate result of a photodissociation event, i.e. the branching ratio between radical desorption, recombination of diffusion (enabling later chemistry). It is not sufficient to explore the fate of radicals at longer times, i.e. whether they continue to diffuse thermally and how they interact with other radicals in the ice (Fig. \ref{fig:ice_chem}). The main role of MD simulations has therefore been to uncover photochemistry mechanisms and to provide input to other kinds of simulations that are better suited at following the chemistry over longer time scales and involve more than one set of photoproducts/radicals.

\subsection{Monte Carlo Models}

The next level of chemical theory employed to elucidate ice photochemistry is microscopic Monte Carlo modeling. In these precise kinetic models, the motions of individual atoms, radicals and molecules are followed on top of and inside of ice mantles. The models naturally take into account and record the locations of each species within the ice matrix. The models can further treat the surface and bulk binding energies accurately as a function of local binding conditions, typically by assuming a parametric dependence on the kind and number of nearest and next-to-nearest neighbors. The theory and astrochemical applications of microscopic and macroscopic Monte Carlo methods in astrochemistry was recently reviewed\cite{Cuppen13}. 

Important for this review, Microscopic Monte Carlo models provide perhaps the best way to study the diffusion step that connects UV photodissociation with a chemical reaction that is productive, i.e. not simply a recombination of the produced radicals. The past couple of years have seen a number of studies on this topic\cite{Oberg09e,Karssemeijer14,Karssemeijer14b}. The inputs to these models are a surface structure, a list of possible events at any time step and barriers for these events which governs their probability. A very important outcome of these studies is the realization of competition between diffusion and reaction when both have comparable barriers. This has been incorporated into modern rate equation models that can study the photochemistry using larger chemical networks. While the Microscopic Monte Carlo techniques could in theory be used to study ice photochemistry for astrochemical relevant ices, this has not yet been done beyond simulations aimed at exploring water ice formation\cite{Cuppen07}. 

In macroscopic Monte Carlo models,  the master equation for the combined gas-phase and surface
chemistry is solved using Monte Carlo techniques \cite{Tielens82, Charnley98}. These models employ averaged grain-surface reaction rates, and couple them with gas-phase chemistry, producing exact  gas-phase and grain-surface populations for each chemical species, albeit without the structural information inherent in the microscopic models. In terms of computational complexity they span the gap between microscopic Monte Carlo and rate equations, but have so far only incorporated limited ice photochemistry \cite{Vasyunin09,Vasyunin13}.

\subsection{Rate equation models}

Astrochemical models that incorporate comprehensive ice photochemistry almost exclusively use the rate equation approach or some modification of it\cite{Garrod08}. In these models a set of ordinary differential equations that describes the gas-and grain-phase chemical kinetics is solved to produce time dependent abundance information for each chemical species as the gas and ice are exposed to constant or changing physical conditions. These models are responsible for demonstrating that much of the chemical complexity observed in the gas-phase in astrophysical environments, especially around protostars could have their origin in ice photochemistry. The application of the rate equation approach to ice photochemistry in astrophysical environments was recently reviewed\cite{Garrod13b}.

During the past decade rate equation models have evolved in sophistication to compete in accuracy with macroscopic Monte Carlo methods. Modified rate equation models take into account both the competition between diffusion and reaction, and the stochastic nature of grain surface reactions when there are on average less than one reactive species per grain\cite{Caselli98,Garrod09}. A second development is the treatment of ice surfaces separate from ice layers\cite{Hasegawa93}, and so called three-phase models (ice mantle, ice surface and gas-phase) are now standard\cite{Vasyunin13,Garrod13}.

To model ice photochemistry, rate equation models, as well as Monte Carlo-based models, use desorption, diffusion and reaction barriers to determine formation rates of new molecules. These barriers cannot be calculated from first principles for complex or amorphous ice systems and, with a few exceptions, are not constrained experimentally.  Instead modelers use `reasonable' assumptions on the relationship between diffusion and better known desorption barriers, on photodissociation branching ratios, etc. Providing better constraints on these barriers and branching ratios is the key objective of many ice photochemistry studies.

\section{Experimental Methods}

\begin{figure}[htbp]
\begin{center}
\includegraphics[width=5in]{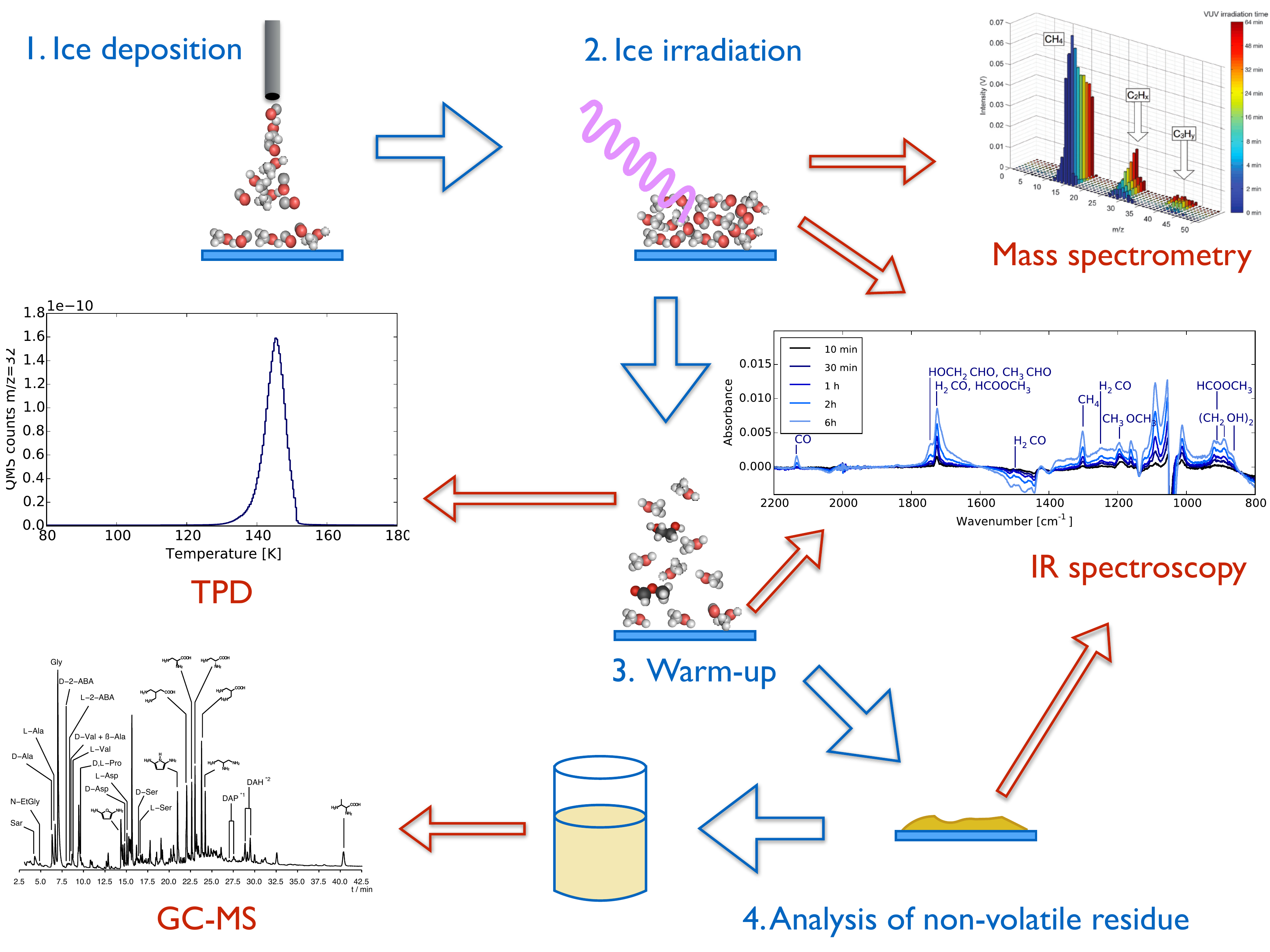}
\caption{Illustration of the different steps involved in a typical ice photochemistry experiment from ice deposition, through irradiation and warm-up, during which the ice chemistry products are analyzed using spectroscopic and mass spectrometric methods, to ice sublimation and residue analysis. The latter step is only part of a subset of ice photochemistry experiments, i.e. many ice photochemistry experiments comprise only {\it in situ} analysis of the ice product composition. Figures illustrating mass spectrometry and GC-MS are reproduced with permission from Ref. 129 (copyright 2015 Royal Society of Chemistry), and Ref. 145 (copyright 2002 Nature Publishing Group).}
\label{fig:exp}
\end{center}
\end{figure}

In the context of astrochemistry, the relevant experimental conditions for ice photochemistry experiments involves a high or ultra-high  vacuum (HV or UHV), cryogenic temperatures similar to those found in molecular clouds, a thin ice or ice mixture, a `reasonable' UV source that can expose the ice film to UV photons {\it in situ}. Spectroscopic and/or mass spectrometric analysis on the ice and/or desorbing molecules are performed during UV exposure and/or after warm-up of the photolyzed ice\cite{Allodi13}. Figure \ref{fig:exp} outlines the different steps of a typical astrochemically motivated ice photolysis experiment.  To achieve the desired experimental conditions and to efficiently analyze the mechanism, kinetics, and products of an ice photochemistry experiments require a combination of experimental techniques. These techniques are the topic of this section.

\subsection{Vacuum Techniques}

\begin{figure}[htbp]
\begin{center}
\includegraphics[width=5in]{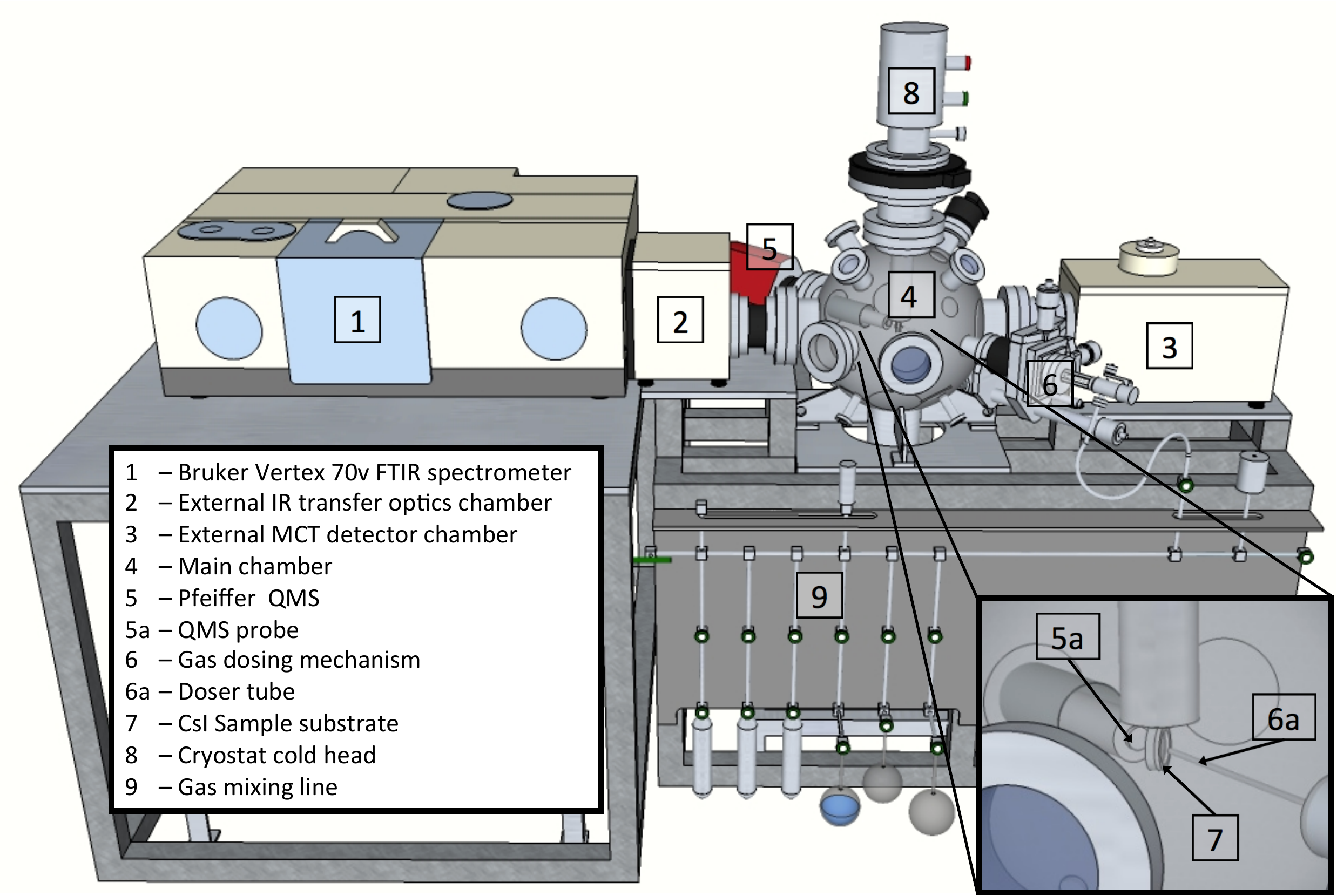}
\caption{An example of an UHV experimental set-up designed for the study of astrochemical ices. Key features include the ice deposition tube connected to a gas-mixing system, the cryogenically cooled substrate where ices are deposited, the infrared spectrometer and the quadropole mass spectrometer. A UV source is connected to one of the open source during photochemistry ice experiments. Reproduced with permission from Ref. 86. Copyright 2015 AAS.}
\label{fig:chamber}
\end{center}
\end{figure}

Fig. \ref{fig:chamber} shows a drawing of an astrochemistry ice experiment with the vacuum chamber at the center\cite{Lauck15}. High or Ultra-High Vacuum (HV and UHV) is required in the astrochemical relevant ice photochemistry experiments to avoid contamination of the initial ice, and reactions between UV photodissociation products. The irradiated  ices in these experiments are kept at low temperatures (10--100~K) to imitate interstellar conditions. At such low temperatures most gases, including N$_2$ and O$_2$, i.e. air freezes out. Without good vacuum conditions it is thus not possible to control the composition of the ice. Ultra-high vacuum (UHV) is especially important for experiments that require thin ices (less than a few 100 monolayers (ML) thick) and/or long irradiation times. HV set-ups are generally only used for experiments that involve thick ices. To achieve ultra-high vacuum ($\sim$10$^{-10}$ mbar) requires a combination of diffusion, turbomolecular and or ionic/gatter pumps, and the possibility to bake-out the vacuum chamber.  A better vacuum than 10$^{-10}$  mbar is typically achieved in UHV set-ups during ice experiments through cryopumping by cooled surfaces. 

\subsection{The Ice Substrate}

Ices are grown in the vacuum chamber through direct deposition of a vapor (mixture) onto a cryo-cooled substrate. The substrate can be constructed from a range of materials and the choice depends on the purpose of the experiment. Experiments that rely on spectroscopy for ice chemistry analysis tend to favor either a (IR) transparent window that enables transmission spectroscopy\cite{MunozCaro10}, or a flat metal surface that enables reflection-absorption spectroscopy\cite{Oberg07b}. Such materials clearly do not mimic interstellar grain surfaces very well and are not suitable for explorations of ice chemistry in the first few ice layers that build up on grain surfaces. For such experiments a silicate or graphite substrate is instead preferred\cite{Vidali06,Noble12,Suhasaria15}. Beyond the substrate-ice interface the exact material of the substrate seems to matter less for most ice photochemistry applications as long as the VUV photons reaching the sample do not produce a significant amount of photoelectrons\cite{Bertin12}. 

In a typical experiment the substrate is cooled down to $\sim$10~K during ice deposition, which requires a combination of a Helium cryostat, comprehensive heat-shielding of the sample, and excellent thermal contact between the substrate holder and the cryofinger. Once the substrate is cooled down, the ice is grown {\it in situ} through vapor deposition. The vapor is guided into the chamber through a thin pipe controlled by a leak valve and connected to a gas sample or gas mixing line (Fig. \ref{fig:chamber}). The angle at which the ice is deposited, the deposition rate and the substrate temperature together determine the resulting ice morphology\cite{Kimmel01,Gerakines15}.  In particular, deposition along the surface normal produces the most compact ices, while deposition at a high inclination and background deposition produce high-porosity ices. 

The composition of the ice is determined by the composition of the vapor that is directed into the deposition tube. Ice mixtures are built up either through co-deposition of two species through separate tubes directed onto the same sample, or by deposition a pre-made vapor mixture. Many ice experiments include intricate gas handling systems to enable depositions of both pure and mixed ices, in separate experiments and in sequence to construct layered ices. 

The ice thickness is regulated by the deposition rate and time. The deposition rate can be (and often is) estimated from the increase in pressure during deposition, through this pressure increase can only be converted to an ice growth rate straightforwardly when the ice is built-up through background dosing. During and after deposition, the ice thickness can also be measured spectroscopically using tabulated band strengths for different ice features. An alternative approach is to use a quartz balance\cite{Westley95}. The thicknesses of thick ices (1000s of MLs) can also be determined directly using laser interference\cite{Hudgins93}. Ice thicknesses in astrochemical motivated ice photochemistry experiments can vary between a few ML and 10s of thousands of layers\cite{Oberg09e,MunozCaro03}. The choice of ice thickness depends on the experimental objective. Thin ices mimic interstellar conditions better -- a typical interstellar ice thickness is $<$100~ML -- and are required for experiments  where both surface and bulk processes are important to quantify. Thick ices are required to measure the formation of very complex molecules that form at trace abundances.

\subsection{UV Light Sources}

Astrochemically motivated ice experiments have employed a range of UV sources, including broad band UV lamps, monochromatic lamps, UV lasers and synchrotron radiation. The choice of UV source depends both on the overall experimental objective, e.g. whether the aim is to simulate the ISM or to elucidate a specific mechanism, and on the required flux. Most experiments have employed lamps or other sources with UV photon energies below 11 eV. This is below the ionization threshold of most small molecules, which typically constitute the initial ice matrix. The ion production is therefore expected to be low compared to the radical production. The dominance of radicals over ions is not necessarily preserved for ices containing larger molecules, e.g. PAHs, or for experiments carried out at synchrotron facilities with access to higher energy photons. 

Broad-band hydrogen and deuterium UV lamps, and Lyman-$\alpha$-dominated hydrogen lamps are the most commonly used UV sources in astrochemistry ice experiments. The lamps are cheap, easy to operate, and produce high fluxes. The precise hydrogen lamp VUV spectra, especially the relative contribution of Lyman-$\alpha$ to the total photon flux, is sensitive to lamp operation conditions\cite{Chen14}. This implies that the lamp spectra can be controlled to e.g. be Lyman-$\alpha$-dominated. Lyman-$\alpha$ emission can also be suppressed  by putting a quartz window between the lamp and the UHV chamber and ice substrate. Perhaps most importantly, hydrogen lamps can be set up to emit a combination of continuum and Lyman-$\alpha$ radiation, which simulates some ISM spectra well. Hydrogen lamps are thus versatile VUV sources. The sensitivity of hydrogen lamp VUV spectra on operating conditions can make inter-lab comparisons difficult, however. 

Monochromatic lamps do not provide realistic interstellar radiation fields, with the exception of Lyman-$\alpha$ lamps. What they provide instead is better control over the ice photochemistry experiments, enabling excitation of specific molecular transitions. Existing monochromatic VUV lamps, based on excimers or a combination of H$_2$/D$_2$ lamp with a monochromator, provide rather low fluxes, however, and as a result most ice photochemistry applications would require very long irradiation times, increasing the level of ice sample contamination by residual gases in the chamber compared to broadband lamp experiments. VUV lasers offers a potentially higher flux monochromatic option. They have been used in ice photocdesorption experiments to measure H$_{2}$O and O($^{3}$P$_{J}$), and benzene photodesorption from water ice (mixtures), and their use could certainly be expanded to more ice photochemistry studies\cite{Thrower08,DeSimone15}. 

Finally, synchrotron beamlines in the VUV provides a monochromatic tunable source of UV photons ideal to study photochemical mechanisms in ices. Exquisite spectral resolution can be reached through the use of gratings, enabling excitations of specific molecular transitions, whilst higher fluences can be obtained by the use of the undulator output only. Several recent studies exist on ice photoabsorption, photodesorption, and photochemistry using the VUV beamline DESIRS at the synchrotron facilities SOLEIL in Saint-Aubin, FRANCE\cite{Fayolle11b,Fayolle13,deMarcellus11,Modica14} and the high-flux beamline at the NSRRC synchrotron facility in Hsinchu, Taiwan\cite{Chen15,Lo15}

\subsection{{\it in situ} Spectroscopy of Ices}

Infrared spectroscopy (IR) of the ice substrate is perhaps the most common analysis tool employed in ice photochemistry experiments. The great advantage of IR spectroscopy (as well as UV/Vis, far-IR and THz spectroscopy) is that the ice composition can be determined non-destructively {\it in situ} as it evolves during radiation and warm-up. UV fluence and time dependencies of a photochemical system can thus be directly constrained. IR spectroscopic measurements are non-invasive, since they do not change the structure or chemical composition of the ice. IR spectroscopy is also flexible in terms of its dynamic range; in transmission mode, IR spectroscopy can be used to study ice coverages from sub-ML to 1000s of layers of ice. Higher sensitivity for the sub-ML coverage regime can be achieved in reflection-absorption mode.

Disadvantages with IR spectroscopy are lack of sensitivity compared to mass spectrometric techniques, and lack of molecule specificity in complex ice mixtures. To mitigate these disadvantages, ice spectroscopy during the radiation and warm-up steps of a photochemistry experiment is generally complemented by {\it in situ} mass spectrometry of sublimated ices and/or {\it ex situ} mass spectrometry of ice photochemistry residues. Another limitation with IR spectroscopy is that it is limited to IR active species. Most abundant interstellar ices are IR active. Indeed that is how their presence in interstellar environments are known\cite{Oberg11c}. There are several suspected ISM ice constituents that are IR-inactive, however, most importantly H$_2$, N$_2$ and O$_2$. A final limitation of IR spectroscopy is that IR spectra and band strengths of unstable species are generally not well known. Direct, quantitive  constraints on radical production rates are therefore difficult to obtain using IR spectroscopy.

An IR spectrometer can be configured to observe an ice in transmission or reflection-absorption mode. The former is less sensitive, but produces spectra that are easier to interpret in terms of translating an absorption signal to an ice column. Reflection absorption IR spectroscopy (RAIRS) is very sensitive, but its response to ice column is only linear in very narrow ice thickness regimes\cite{Teolis07}.  In particular, RAIRS is extremely sensitive to surface orientation, resulting in substantially different IR intensities for molecules interacting with the substrate directly. RAIRS is therefore difficult to use to quantify ice columns. These complications can be mitigated by using angles that result in spectroscopy that is better described as absorption-reflection-absorption, while still enhancing the sensitivity. Each technique (transmission and RAIRS) also presents their own limitations on the substrate material. Whether transmission spectroscopy and RAIRS is selected therefore depends on the application, and especially how crucial it is to derive accurate ice column densities. 

There are two large libraries of ice spectra of astrochemical relevance that report ice band strengths and/or optical constants that  be used to identify and quantify products of ice photochemistry experiments. The Sackler Laboratory Ice Database is located at http://icedb.strw.leidenuniv.nl, and the The Cosmic Ice Laboratory at\\ http://science.gsfc.nasa.gov/691/cosmicice/spectra.html (IR spectra) and\\ http://science.gsfc.nasa.gov/691/cosmicice/constants.html (optical constants). Ice spectra and optical constants are also found in a large number of astrochemical papers\cite{Hudgins93,Gerakines95,Ehrenfreund97,vanBroekhuizen06,Moore07,Oberg07a}.

In addition to IR absorption spectroscopy, Raman, VUV, UV-Vis spectroscopy, far-IR and THz time-domain spectroscopy have all been used to characterize ice compositions in astrochemistry experiments\cite{Ferini04,Bouwman09,Allodi13}. Raman spectroscopy could be used to study the handful of astrochemically important molecules that are IR-inactive. It has not been applied to any ice photochemistry experiments, but used successfully in other ice studies\cite{Bennett13}. Most ices absorb at VUV wavelengths and the resulting spectra are sensitive to the local ice environment\cite{Cruz-Diaz14a} and can be used for kinetic photochemistry studies\cite{Jones14}. UV/Vis has mainly been used to explore the photochemistry of polycyclic aromatic hydrocarbons (PAHs) in ice matrices; by contrast to most smaller molecules, PAHs present strong absorption bands at UV/Vis wavelengths. Small radicals often absorb at these wavelengths as well, and this technique could thus be applied to ice photochemistry experiments more generally to characterize radical production rates and unravel dissociation and reaction pathways\cite{Grim87}. Far-IR and THz spectroscopy also has great potential to probe ice compositions and structures \cite{Hudgins93,Ioppolo14}, though it has yet to be applied to study ice photochemistry.

\subsection{{\it in situ} Mass Spectrometric Analysis Methods}

Mass spectrometry is used in ice experiments to monitor vacuum quality, ice deposition and sublimation, and to detect and quantify new product during and after a photochemistry experiment. Most ice experiments incorporate a quadropole mass spectrometer (QMS) for monitoring of the background gas composition and for temperature programmed desorption (TPD) studies. TPD experiments provide a simple and sensitive approach to analyze the final products of ice photochemistry experiments. In TPD experiments, the substrate is heated using a linear temperature ramp (Fig. \ref{fig:exp})\cite{Attard04}. The gas phase composition is monitored as a function of time (and temperature) with the QMS. The observed desorption curve of a species depends on the number of molecules on the ice, the ice structure, the desorption order, the desorption attempt frequency and the desorption barrier. If reference TPD curves are available and the target molecule fractionation pattern in the QMS is known it is often possible to uniquely identify and quantify a desorbing species based on the shape and location (in time) of the desorption curve and the detected mass fragment pattern\cite{Oberg09d}. It is important that these reference TPDs represent the relevant ice environment from which the molecule in question is desorbing. Binding energies depend on the local ice environment\cite{Collings04} and reference TPD curves should thus be recorded using the target molecule initially embedded in the dominant ice matrix species of the ice photochemistry experiment.  

Most TPD set-ups are limited to analyze moderately complex ice mixtures, where the fragmentation patterns of individual molecules can be distinguished.
Detections of more complex species in TPD studies have been made using highly sensitive reflectron Time-of-Flight (ToF) mass spectrometry (ReTOF) coupled with single photon fragment-free photo ionization (PI) at 10.49 eV\cite{Maity14}. This elegant, but experimentally more challenging technique has been applied electron-irradiated ices, but not yet to ice photochemistry experiments. 

An intrinsic limitation of TPD experiments is that they are limited to analyze the final ice composition at the time and temperature of desorption. This may be quite different from the composition at the end of irradiation at lower temperatures, and provides no information on how the chemical composition evolves with time during UV exposure. This limitation has been addressed by coupling non-thermal desorption techniques with mass spectrometry that employs a suitable ionization source (or none at all).  In such experiments energetic particles or photons are used to desorb some of the ice at low temperatures, and the resulting gas molecules are analyzed using the mass spectrometer. Reactive ion scattering (RIS) and low-energy sputtering (LES) were applied to ice photochemistry experiments and resulted in the successful detection of glycine production {\it in situ}\cite{Lee09}. In these experiments a low-energy Cs$^+$ beam collides with the ice substrate and scatters ions and neutrals from the surface layer that are detected and mass analyzed using a QMS with its ionizer filament switched off. In a different set of experiments a two-color laser-desorption laser-ionization time-of-flight mass spectroscopic method (2C-MALDI-TOF) was developed to analyze irradiated ice products\cite{Gudipati12,Henderson15}. A different laser desorption ToF mass spectrometric method was recently developed and used to analyze the photoproducts in a hydrocarbon ice photochemistry experiment\cite{Paardekooper14}. These experiments are important developments, enabling low-temperature {\it in situ} analysis of larger ice molecules than is possible using IR spectroscopy. It is important to note, however, that laser ablation and (laser) ionization mass spectrometry provides limited quantitative information of the solid under investigation due to signal saturation (of abundant species) and varying ionization cross-sections of species at any given wavelength. 

Temperature programmed desorption or non-thermal desorption studies of ice can be combined with resonance-enhanced multiphoton ionization
(REMPI) to obtain information about the rotational and vibrational states of the desorbed ice molecules, providing important constraints about the formation and desorption mechanisms of target molecules\cite{Fukutani05,Yabushita09,Hama11,Gudipati12}.

\subsection{Post-processing Analysis Methods}

The largest and most complex products of ice photochemistry experiments are so far nor possible to identify and quantify {\it in situ} due to a combination of very low abundances and low volatility. These photoproducts form a residue on the substrate following sublimation of the initial ice and any volatile photoproducts. The residue is extracted and analyzed {\it ex situ} using gas chromatography
mass spectrometry (GC-MS) and other high sensitivity mass spectrometric techniques, and/or high-precision liquid chromatography (HPLC)\cite{MunozCaro02,Bernstein02,Chen08,Danger13}.

The first step, the residue extraction is done by gently scraping the substrate or by dissolving the residue while still on the substrate. In either case, the extracted residue is dissolved in some combination of water and organic solvents. In most experiments the dissolved residue is then hydrolyzed using HCl. The resulting material is then dried and injected into the GC-MS, or redissolved and injected into an HPLC. In some experiments there is an additional derivatization step between hydrolysis and analysis.

These sequence of sample treatments is necessary to detect very small amounts of highly refractory organics, and especially to release amino acids from polymers. The required heat-up to room temperature and hydrolysis steps makes it difficult to ascertain, however, when during the ice experiment the target molecule formed, i.e. during irradiation, during warm-up, or during extraction. A recent study has also shown that substantial chemical degradation may occur during the heat-up in the GC-MS itself, further complicating interpretation of these experiments\cite{Fang15}

\section{Ice Photodissociation Rates and Cross Sections}

The most fundamental step of any photochemistry experiment is the photodissociation of a parent molecule into more reactive species. The total photodissociation rate or cross section is the topic of this section. The dissociation branching ratio and the energy-dependence of both cross sections and branching ratios are treated in the subsequent section. 

There is surprisingly little data on ice photodissociation rates and cross sections in the literature, and astrochemical models depend on either gas-phase photodissociation cross sections, or laboratory data acquired in a handful of experiments over the past two decades. As demonstrated below, the former approach is likely to overestimate radical production in interstellar ices because it does not account for the cage effect in ices\cite{Laffon10}. Following a photodissociation event in an ice the dissociation fragments have a non-negligible  probability to reform the parent molecule, since diffusion away from the dissociation site is inhibited by surrounding molecules. Measured ice photodissociation cross sections are therefore `effective' cross sections, which implicitly take into account that some photodissociation events do not result in a net loss of the parent molecule.

\begin{figure}[htbp]
\begin{center}
\includegraphics[width=5in]{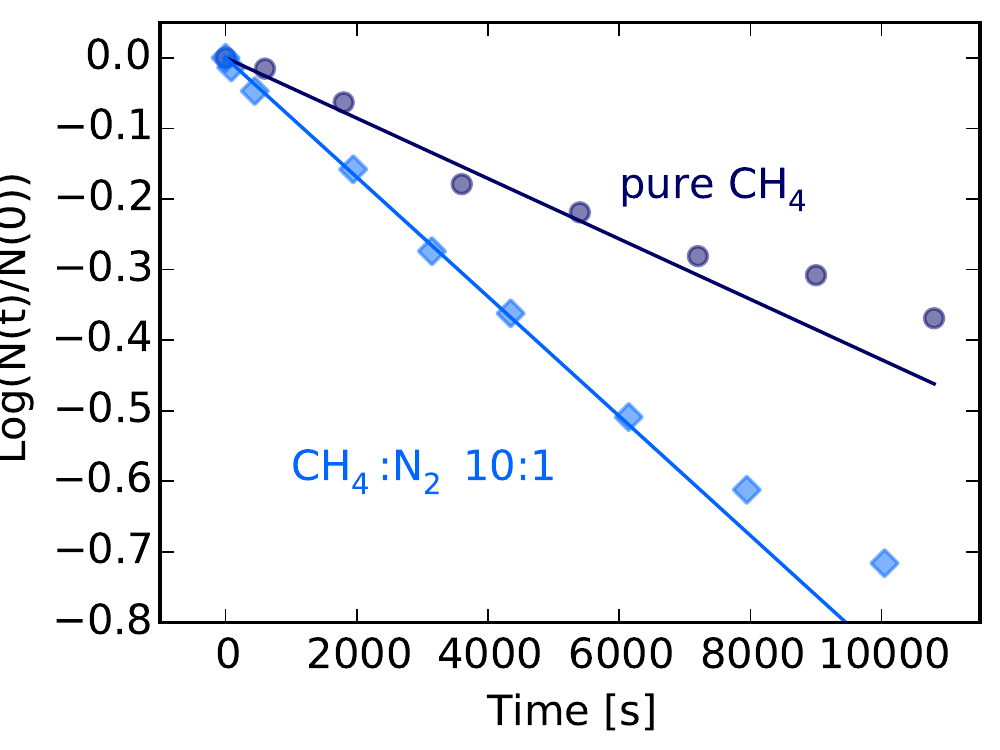}
\caption{Photodissociation time series of pure CH$_4$ ice and CH$_4$ embedded in a N$_2$ matrix\cite{Cottin03}. The circles and diamonds are experimental data. The lines mark the expected $N(t)$ when fitting Eq. 1. to the first few experimental points. Adapted with permission from Ref. 135. Copyright 2003 AAS.}
\label{fig:pd}
\end{center}
\end{figure}

The UV destruction of an optically thin sample at early times, when the formation of the original molecule through radical reactions is negligible (except for the fast recombination at the site of dissociation due to the cage effect), is described by

\begin{equation}
N(t) = N(0){\rm exp}(-\phi\times t\times\sigma_{ph}),
\end{equation}

where N is the abundances of the species (in molecules cm$^{-2}$), N(0) is the initial abundances, $\sigma_{ph}$ the photolysis cross section (in units of cm$^2$), $\phi$ the flux of UV photons (in units of photons cm$^{-2}$ s$^{-1}$), and t is the irradiation time (in s). $\sigma_{ph}$ can then be determined by measuring the decay of the abundance of a species as a function of time and UV flux.  The column of an ice is most straightforwardly measured using infrared spectroscopy. The optical depth or absorption of an ice feature is proportional to the ice column density (though it also depends on the ice structure and composition), and a reduction in absorption  as a function of time can be used  together with tabulated ice band strengths to determine $N(t)$. The time series can be fit by Eq. 1 to extract the photodissociation cross section. Figure \ref{fig:pd} shows photodissociation curves of CH$_4$ ice and two best fit photodissociation cross sections.

The first extensive survey of pure ice photolysis was carried out on pure H$_2$O, NH$_3$, CH$_4$, CO, CO$_2$, O$_2$, N$_2$, H$_2$CO, and CH$_3$OH ices that were exposed o a hydrogen discharge lamp at 10~K\cite{Gerakines96}. The lamp flux was estimated to $\sim$10$^{15}$ photons cm$^{-2}$ at the substrate and the lamp spectra consisted of five bands centered on 1220 (Lyman-$\alpha$), 1360, 1450, 1600, and 2800 \AA. Important for the evaluation of these experiments, the ices were  $\sim$0.1$\mu$m thick. This is thick enough that not all of the ice will receive an equal dose of UV radiation due to substantial absorption of VUV photons in the upper layers of the ice. The case is especially severe for water, NH$_3$, H$_2$CO and CH$_3$OH, for which 90\% of all photons may have been absorbed in the topmost $\sim$0.1$\mu$m ice, based on measured gas and (water) ice photodissociation cross sections between 115 and 155nm\cite{Cottin03}. Newer experiments demonstrate, consistent with previous estimates, that 95\% of all UV photons are absorbed in the top 300-800 ML of such ices\cite{Cruz-Diaz14a}. The optically thin limit is thus not guaranteed in these experiments. The authors still used the equation for optically thin ices and obtained photodissociation cross sections for CH$_4$, CO$_2$, H$_2$CO and CH$_3$OH and an upper limit for CO photodissociation (Table \ref{tab:pd}). 

The obtained ice photolysis cross sections can be compared with gas-phase photodissociation cross section at Lyman-$\alpha$\cite{vanDishoeck06}. Table \ref{tab:pd} shows that the reported ice photodissociation cross sections are more than an order of magnitude lower compared to the corresponding gas phase cross sections, except for CO$_2$, which has a very low Lyman-$\alpha$ cross section in the gas-phase. The ice CO$_2$ photodissociation cross section is still low compared to what would be expected for the employed VUV lamp, since CO$_2$ gas has a large photodissociation cross section in other parts of the VUV spectrum\cite{vanDishoeck06}. There are several potential explanations for why the measured ice phase photodissociation cross sections are so low: 1. the UV flux in the ice experiments was overestimated (VUV lamp calibrations are notoriously uncertain), 2. the ices are partially optically thick to VUV radiation, 3. there is an intrinsic difference in UV absorption cross sections between gas and ice, 4. there is an intrinsic difference in the dissociation quantum yield per absorbed photon in gas and ice, and 5. a majority of produced dissociation fragments in the ice combine quickly to reform the parent molecule (the cage effect). 

\begin{table}[htdp]
\caption{Photodissociation cross sections of common interstellar ice mantle species, in the gas phase (at Lyman-$\alpha$) and in ice using a hydrogen discharge lamp.}
\begin{center}
\begin{tabular}{l c c c c c }
\hline\hline
Species 	&Gas-phase Ly-$\alpha$\cite{vanDishoeck06}	&Pure ice 10--20~K		&X:N$_2$ 10~K$^{a}$	
&X:H$_2$O 10--20~K$^{a}$ \\
&[10$^{-19}$ cm$^{2}$]		&[10$^{-19}$ cm$^{2}$]	&[10$^{-19}$ cm$^{2}$]&[10$^{-19}$ cm$^{2}$]\\
\hline
H$_2$O 	&120	&\\
NH$_3$ 	&100	&0.32[0.10]\cite{Cottin03}, 14[7]\cite{Oberg10b}	&2.3[0.8]\cite{Cottin03}		&3.2[1.0]\cite{Cottin03}, 50[25]\cite{Oberg10b}\\
CH$_4$ 	&180	&7.2\cite{Gerakines96}, 0.91[0.28]\cite{Cottin03}, 5.0[3.0]\cite{Oberg10b}	&1.8[0.5]\cite{Cottin03}		&1.5[0.5]\cite{Cottin03}, 39[20]\cite{Oberg10b}\\
CO & --							&$<$0.8\cite{Gerakines96}, $<$0.01\cite{Cottin03}\\
CO$_2$ 	&0.6	&5.6\cite{Gerakines96}, 3.8[1.6]\cite{Cottin03}	&2.6[0.8]\cite{Cottin03}	&3.3[1.2]\cite{Cottin03}\\
N$_2$ 	&--				&\\
H$_2$CO &100			&6.2\cite{Gerakines96}\\
CH$_3$OH&140	&5.0[2.0]\cite{Cottin03}, 26[9]\cite{Oberg09d}	&15[6]\cite{Cottin03}		&2.7[0.9]\cite{Cottin03}\\
\hline
\end{tabular}
$^{a}$The reported cross sections are for X, in a X:Y mixture where the mixing ratio varies between 1:4 and 1:10 for the different experimental studies.
\end{center}
\label{tab:pd}
\end{table}%

The first concern is addressed by a second large survey of astrochemically relevant ice photodissociation cross sections targeting CH$_4$, CH$_3$OH, NH$_3$, CO$_2$, CO and HNCO in pure form and mixed with N$_2$ and H$_2$O in 1:10 proportions\cite{Cottin03}. In this study, the flux of the VUV hydrogen discharge lamp was well-calibrated using photodiode measurements. Similarly to previous studies, the ices were thick ($>$0.1$\mu$m). Following a kinetic analysis the authors still treated the ices as optically thin. Three ices were part of both surveys and among these the results are consistent for CO and CO$_2$, but not for CH$_4$. Regardless of the origin of the CH$_4$ discrepancy it is clear that the low ice photolysis cross sections found in the original study cannot be generally explained by an overestimate of VUV flux in the ice experiments.

Three of the above ices (CH$_3$OH, CH$_4$ and NH$_3$)\cite{Gerakines96,Cottin03} have also been photolyzed in the thin ice regime (with ice thicknesses of 10--20~ML), i.e. in experiments where optically thin ices are guaranteed\cite{Oberg09d,Oberg10b}. The resulting photodissociation cross sections are consistently higher compared to the thicker ice experiments (Table \ref{tab:pd}). The cross sections are still approximately one order of magnitude lower than corresponding gas-phase photodissociation cross sections, however. Ice opacity alone cannot thus explain differences between literature gas and ice cross sections.

The third potential explanation, an intrinsic difference in gas and ice VUV absorption cross sections, has been addressed by a number of studies on UV absorption spectra of ices\cite{Baratta02,Mason06,Cruz-Diaz14a,Cruz-Diaz14b,Cruz-Diaz14c}. VUV photo-absorption spectra of 20--25~K amorphous ices of H$_2$O, CO, CO$_2$, NH$_3$ show that the spectral shapes are substantially different in the condensed phase (shifted and broadened) compared to the gas-phase. However, depth dependent photolysis experiments result in absorption coefficients for H$_2$O, CH$_4$ and CH$_3$OH that are of the same order of magnitude for ices as for gas at the position of Lyman-$\alpha$\cite{Baratta02}. More recently measurements of the absorption cross sections across the VUV range for CO, H$_2$O, CH$_3$OH, NH$_3$ ices (and additional species) at 8~K demonstrated that the VUV-integrated photoabsorption cross sections for all ices are at most 50\% different from the gas-phase values\cite{Cruz-Diaz14a,Cruz-Diaz14b}, i.e. there are no order-of-magnitude differences between gas and ice VUV absorption cross sections. 

With the first three potential explanations ruled out, the two remaining ones are a reduced photodissociation quantum yield per absorbed photon, and fast recombination of radicals in the ice. The latter explanation can be tested experimentally, since the importance of fast radical recombinations upon dissociation should be reduced in ice matrices where diffusion is easier, and thus the cage effect is reduced. For example, warmer ices should present larger photodissociation cross sections than colder ices, and molecules embedded in matrices where diffusion is easier should present larger photodissociation cross sections compared to molecules in more strongly bound ices. Experiments on the CH$_3$OH photodissociation cross section at four temperatures between 20 and 75~K shows that the cross section increases with temperature such that it is a factor of two larger at 75~K compared to 20~K\cite{Oberg09d}. Table \ref{tab:pd} also shows that when molecules are placed in a comparatively loosely bound N$_2$ matrix, the photodissociation cross section increases by factors of 2--8 compared to the values measured for pure ices, with the exception of CO$_2$, which is unaffected\cite{Cottin03}. The measured photodissociation cross sections are also increased if molecules are placed H$_2$O ice matrices\cite{Cottin03,Oberg10b}, but this is probably a chemical effect, i.e. dissociation products have more potential reaction partners that do not result in a reformation of the parent molecule\cite{Laffon10}. Taken together, these experiments are all suggestive of that photodissociation in ice and gas are of the same order of magnitude, but that the effective cross sections are much lower due to fast back-reactions in pure ices. 

There is a second family of photodissociation studies whose main concern is the survival time scale of different prebiotic molecules in astrophysical environments\cite{Ehrenfreund01,Bernstein04}. UV irradiation experiments on amino acids embedded in Ar, N$_2$ and H$_2$O ice showed that, similar to smaller organics, amino acids have a reduced photodissociation cross section in ice matrices compared to the gas. The exact nature of the ice matrix affected the photodissociation cross section by up to a factor of 3. In general organic acids  appear particularly unstable to VUV radiation compared to other, closely related organic molecules, such as nitriles of similar sizes\cite{Bernstein04}.

\section{Pure Ice Photochemistry}

In the ISM, ices are always to some extent mixed. Experiments on pure ices thus do not mimic interstellar ice chemistry very well. Such experiments can still be extremely useful, however, to unravel the mechanisms through which ice photochemistry precedes, since interpretation is facilitated by having a single parent molecule. This section reviews photochemistry experiments on pure ices and the constraints they have provided on the photodissociation cross sections and energy barriers that regulate ice photochemistry in both laboratory and interstellar settings. 

This section is organized as follows. \S5.1 introduces the experiments and the data that the subsequent analysis is based on. \S5.2 reviews how spectroscopic time series during UV irradiation can be used to deduce different generations of photoproducts. \S5.3 discusses constraints on photodissociation branching ratios of CH$_4$ and CH$_3$OH ice from product composition analysis. \S5.4 finally treats the diffusion of radicals in ice and what qualitative constraints can be obtained directly from product composition time series during irradiation at fixed temperatures, and during warm-up after the UV source has been turned off. 

\subsection{Description of Experiments}

\begin{figure}[htbp]
\begin{center}
\includegraphics[width=6in]{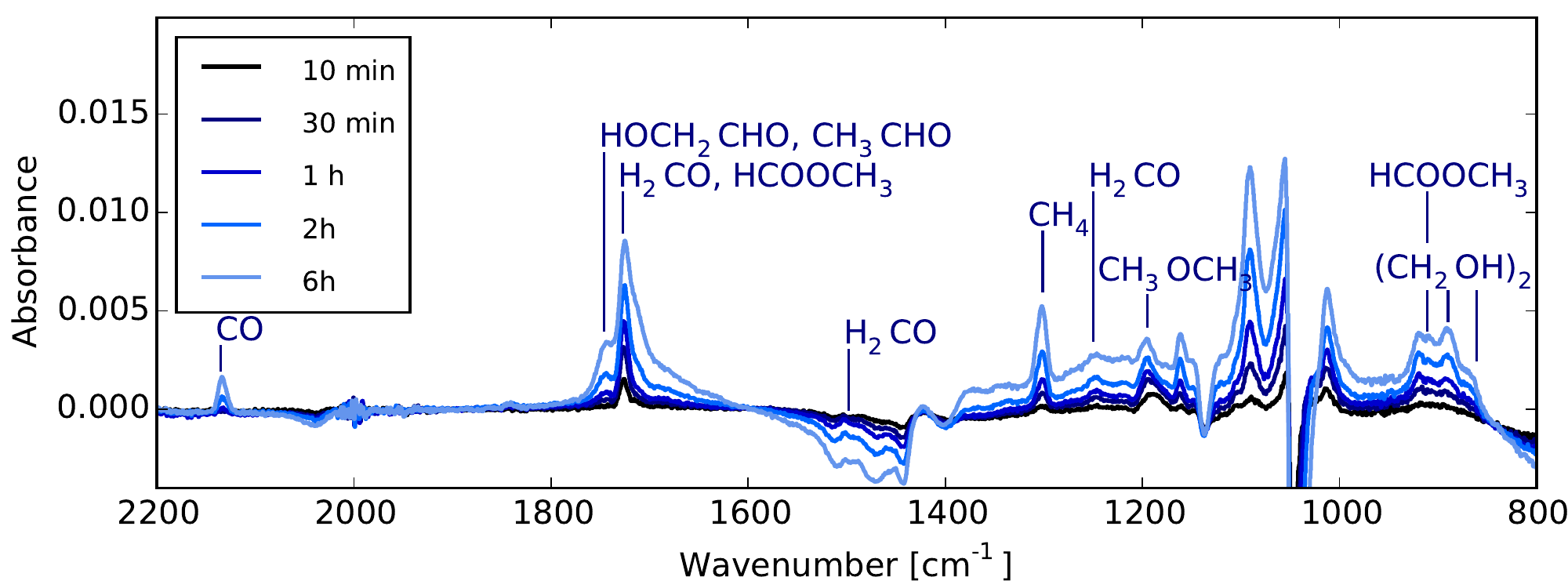}
\caption{Difference spectra (i.e. spectra at a time $t$ $-$ spectra at $t=0$) of photolyzed pure CH$_3$OH ice at 50~K after different VUV exposure times\cite{Oberg09d}. Some of the identified photoproducts have been marked and growth of their spectral features as a function of irradiation time are clearly observed.}
\label{fig:pd_spec}
\end{center}
\end{figure}

The formation kinetics of ice photoproducts are generally characterized using abundance time series, where the amount of product is measured as a function of time and UV flux. Similarly to photodissociation cross sections, most constraints on the kinetics or formation rates of photoproducts are extracted from time series of infrared spectra. Figure \ref{fig:pd_spec} shows a example of a time series of spectra following UV irradiation of a thin (25~ML) CH$_3$OH ice at 50~K. Similar time series have been reported for UV irradiation of pure H$_2$O, NH$_3$, CH$_4$, CO, CO$_2$, O$_2$, N$_2$, H$_2$CO, and CH$_3$OH ice\cite{Gerakines96}, CH$_3$NH$_2$ ice \cite{Moon08}, CH$_3$OH ice\cite{Oberg09d}, CH$_4$ ice\cite{Bossa15}, CH$_3$NO$_2$ ice\cite{Maksyutenko15} and HOCH$_2$CN ice\cite{Danger13}.

As Fig. \ref{fig:pd_spec} demonstrates, even pure ice photolysis can result in complicated IR spectra within minutes of the onset of  VUV irradiation at moderate fluxes ($\sim$10$^{13}$ cm$^{-2}$ s$^{-1}$). Based on the IR spectra, UV irradiation of low-temperature (20-70~K) CH$_3$OH ices results in increasing amounts of CO, H$_2$CO, CH$_4$, CH$_3$OCH$_3$, HCOOCH$_3$, (CH$_2$OH)$_2$ and other complex organics with time. For thin ices, the integrated line absorptions or optical depths are proportional to the ice column density. If band strengths are known these kind of time series can be used to determine column densities of photoproducts as a function of irradiation time and further as a function of UV fluence (if the VUV flux is known). Such growth curves can then be used to characterize the kinetics of the photochemistry at different experimental times. It is important to note that to obtain quantitative data on the rate coefficients and energy barriers that regulate the ice chemistry, the complete ice system needs to be modeled quantitatively. This is a complex computational problem, which has not yet been solved, but for a handful of very simple systems. Most constraints reported in this review will therefore be more qualitative in nature.

\subsection{Photochemical Generations and Basic Kinetics}

Qualitatively, different photoproduct formation kinetics are expected at early times, when the chemistry is dominated by photodissociation of the original ice and radical reactions between the first generation of dissociation fragments, and at later times, when dissociation of first-generation photoproducts is non-negligible. Different kinds of growth curves are expected for molecules that form from direct photodissociation of the original ice (e.g. H$_2$CO from CH$_3$OH) and molecules that require the photolysis of two ice molecules to form. The former should display first order kinetics, and the latter second order kinetics. A potential example of the latter is CH$_4$ formation in CH$_3$OH ice: CH$_3$OH + h$\nu$ $\rightarrow$ CH$_3$ + OH and CH$_3$OH + h$\nu$ $\rightarrow$ CH$_2$OH + H, followed by CH$_3$+H$\rightarrow$CH$_4$, though {\it a priori} direct photodissociation of CH$_3$OH to CH$_4$ cannot be excluded. Molecules that form from dissociation of first and second generation photoproducts should display a yet different growth curve with a sharply increasing yield with time. For example CO in CH$_3$OH photolysis experiments requires multiple photodissociations of a single molecule to form: CH$_3$OH $\rightarrow$ H$_2$CO $\rightarrow$ HCO $\rightarrow$ CO. Finally different growth curves (and temperature dependencies) are expected for molecules that depend on diffusion of atoms, radicals and molecules other than H atoms. H atom diffusion is expected to be fast at most relevant ice temperature and thus not a rate limiting step, while diffusion of radicals and heavier atoms is expected to be slow below 30~K, and then become successively easier in warmer ices.

\begin{figure}[htbp]
\begin{center}
\includegraphics[width=5in]{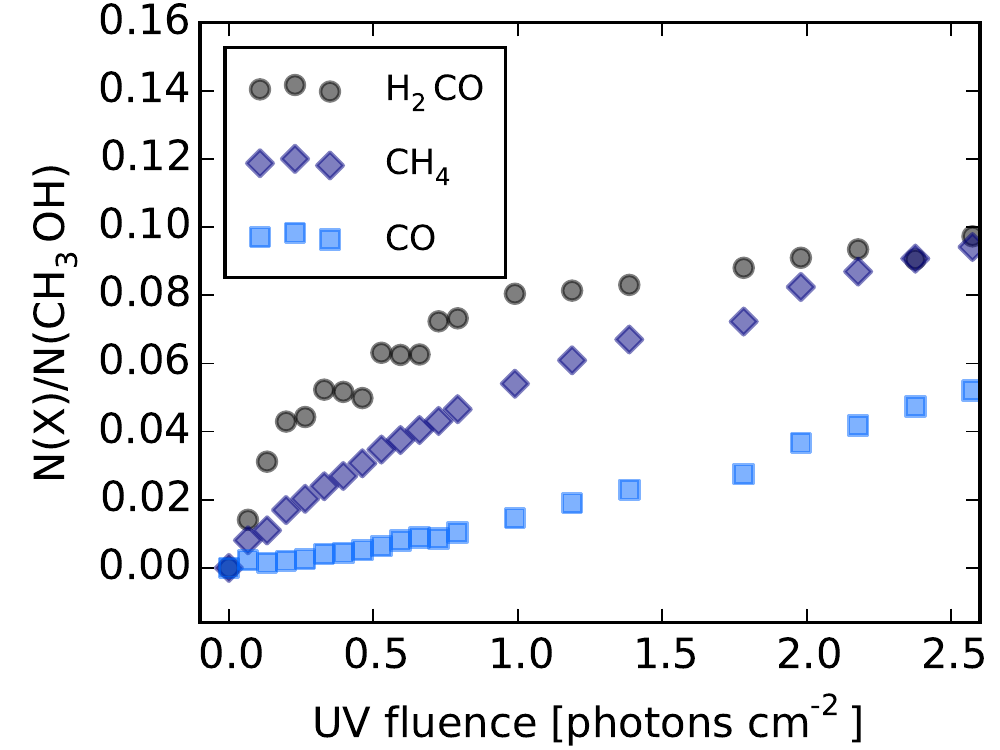}
\caption{Growth curves of H$_2$CO (black circles), CH$_4$ (navy diamonds) and CO (blue squares) during irradiation of a pure CH$_3$OH ice at 20~K\cite{Oberg09d}. The abundances are presented as fractions of the initial CH$_3$OH ice column density.}
\label{fig:pd_kin}
\end{center}
\end{figure}

Figure \ref{fig:pd_kin} shows that small molecules with different formation kinetics do indeed present different growth curves. During irradiation of a 20~K CH$_3$OH ice, H$_2$CO production is very fast at early times, and then levels off fairly quickly as a balance between CH$_3$OH and H$_2$CO photodissociation is reached. This is the expected growth curve for a species that is both formed and destroyed through a first order process. CH$_4$ production is initially much slower, but catches up with H$_2$CO toward the end of the experiment as would be expected for a species that forms through second order kinetics and is mainly destroyed through photodissociation. Finally, CO formation is initially very slow due to lack of first and second generation products at early times, and speeds up at later times as larger abundances of CO precursors (HCO and H$_2$CO) become available (Fig. \ref{fig:reac_dia}).

A similar, but more extreme diversity in the formation kinetics of small molecules has been seen in other experiments that incorporate higher UV fluxes\cite{Gerakines96}. In the case of H$_2$O photodissociation, an immediate formation of OH (photodissociation product) is observed, while the formation of H$_2$O$_2$ (first generation photoproduct) is delayed and HO$_2$ (second generation photoproduct) even more so. During methane ice irradiation ethane forms early, while propane and ethylene follow after longer UV exposures. During CO$_2$ ice photolysis, CO and CO$_3$ are observed immediately after onset of irradiation while O$_3$ requires a large VUV fluence to form at observable abundances, and its formation is coincident with CO$_3$ destruction.

A recent clear example of the different kinetics of first, second and third generation products comes from the study of CH$_4$ ice photochemistry, and especially  the formation kinetics of CH$_3$CH$_3$, CH$_2$CH$_2$ and HCCH\cite{Bossa15}. Though all three molecules can form through a mixture of reactions that involve first, and later generation photodissociation products, the different shapes of their growth curves suggest that the less saturated hydrocarbons are later generations of photoproducts. Another recent example is the photolysis of CH$_3$NO$_2$ where H$_2$CO precedes CO formation, which in its turn precedes CO$_2$ formation, all with different growth curve shapes, in accordance with theoretical expectations\cite{Maksyutenko15}.

\subsection{Photodissociation Branching Ratios}

\begin{figure}[htbp]
\begin{center}
\includegraphics[width=6in]{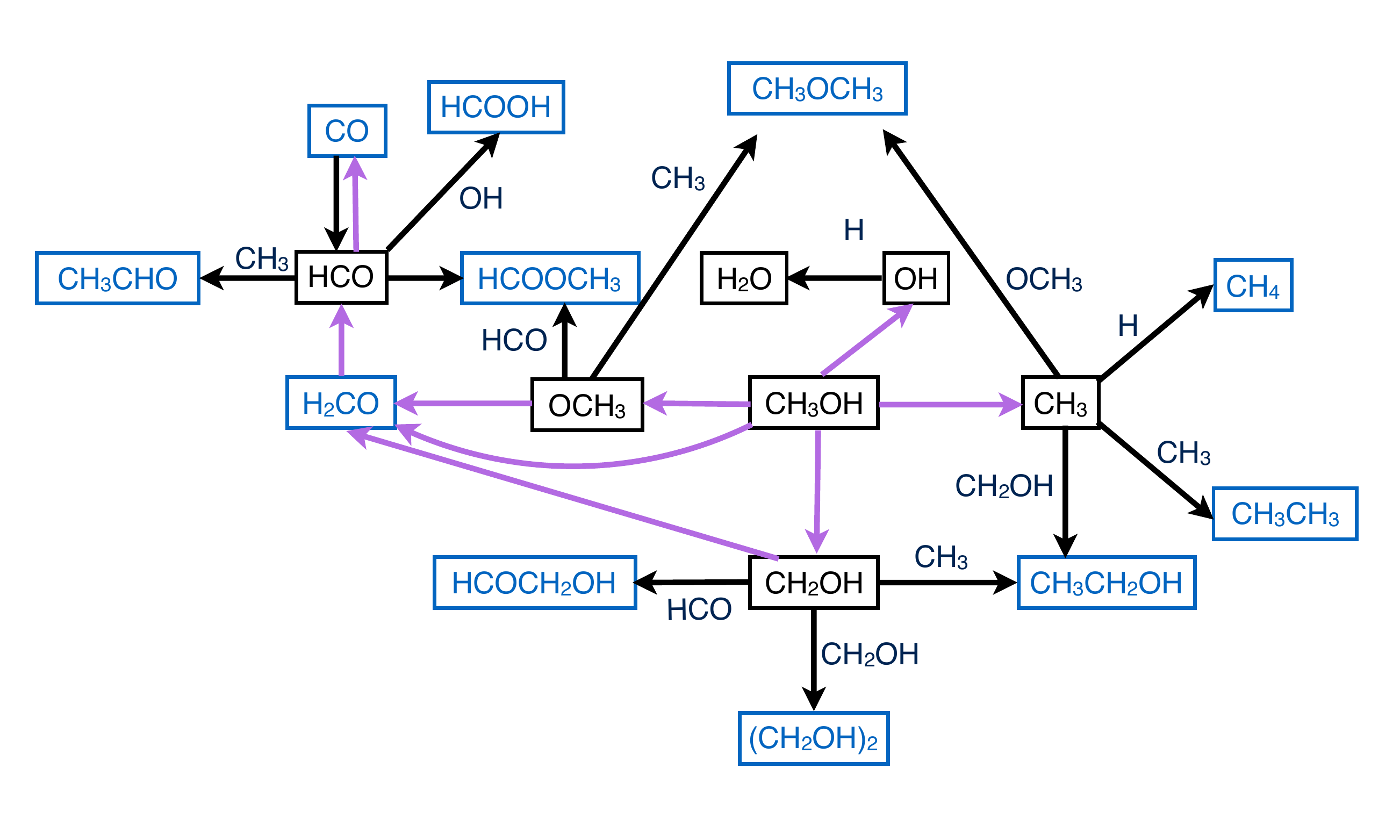}
\caption{Proposed reaction diagram for CH$_3$OH photochemistry displaying major dissociation and radical combination reactions. Note that other classes of reactions that may be active in the ice, including insertion and hydrogen abstraction reactions are not shown. Photodissociation steps are marked with purple arrows. Radical reaction steps are indicated with black arrows and the reaction partner is listed. The chemistry is regulated by a combination of the total (VUV energy dependent) photodissociation cross section, the radical branching ratio during photolysis, the absolute and relative diffusion rates of radicals in the ice, and how these rates compare to one another. Adapted with permission from Ref. 119. Copyright 2009 EDP Sciences.
\label{fig:reac_dia}}
\end{center}
\end{figure}

A key mechanistic and kinetic question to answer when analyzing the experiments presented above (and thus assessing how to extrapolate them to astrophysical settings) is what radicals a particular molecule dissociates to when absorbing a photon of a particular energy while residing in a particular ice matrix. The resulting photodissociation branching ratio will determine the relative production rates of radicals in the ice and thus which products can form in abundance (Fig. \ref{fig:reac_dia}). In this sub-section we focus on how the product composition in VUV lamp photolysis experiments can be used to elucidate the photodissociation branching ratios of the parent molecule. 

In pure ice photolysis the product composition at early experimental times (when all radicals can be assumed to be generated from dissociation of the original ice) depend on a combination of photodissociation branching ratios and radical-specific diffusion barriers. Together these characteristics determine the formation and combination rate of reactants into new molecules. Isolating the effects of radical production from radical diffusion on the final product composition can be challenging. The general approach has been to consider either radicals with similar (expected) diffusion characteristics, or creating an ice environment where the importance of differences in diffusion barriers is minimized by e.g. increasing the concentration of key radicals. 

\begin{figure}[htbp]
\begin{center}
\includegraphics[width=6in]{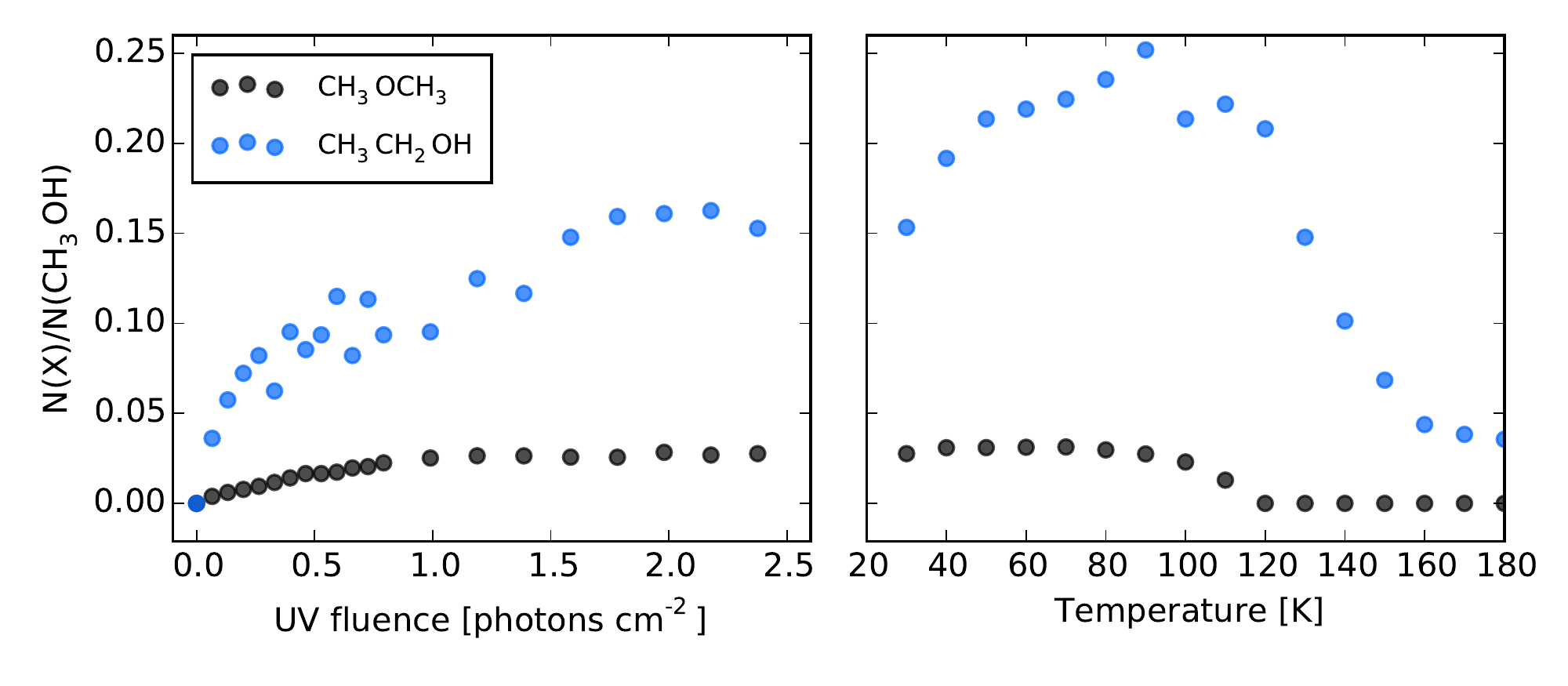}
\caption{Growth curves of CH$_3$OCH$_3$ and CH$_3$CH$_2$OH during VUV irradiation of a CH$_3$OH:CH$_4$ ice mixture at 30~K (left panel) and during warm-up of the same ice mixture after VUV irradiation has been turned off. The abundances are presented as fractions of the initial CH$_3$OH ice column density\cite{Oberg09d}.}
\label{fig:pd_radical}
\end{center}
\end{figure}

This general analysis approach was first applied to photolysis of pure CH$_3$OH ice to constrain the relative production rates of the OCH$_3$ and CH$_2$OH radicals from CH$_3$OH ice photolysis. In the pure CH$_3$OH ice photolysis experiments it was noted that CH$_3$CH$_2$OH consistently formed at a higher rate compared to CH$_3$OCH$_3$ regardless of the ice temperature or photon flux. The proposed main formation pathways of CH$_3$OCH$_3$ and CH$_3$CH$_2$OH are CH$_3$+OCH$_3$ and CH$_3$+CH$_2$OH\cite{Garrod08}. As a first approximation the relative production rates of CH$_3$OCH$_3$ and CH$_3$CH$_2$OH during CH$_3$OH ice photolysis should thus depend on the relative production rates and diffusion barriers of CH$_3$O and CH$_2$OH. The experiment was repeated using a CH$_3$OH:CH$_4$ 1:1 ice mixture to create an overabundance of CH$_3$ radicals in the ice and thus minimize the importance of diffusion for the production of the two complex molecules. The relative production rates of CH$_3$OCH$_3$ and CH$_3$CH$_2$OH should then be regulated by the relative abundances of the CH$_2$OH and OCH$_3$ radicals in the ice. Figure \ref{fig:pd_radical} presents the resulting formation of CH$_3$OCH$_3$ and CH$_3$CH$_2$OH during VUV irradiation of the ice mixture. CH$_3$CH$_2$OH forms at 5x higher rate than CH$_3$OCH$_3$, indicative of that during CH$_3$OH photolysis the effective branching ratio of CH$_2$OH over OCH$_3$ is $\sim$5. This is very different compared to gas-phase dissociation branching ratio\cite{Lucas15} and it is currently unclear how matrix dependent this branching ratio is. 

The CH$_4$ photodissociation branching ratio has been similarly determined by measuring the relative yields of different carbon chains in CH$_4$ ice photochemistry experiments\cite{Oberg10b,Bossa15}. Based on infrared spectroscopy of the ice during irradiation, pure CH$_4$ ice photolyzes into C$_2$H$_6$ and C$_2$H$_4$ at a ratio of $\sim$9 at early experimental times. At this early time C$_2$H$_6$ should form pre-dominantly from CH$_3$+CH$_3$ and C$_2$H$_4$ from CH$_2$+CH$_2$, though photodissociation of C$_2$H$_6$ cannot be ruled out as a source of C$_2$H$_4$. Ignoring the photodissociation pathway, which was done in the study, results in a lower limit on the CH$_3$/CH$_2$ production ratio. Further assuming that the diffusion barriers of CH$_3$ and CH$_2$ are comparable, and accounting for that each product requires two radicals to form, a lower limit of the dissociation branching ratio is inferred to be CH$_3$/CH$_2\geq3$. It should be noted, however, that while astrochemistry codes estimates binding energies to increase with number of hydrogens bound to a heavy element\cite{Garrod08}, it is unknown how the relative stability of radicals such as CH$_3$ and CH$_2$ may affect their relative diffusion barriers. 

The same photochemistry of pure CH$_4$ ice was subsequently  studied using highly sensitive Laser Desorption Post-Ionization Time-Of-Flight Mass Spectrometry (LDPI TOF-MS). The stable end-products were analyzed quantitatively using a set of coupled reactions and rate constants\cite{Bossa15}. In particular the analysis included photodissociation of C$_2$H$_6$ into C$_2$H$_4$ and C$_2$H$_2$ as production pathways of the latter two molecules. The resulting branching ratio of CH$_3$:CH$_2$:CH is $95[5]:4[1]:2[1]$, consistent with the previous limit. These yields are substantially different from gas phase studies, reinforcing the importance of the matrix environment and the high efficiency of reverse reactions in ices for setting photodissociation branching ratios.

The important role of the ice matrix in setting the photodissociation branching ratio is confirmed by experiments on more complex organic ices. For example, photolysis of CH$_3$NO$_2$ ice shows that photodissociation in the solid state is a more complex process compared to photodissociation in the collision free gas limit, and that ice `caging' contributes to the observed photodissociation branching ratio\cite{Maksyutenko15}. Deviations in photodissociation branching ratios thus appear to be a general feature of ice photochemistry, implying that using gas-phase photodissociation branching ratios in models of ice photochemistry is a bad approximation. 

\begin{figure}[htbp]
\begin{center}
\includegraphics[width=5in]{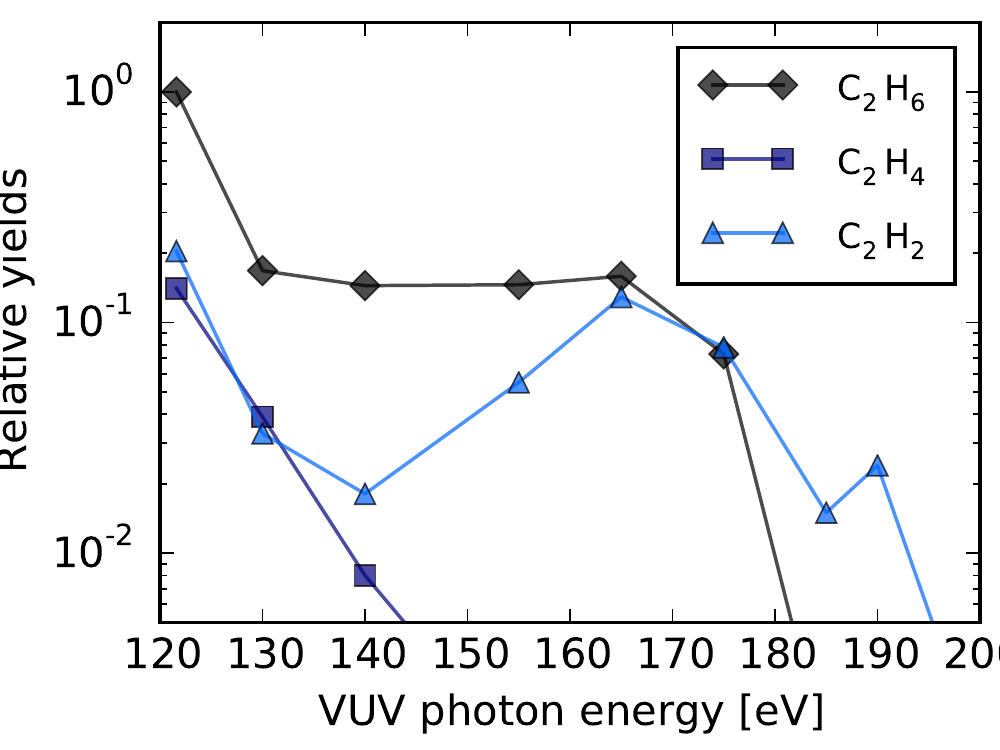}
\caption{The relative yields of C$_2$H$_2$, C$_2$H$_4$ and C$_2$H$_6$ at the end of pure CH$_4$ ice photolysis experiments using different photon energies\cite{Lo15}.}
\label{fig:pd_ch4}
\end{center}
\end{figure}

To address how VUV excitation affects photodissociation rates and branching ratios require a tunable VUV source, or multiple monochromatic VUV sources. In a recent experiment,  pure CH$_4$ ice was irradiated at 3~K with VUV photons from a synchrotron at 121.6, 130, 140, 155, 165, 175, 185, 190 and 200 nm\cite{Lo15}. CH$_3$, C$_2$H$_2$, C$_2$H$_4$ and C$_2$H$_6$ were all identified using infrared spectroscopy. The longest UV wavelengths at which photoproducts were generated were 140, 140, 175 and 190 nm, for CH$_3$, C$_2$H$_2$, C$_2$H$_4$ and C$_2$H$_6$, respectively\cite{Lo15}. Figure \ref{fig:pd_ch4} shows the normalized production rate of  C$_2$H$_2$, C$_2$H$_4$ and C$_2$H$_6$ as a function of wavelength. Interpretation of the experiment is complicated by different photon exposures at different wavelengths, but the changing ratios indicates a changing photodissociation branching ratio with VUV energy, which should be possible to quantify with detailed modeling of the product formation curves as functions of photon dose and energy.

\subsection{Diffusion of Radicals}

Following radical generation, radicals combine into new molecules if other radicals are immediately available or can be reached through diffusion. 
It is important to note that in addition to diffusion barriers the kinetics may be controlled by reaction barriers. If there is no reaction barrier and the reactants are correctly oriented, the reaction is expected to proceed with a high efficiency, and a 100\% efficiency is typically assumed in astrochemistry ice models. Most radical-radical reactions are assumed to fall in this category, as are reactions with ions. Reactions between radicals and atoms may or may not have a barrier and need to be investigated on a case-by-case basis. Reactions between radicals/atoms and molecules often do have barriers. The importance of reaction barriers for the reaction kinetics depend on their absolute magnitude, on their magnitude relative to diffusion barriers, and on the importance of tunneling. It is very important to note that reactions occur in competition with diffusion and desorption, and that if diffusion barriers are large compared to reaction barriers, two reactants will remain adsorbed next to one another until they react\cite{Cuppen13}. If the reaction barrier is large compared to the diffusion barriers, the reaction probability will decrease with barrier height. In most ice photochemistry experiments, the reaction and diffusion barriers have not been modeled separately, limiting our understanding of the relative barrier heights of diffusion and reaction for most ice systems. In this sub-section we will limit the discussion to radical-radical reactions that should be governed by photodissociation and diffusion/reorientation kinetics alone.

One method to constrain diffusion barriers is to observe molecule formation during warm-up of an ice that is photolyzed at a low enough temperatures that many of the photo produced radicals become frozen into the matrix. During warm-up diffusion barriers of increasing magnitude become accessible. When radicals that were previously frozen in start diffusing new molecules should form. This formation should proceed until all radicals are consumed. Products that rely on radicals with low diffusion barriers should thus both begin forming and stop forming at lower temperatures compared to radicals with high diffusion barriers. Figure \ref{fig:pd_radical}b shows the formation of CH$_3$CH$_2$OH and CH$_3$OCH$_3$ during warm-up of a 30~K, photolyzed CH$_3$OH:CH$_4$ ice mixture to 180~K. There is no measurable CH$_3$OCH$_3$ production in the ice after 50~K, indicative of that all relevant radicals (most likely CH$_3$ and OCH$_3$) that were produced in the ice at 30~K could overcome diffusion and steric barriers between 30 and 50~K and have thus been consumed. By contrast CH$_3$CH$_2$OH production continues until 90~K. This indicates that the combined diffusion and steric barriers for CH$_2$OH in CH$_3$OH ice is higher than those for OCH$_3$. Furthermore the different formation kinetics of CH$_3$CH$_2$OH and CH$_3$OCH$_3$ demonstrate that for a reaction to go to completion both radicals involved in the reaction must be mobile. If it was sufficient for the more volatile CH$_3$ radical to be mobile, then CH$_3$CH$_2$OH and CH$_3$OCH$_3$ production would have begun to form and ceased to form at the same temperature. Steric interference in ice matrices is clearly important for regulating ice photochemistry kinetics\cite{Oberg09d}. 

\begin{figure}[htbp]
\begin{center}
\includegraphics[width=5in]{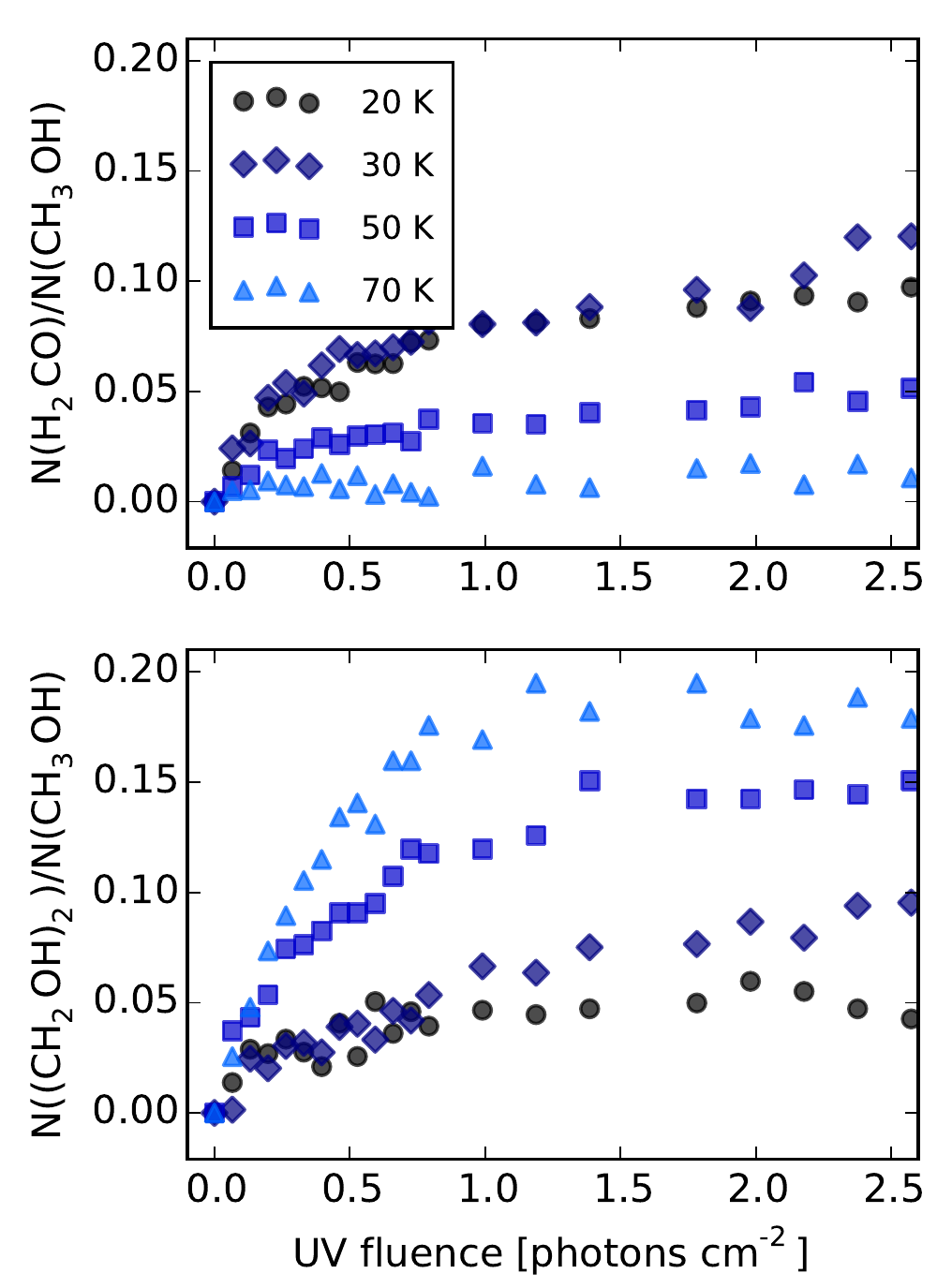}
\caption{Growth curves of H$_2$CO (upper panel) and (CH$_2$OH)$_2$ (lower panel) during irradiation of a pure CH$_3$OH ice at different temperatures between 20 and 70~K. The abundances are presented as fractions of the initial CH$_3$OH ice column density.}
\label{fig:pd_t}
\end{center}
\end{figure}

A second method to constrain relative diffusion barriers of radicals is to photolyze the same ice at different temperatures and take note of how the product composition changes with ice temperature. In general, species that form from two or more heavy species (i.e. not H) are expected to form more efficiently at higher temperatures where diffusion is faster. Figure \ref{fig:pd_t} displays how differently production of H$_2$CO and (CH$_2$OH)$_2$ responds to ice temperature during irradiation. H$_2$CO can form through direct photodissociation, i.e. no radical diffusion required. By contrast, (CH$_2$OH)$_2$ depends on CH$_3$OH photodissociation as well as CH$_2$OH diffusion in the CH$_3$OH ice. The H$_2$CO formation rate curves are similar at 20 and 30~K, and then decreases between 30 and 70~K as radicals become more mobile in the ice, opening up new destruction channels for H$_2$CO, including  H$_2$CO desorption. By contrast the (CH$_2$OH)$_2$ formation rate increases steadily between 20 and 70~K. Kinetics studies can thus be used to identify whether the production pathway of a molecule is regulated by a barrier (typically a diffusion barrier in the case of radical chemistry in ices) or not.

The expectation of increased production for molecules forming from two or more heavy radicals has two important caveats, however. First, if the ice is too warm, desorption of the reactants can become significant and reduce the formation rate. Second, if a molecule forms in competition with other molecules, the maximum production rate may be achieved at a lower than expected temperature. For example, in a CH$_3$OH ice photochemistry experiment, generated CH$_2$OH could react with a CH$_3$ radical to form CH$_3$CH$_2$OH or with another CH$_2$OH radical to form (CH$_2$OH)$_2$. If CH$_2$OH is formed in excess (as it is based on photodissociation branching ratio constraints) and presents a higher diffusion barrier than CH$_3$, (CH$_2$OH)$_2$ production will benefit more than  CH$_3$CH$_2$OH from photolysis at an elevated temperature. This is exactly what is observed\cite{Oberg09d}, and photochemistry experiments at different temperatures should thus be possible to use to quantitatively constrain diffusion barriers of different radicals. This has yet to be achieved, however.

\section{Photochemistry in Simple/Binary Ice Mixtures}

Experiments on simple or binary ice mixtures tend to be designed with similar goals to pure ice photochemistry experiments, i.e. to elucidate mechanisms and constrain key barriers and rates rather than attempting to obtain `realistic' product compositions. This section reviews three different kinds of simple mixture experiments: binary mixtures of a reactive species with water ice, binary and tertiary mixtures of a reactive species with inert or almost inert ices, and mixtures of polycyclic aromatic hydrocarbons (PAHs) with different matrix species.

\subsection{Photochemistry in Water Ice Mixtures}

Ice mixture photochemistry experiments commonly include water as the main ice component or ice matrix. This is motivated by the dominance of water ice in most interstellar ice mantles. The presence of water ice complicates the kinetics of ice photochemistry in two ways compared to pure ices. First water itself dissociates when exposed to VUV, producing reactive radicals that will participate in ice radical chemistry. Second the diffusion environment in an water matrix depends on ice morphology\cite{Mispelaer13}, ice porosity\cite{Lauck15} and water ice mixing ratio\cite{Oberg10b}. Most H$_2$O-ice mixture photochemistry experiments involve complex mixtures and focus on the formation of large, non-volatile organics (see e.g. \S7). These provide few constraints on the impact of water on the photochemistry. To understand the role of the water ice matrix for ice photochemistry we instead rely on  a handful of experiments that present much simple water ice mixtures and a more detailed analysis of the formation kinetics. Perhaps by chance, most of them involve simple mixtures of a small hydrocarbon and water, i.e.  photolyzed mixtures of CH$_4$:H$_2$O \cite{Hodyss09,Oberg10b}, and C$_2$H$_2$:H$_2$O ices\cite{Cuylle14}.

The importance of considering water ice a chemically active matrix has been understood since the beginning of ice mixture photochemistry experiments, and is especially apparent in experiments where a single molecule without any oxygen, such as CH$_4$ or C$_2$H$_2$ is incorporated into a water matrix\cite{Hodyss09,Oberg10b,Cuylle14}. When a 9:1 H$_2$O:CH$_4$ ice is exposed to UV, the most abundant photoproduct is CH$_3$OH\cite{Hodyss09}. CH$_3$OH probably forms through the dissociation of a CH$_4$ and of a H$_2$O molecule to form CH$_3$ and OH, followed by a radical reaction. The kinetics in this particular ice mixture has not been sufficiently studied, however, to exclude a molecule-radical reaction such as CH$_3$+H$_2$O$\rightarrow$CH$_3$OH+H. Despite the high abundance of CH$_3$OH among the observed photoproducts, a kinetic analysis of the experimental results indicates that CH$_3$OH will not be the dominant product at steady-state. Rather CO$_2$ should become the most important carbon carrier if the ice is exposed to a sufficiently large VUV fluence\cite{Hodyss09,Cuylle14}. The lack of steady-state or equilibrium chemistry during ice photolysis is what makes modeling ice photochemistry in both laboratory and astrophysical settings a difficult and computationally challenging problem.

In a similar experiment various H$_2$O:CH$_4$ ice mixing ratios (1:3, 2:1 and 5:1) were used to show that the relative formation fractions of C$_2$H$_6$, CH$_3$OH and CH$_3$CH$_2$OH were consistent with C$_2$H$_6$ forming from two CH$_4$ dissociation products, CH$_3$OH from one H$_2$O and one CH$_4$ dissociation product and CH$_3$CH$_2$OH  from two CH$_4$ and one H$_2$O dissociation fragment, as would be expected in the simplest possible formation scheme of these molecules\cite{Oberg10b}. The experiments also show a later onset for H$_2$CO formation compared to CH$_3$OH, which indicates that in these experiments H$_2$CO is a second generation ice species, possibly a CH$_3$OH photodissociation product. 

The second important consideration for photochemistry kinetics in water ices is the degree at which radical diffusion is possible. This was explored using the same photochemistry experiments as above, on pure CH$_4$ ice and H$_2$O:CH$_4$ ice mixtures (1:3, 2:1 and 5:1) at 20~K, followed by a slow heating ramp\cite{Oberg10b}. The first important result is that the different formation rates of new products at 20~K varies between the ice mixtures, even when normalized to the CH$_4$ photodissociation rate in each experiment. When a pure CH$_4$ ice is irradiated at this temperature, the CH$_4$ dissociation cross section is similar to the combined formation cross section of the most abundant products C$_2$H$_6$ and C$_2$H$_4$. As the water ice concentration is increased from 0, to 25, to 67, to 80\% in the ice mixture, the fraction of dissociated CH$_4$ molecules that become directly incorporated into molecules at 20~K decreases from 100, to $\sim$40, to 30 and $<$20\%, respectively. This implies that in water-dominated ice mixtures, most photoproduced radicals are initially frozen into the ice. In other words at 20~K only a subset of the produced radicals can diffuse a sufficient distance to react at the timescale of the irradiation experiment (several hours), and this subset decreases with decreasing CH$_4$ concentration. 

In all H$_2$O:CH$_4$ ice mixture experiments, new molecules form during the slow heating ramp that followed the VUV irradiation at 20~K. The observed molecules could all be associated with simple radical-radical reactions implying that the increased ice temperature enables radicals to sufficiently diffuse and reorient to find reaction partners. During the heating ramp, CH$_3$CHO forms between 20 and 70~K in water dominated ices, indicative of that both the radicals (CH$_3$ and HCO) needed to produce this molecule are mobile in the ice prior to global ice restructuring at 130~K. By contrast, CH$_3$OH is only observed to form around 130~K, when water ice is re-structured.  At least one of CH$_3$OH building-blocks, most likely OH, must have a very high diffusion barrier in H$_2$O ice to explain the lack of CH$_3$OH production at lower temperatures. This is consistent with the expectation that OH will form strong hydrogen bonds in a water ice.

High diffusion barriers in H$_2$O ice clearly inhibits photochemistry at low temperatures. It does not completely quench it, however, at least not in a laboratory setting where radical production is high.  UV photolysis of CH$_3$OH in water ice results in the production of new molecules as large as ethylene glycol at temperatures as low as 3~K\cite{Krim09}. There are several lessons that can be learnt form this small collection of experiments. First, neither laboratory ices not interstellar ones are well described by steady-state or equilibrium chemistry. A detailed understanding of the kinetics is thus required. Second, the kinetics are strongly affected by the ice diffusion environment, and this environment is difficult to predict even for very simple ice systems. Given these difficulties, it is not strange that there is yet no consensus on how to best model ice photochemistry, neither in laboratory nor in astrophysical settings.

\subsection{Photochemistry in CO, N$_2$ and Noble Gas Ice Matrices}

In an astrochemical context photochemistry experiments on ice mixtures without water is typically motivated by the existence of CO-rich ice phases in interstellar grain mantles and CO- and N$_2$-rich ices in outer Solar System, or by a desire to explore the kinetics of isolated molecules in inert ice matrices. CO, N$_2$ and noble gas ice matrices share the common characteristic that none of them are readily dissociated by typical laboratory VUV sources, which simplifies interpretation compared to experiments that employ chemically active ice matrices, such as water ice. Outside of astrochemistry the photo production of radicals  in CO, N$_2$ and noble gas matrices has been studied for many decades\cite{Milligan64,Jacox65,Milligan67}, to spectroscopically identify and characterize unstable species; matrix isolation enables the long-time survival of highly reactive radicals and results in very narrow infrared line widths, which together makes matrix isolation an excellent tool for radical characterization. In this sub-section we review the outcome of photochemistry experiments on CH$_3$OH in CO ices\cite{Oberg09d}, CH$_4$ in  N$_2$ and CO ices\cite{Hodyss11}, H$_2$S in CO (and CO$_2$) ice\cite{Chen15}, ethylene, methane and acetylene in N$_2$ ice\cite{Wu12,Wu14,Chen15b}, and H$_2$CO in noble gas matrices\cite{Butscher15}. Some of these experiments involve UV energy resolved data, which provides additional constraints on the formation mechanisms.  In addition to these photochemistry experiments, there are series of experiments on ice photodesorption which somewhat address VUV energy-resolved photochemistry in simple ices and ice mixtures\cite{Fayolle13,Fillion14,Chen14}, but they are not further considered in this review.

\subsubsection{Photochemistry in CO Ice Matrices}

Of the `inert' ice matrices considered in this section, CO is the chemically most active. It both presents some channels to dissociation by VUV photons (through reactions between excited CO and ground state CO to form CO$_2$+C), and can react with photo produced atoms such as H, and perhaps also with some radicals, at cryogenic temperatures. This complicates interpretation of experiments compared to when truly inert ice matrices are used. The main reason that CO ice mixtures are still explored is that there are CO-rich phases in both the Solar System and into ISM. In particular, the most important starting material for COM formation in the ISM, CH$_3$OH\cite{Garrod08} is probably present in a CO ice environment (see \S1). This is important when considering the outcome of CH$_3$OH ice photolysis and how to apply it to an astrophysical setting. 

Based on experiments, the photochemistry product composition and kinetics in VUV-irradiated CH$_3$OH ices depend on the amount of CO mixed in with the CH$_3$OH\cite{Oberg09d}. An increasing CO abundance results in an increasing production of HCO-X type complex molecules such as CH$_3$CHO, HCOOCH$_3$ and CH$_2$OHCHO and a corresponding decreasing production of complex molecules that lack the HCO group, e.g. CH$_3$OCH$_3$ and CH$_3$CH$_2$OH. In CH$_3$OH:CO ice mixtures where the mixing ratio is unity or where CH$_3$OH is more abundant than CO, the total production rate of COMs at 20~K appears similar in the ice mixture compared to pure CH$_3$OH ice. In both experimental setups COMs form abundantly during radiation at 20~K, followed by additional formation during warm-up when radical diffusion and/or re-orientation become possible.

If the CO content is increased such that CO:CH$_3$OH ice mixing ratio  is 10:1, production of complex molecules is quenched during irradiation at 20~K.
 When the ice mixture is warmed-up, new molecules form abundantly from radical combinations, demonstrating that radicals were produced at 20~K in this experimental set-up at a similar rate to 1:1 mixtures and pure CH$_3$OH ice experiments.  In the CO:CH$_3$OH 10:1 ice system, however, few of the photo produced radicals seem to form close enough to one another to allow for radical combination reactions in the cold ice where thermal diffusion is low or perhaps non-existent. Warm-up to temperatures where diffusion is possible is thus needed for the chemistry proceed. It should be noted that an ice morphology with a low radical concentration is most likely more representative for the ISM than the typical laboratory ice setup. This experiment thus highlights the importance of understanding radical diffusion to correctly predict COM formation in space.

In an experiment specifically  aimed at exploring diffusion of radicals in ice, CO was deposited together with H and OH, formed from photodissociation of H$_2$O in a microwave discharge tube, to form an ice \cite{Oba10}. Reactions in this system depend only on the thermal diffusion and reaction kinetics of CO and OH. That is, there should be no substantial non-thermal diffusion. By contrast, non-thermal diffusion might be important for ice photochemistry when the radicals are produced energetically in the ice. In the experiments, the H$_2$CO$_3$/CO$_2$ production ratio increased with increasing OH/CO ratio and temperature. The proposed formation scenario of H$_2$CO$_3$ was CO+OH$\rightarrow$HOCO, and HOCO+OH$\rightarrow$H$_2$CO$_3$, and H$_2$CO$_3$ thus requires two OH radicals to form. The observed temperature dependence of the H$_2$CO$_3$/CO$_2$ production ratio as well as of the individual products suggest that OH diffusion is the rate limiting step and proceeds with a substantial barrier, similar to what was observed for CH$_3$OH formation from CH$_3$+OH above.

The role of the CO matrix has also been explored in experiments that employ different VUV sources. H$_2$S:CO 1:20 20~K ices were irradiated with four different sources: 30.4~nm and 58.4~nm light from a synchrotron, and UV irradiation from a microwave-discharge hydrogen-flow lamp that included or excluded Lyman-$\alpha$ photons\cite{Chen15}. Importantly, CO is not dissociated directly by the lamp. Some dissociation can occur through CO excitation followed by reaction with a neighboring CO to form CO$_2$, but the reaction rate has been observed to be low compared to direct dissociation rates\cite{Gerakines96,Oberg07b,MunozCaro10,Chen14}. By contrast CO can be directly dissociated by the synchrotron radiation.  Any product requiring O atoms to form in the lamp experiments, such as SO$_2$, must thus obtain an O from a a second or third generation photoproduct. 

The product compositions were similar in  the two lamp irradiation experiments, which should not be surprising since all H$_2$S  can be dissociated by both lamps.Iin the these experiments C$_2$S production seemed to follow OCS production. By contrast, CS$_2$ was formed from the beginning in the EUV experiments, consistent with the expectation that EUV radiation can provide an immediate source of C following CO photodissociation while VUV irradiation does not. The total production rate was highest in the lamp experiment that included Lyman-$\alpha$, followed by 58.4~nm radiation, lamp irradiation without Lyman-$\alpha$ and 30.4~nm radiation. The difference between the best and worse source was only a factor of three, however. This suggests that OCS formation is robust in this kind of ice and that CO dissociation does not aid its formation, but also does no great harm. A detailed kinetic analysis is however needed to extrapolate this result to the ISM where timescales are different and non-thermal diffusion of radicals may be less important due to lower radical concentrations.

\subsubsection{Photochemistry in N$_2$ Ice Matrices}

N$_2$ matrices should be more chemically inert than CO ices, since they cannot be photolyzed by VUV lamps. Still N$_2$ may react with some of the photo produced atoms and radicals, similar to CO. In the outer Solar System, N$_2$ is expected to be a major ice constituent, which has motivated some of the existing photochemistry studies on N$_2$ ice mixture films\cite{Hodyss11}. When 14-18~K CH$_4$ is photolyzed in a CO:N$_2$ ice mixture with a UV lamp, C$_2$H$_6$ and C$_2$H$_4$ are the main products, similar to in pure CH$_4$ experiments\cite{Hodyss11}. These molecules form during irradiation indicating that either thermal diffusion of the hydrocarbon radicals is possible at T$<$20~K in this kine of ice, or the production rate of radicals is high enough to enable chemistry with the aid of non-thermal diffusion and re-orientation alone. The experimental results have not yet been incorporated into any detailed, quantitative model and it is therefore not possible to say what the importance of this hydrocarbon production pathway is in the Solar System (or beyond) compared to e.g. ice interactions with cosmic rays or electrons.  Importantly, N-containing photoproducts were detected at low abundances in the experiment, demonstrating that even the VUV-robust N$_2$ becomes a chemically active matrix in the presence of photo-produced radicals.

Synchrotron UV radiation has been applied to experiments both to record ice absorption spectra (at high resolution) and to irradiate pure N$_2$ ice, and CH$_4$ dispersed in N$_2$ ice (CH$_4$:N$_2$ = 1:500) at 20 K at VUV wavelengths 130 nm (9.5 eV), 121.6 nm (10.2 eV), and 91.6 nm (13.5 eV)\cite{Wu12}. For the case of pure N$_2$ ice, irradiation at 13.5 eV yielded the most photoproducts (traced by N$_3$ production). Based on absorption spectroscopy the relative absorbance of N$_2$ ice is 0.09 at 10.2 eV and 0.67 at 13.5 eV, and the photochemistry results thus seems qualitatively consistent with excitation efficiencies. Whether there is quantitative agreement is less clear (the photo production rate difference between 9.5/10.2 and 13.5 eV was only a factor of two). In fact, the bond dissociation energy of N$_2$ is 9.76 eV and the photon energy at 9.5~eV should thus be insufficient to dissociate N$_2$ and produce any photochemistry when irradiated at this wavelength.

During irradiation of the N$_2$:CH$_4$ ice mixture, two of the products are HCN and its unstable isomer HNC. These are important molecules in astrochemical reaction networks. More HNC and HCN are formed when the ice is exposed to 9.5 eV radiation compared to 10.2 or 13.5eV radiation. A second observations, is that the HNC/HCN ratio decreases by factor of three during exposure to 13.5~eV radiation and much more moderately in the other two experiments. The former result can be understood when CH$_4$ excitation is considered -- CH$_4$ VUV absorbance peaks at 9.5~eV. It thus seems like N$_2$ dissociation is unimportant, and perhaps even destructive for initial HCN and HNC formation, and that both molecules are formed efficiently from reactions between CH$_4$ photo fragments and N$_2$ molecules. The reason for the decrease in HNC/HCN with time and its dependence on VUV wavelength  is less unclear, but it may proceed through a HNC+H $\rightarrow$ HCN + H kind of reaction.

HCN and HNC also form when C$_2$H$_4$ is dispersed in solid N$_2$ at 10 K and is irradiated with Ly$\alpha$ (121.6~nm)\cite{Chen15b}. By contrast to the N$_2$:CH$_4$ ice experiment, the HNC/HCN ratio increases with irradiation time in the experiment. The reason for this trend reversal is unknown. One clue may by that HCN and HNC do not form in C$_2$H$_2$:N$_2$ photolysis experiments. This result has been taken as evidence for that a critical minimum H atom production rate is needed in the ice for efficient HCN and HNC formation\cite{Wu12,Wu14,Chen15b}. An alternative explanation is that these molecules only form through reactions with N$_2$ and CH$_2$ and CH$_3$, and not through reactions between N$_2$ and CH or C. This hypothesis could be tested by adding H$_2$ to the C$_2$H$_2$ ice experiment and determining the initial product yield (before any more saturated hydrocarbons or radicals have formed).

\subsubsection{Photochemistry in Noble Gas Matrices}

Noble gas matrices do not provide a realistic setting for either ISM or Solar System ice photolysis. They are still used in astrochemically motivated experiments because they provide an almost inert matrix (H atoms can still add to noble gases) within which specific photolysis steps can be studied with minimal interference, e.g. the photodissociation cross section and/or photodissociation branching ratio. Noble gas matrices have most recently been used to study radical combination chemistry, where the radicals were produced from VUV photolysis of H$_2$CO\cite{Butscher15}. During irradiation at 12~K the main products were CO and HCO, as would be expected. After slight thermal annealing, enabling H atom diffusion, the CH$_2$OH radical was observed together with an increase in HCO and other species that can be associated with H addition reactions to CO, H$_2$CO, and the noble gas matrix. The ice was subsequently heated until the matrix and H$_2$CO had sublimated. Production of new molecules due to radical diffusion and reactions was observed, including HOCH$_2$HCO, (CH$_2$OH)$_2$ and CH$_3$OH.

\subsection{PAH Ice Photochemistry}

A special class of ice photochemistry experiments involve polycyclic aromatic hydrocarbons (PAHs). PAHs are known to exist in space at large abundances, though individual PAHs are yet to be uniquely identified. In cold interstellar regions where icy mantles form on grains, PAHs are expected to freeze out and may thus commonly exist in icy grain mantles. This sub-section focuses on simple PAH:ice experiments aimed at elucidating photochemistry kinetics and mechanisms. PAH:ice photochemistry experiments aimed at production of prebiotic molecules are reviewed in \S7.

The kinetics of PAH excitation, ionization, dissociation and chemistry in water ice has been explored by monitoring abundances of PAHs and PAH derivatives with  UV-Vis and infrared spectroscopy during irradiation at 10-125~K\cite{Gudipati03,Gudipati04,Gudipati06,Bouwman09,Bouwman10,Bouwman11a, Bouwman11b}. In general UV irradiation of PAH:H$_2$O ices result in ion production, and the ions appear stable. Taking the pyrene:H$_2$O system as an example: VUV photolysis of pyrene in a water ice results in the production of pyrene cations, hydroxypyrene, hydroxypyrene cations, pyrene anions and pyrenolate anions\cite{Bouwman09}. Compared to other photolysis experiments, these ices are clearly ionized by the UV photons and much of the chemistry appears ion-mediated at low temperatures, i.e. at 10--25~K. The photo-produced ions decay on the timescale of hours to days, which indicates that recombination reactions are slow in water-dominated ices. In warmer water-dominated ices, the pyrene chemistry is instead radical-mediated, similar to all other ices reviewed above.

In addition to ice temperature, the ice matrix composition affects whether the PAH photochemistry is ion or radical mediated. When pyrene is embedded in a CO matrix it appears to present a radical dominated photochemistry\cite{Bouwman10}; the main photolysis products are triplet pyrene and the 1-hydro-1-pyrenyl radical. By contrast, ion formation and charge transfer are both important for the chemistry when PAHs are embedded in NH$_3$ and water:NH$_3$ ice matrices and are exposed to UV radiation\cite{Cuylle12}. Furthermore, the relative amount of water and ammonia determines whether positively or negatively charged PAHs form. In pure water ice, cations are generated through direct ionization, whereas in pure ammonia ice, anions form through electron donation from ammonia-related photoproducts. The solid-state process controlling this latter channel involves electron transfer, rather than acid-base type proton transfer. In the mixed ice, the resulting products depend on the mixing ratio.

PAH concentration is also important for the response of a PAH:matrix ice system to UV irradiation. Two prototypical PAHs (pyrene and coronene) were photolyzed using PAH:H$_2$O concentrations in the range of 1:30 000 to pure PAH at temperatures between 12 and 125 K. Both the photoproduct composition and the ionization efficiency depends on PAH concentration; the ionization efficiency is 60\% in dilute ices, but only 15\% in a 1:1000 ice\cite{Cuylle14}. An increased PAH concentration results in a lower ionization yield due to a combination of an increased recombination rate with increased PAH clustering, and a decreased solvation efficiency of electrons as the OH/e- ratio decreases. OH has a higher electron affinity than H$_2$O and the lower PAH-to-H$_2$O concentration the higher number of photoproduced electrons will become trapped by OH, increasing the ionization efficiency.

Finally there are experiments that have characterized deuterium fractionation chemistry in PAH:ice systems. The D/H level in meteoritic organics is frequently used to infer information about their formation zones. It is therefore interesting that PAHs rapidly exchange H for D when frozen into a D$_2$O ice matrix and exposed to VUV radiation. Based on experiments the PAH becomes enriched in D through three different mechanism: D atom addition, D atom exchange at oxidized edge sites, and D atom exchange at aromatic edge sites\cite{Sandford00}. 

\section{Photochemistry in Multi-Component Ice Mixtures}

This section reviews photo experiments on more complex ice mixtures than has been considered above. These ice mixtures present the advantage that they are closer in composition to ISM ices than pure and binary ices. Photochemical yields in these experiments should therefore be more meaningful. However translating yields from laboratory to interstellar settings is still fraught with danger unless the kinetics and mechanisms are well understood, and they rarely are. Furthermore, most of the time the main analysis is done on the refractory residue that remains after a UV processed ice has sublimated. To obtain sufficient amount of the residue, the ice thicknesses, UV doses and ice temperatures are high, which should alter the kinetics compared to astrophysical settings. It is also often difficult to tell at which stage the molecules formed, i.e. if they are truly ice chemistry products. The experiments in this section should thus be considered to answer the question, could `X' form through UV photo processing of interstellar ice equivalents after `reasonable' UV doses and thermal processing, and not whether it actually will. `X' is frequently a complex molecule of prebiotic interest, such as an amino acid. 

In this section we first review the early experiments that inspired all future studies on ice photochemistry. \S7.2 and \S7.3 reviews experiments that demonstrated the possibility of forming prebiotically interesting molecules through ice photochemistry, including amino acids. \S7.4 presents experimental constraints on the mechanisms through which some of these molecules form. Finally \S7.5 reviews experiments using polarized VUV light sources to induce enantiomeric excess.

\subsection{Early Experiments}

While placed last in this review, photolysis of multi-component ice mixtures came first chronologically in the study of interstellar ice photochemistry. Interstellar ice mantles were first proposed to be production sites of prebiotic molecules during the early days of experimental astrochemistry in the 1970s and 1980s. Photoprocessing of low temperature (10 K) mixtures of ices (water, CO, CO$_2$, NH$_3$ and CH$_4$) subjected to VUV radiation was studied both {\it in situ} through IR spectroscopy, and post-processing using mass spectrometry analysis \cite{Greenberg72,Greenberg76,Hagen79,Greenberg83}. IR spectroscopy of the UV processed and subsequently warmed up ice revealed the presence of carboxylic acids and amino groups. After ice sublimation, a yellow, non-volatile residue remained. This residue was soluble in water and methanol, and contained molecules with molecular weights of 100 and more. 

The general approach introduced by these early experiments has been used by a large number of studies in the 1980s, 1990s and 2000s. Later experiments incorporated more sophisticated analysis methods, modified the ice compositions to fit new observational data, and designed experiments to address the roles of different experimental parameters\cite{Allamandola88,Bernstein95,MunozCaro03}. These changes resulted in detections of new families of molecules, both in the photolyzed ice, including nitriles and iso-nitriles, and in the residue.  GC-MS analysis of the dissolved residue has revealed, among other things, photoproduction of hexamethylenetetramine (HMT), polyoxymethylene (POM), ammonium-carboxylic acid salts, amides, and esters\cite{MunozCaro03}.

\begin{figure}[htbp]
\begin{center}
\includegraphics[width=6in]{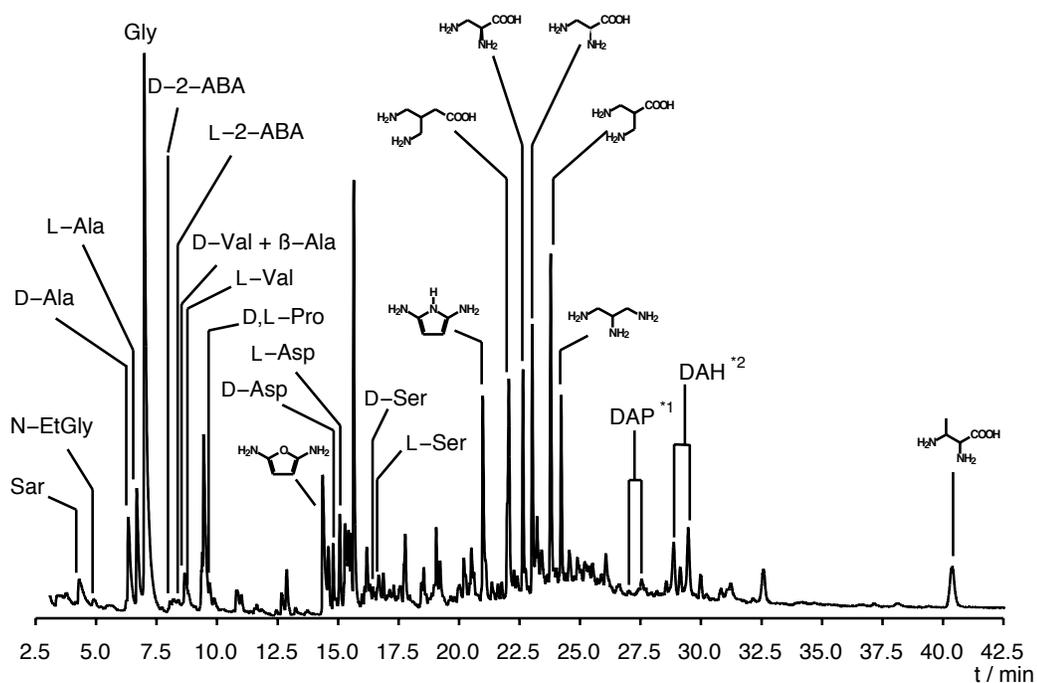}
\caption{A GC-MS spectra of a residue produced by UV irradiation of an ice mixture and subsequent ice warm-up and sublimation. The figure is reproduced with permission and has been edited (increased line thicknesses and labeling) for clarity\cite{MunozCaro02}. Reproduced with permission from Ref. 145. Copyright 2002 Nature Publishing Group.}
\label{fig:aa}
\end{center}
\end{figure}

\subsection{Formation of Amino Acids}

Amino acids were first detected and uniquely identified in the residues of ice UV photochemistry experiments in 2002\cite{MunozCaro02,Bernstein02}. In two independent experiments, two thick interstellar ice analogs (H$_2$O:CH$_3$OH:NH$_3$:CO:CO$_2$ 2:1:1:1:1, and a water ice mixture with  0.5-5\% NH$_3$, 5-10\% CH$_3$OH and 0.5-5\% HCN, respectively) were VUV irradiated during deposition at 12--15~K, and then subsequently heated up to enable analysis of the remaining residue. Both experiments detected many different kinds of amino acids in the dissolved and hydrolyzed ice residue using GC-MS (Figure \ref{fig:aa}). The amino acid yields were defined differently in the two experiments, but seem both consistent with a conversion of a fraction of a percent of the original carbon into amino acids. 

There are also some interesting differences between the two experimental set-ups and the resulting product compositions and deduced formation pathways. In the \\H$_2$O:CH$_3$OH:NH$_3$:CO:CO$_2$ 2:1:1:1:1 ice experiment the initial residue mainly consisted of saturated organic compounds, and included carboxylic groups and HMT\cite{MunozCaro02}. The ice origin of the amino acids was confirmed by experiments where the carbon in CH$_3$OH, CO and CO$_2$ was substituted for $^{13}$. Amino acids were only detected in hydrolyzed residue samples, however, and and detected amino acids may thus form during the hydrolysis step. The second ice experiment included no CO or CO$_2$ in the ice, but did include HCN\cite{Bernstein02}. Prior to hydrolysis, GC-MS and HPLC analysis revealing production of N-formyl glycine, cycloserine (4-amino-3-isoxazolidinone) and glycerol. A small amount of glycine could also be extracted from the residue by hot water alone. Large glycine yields, and any alanine, serine, glycerol, ethanolamine and glyceric acid were only observed after hydrolysis, confirming that the hydrolysis step is important for amino acid production in this kind of experiments. Still, these two experiments demonstrate that the building blocks of amino acids and perhaps amino acids themselves are readily formed in ices that are exposed to large doses of UV radiation. Furthermore, deviations from the original experiments in terms of irradiation time, ice temperature, ice mixture composition, photon dose per molecule, and ice substrate did not quench amino acid production, though some differences in product composition between different experiments were observed\cite{Nuevo08}. The formation of amino acids in this type of ice experiment thus appears robust.

An important question raised by these experiments is whether amino acid production through ice chemistry alone is possible, or whether  individual amino acid molecules only form in the residue or during residue hydrolysis through degradation of macromolecules, such as HCN polymers. Theoretical calculations suggest that amino acid formation in the ice is possible through radical-radical reactions following ice photolysis\cite{Woon02}. In particular, the carboxylic group (COOH) of the amino acids can form through CO+OH reactions in the ice, where the CO could come from CH$_3$OH dehydrogenation if no CO was originally present in the ice.  Sequential hydrogenation of HCN forms CH$_2$NH$_2$ and CH$_2$NH$_2$+COOH forms glycine\cite{Woon02}. This glycine formation scenario could be tested by isotopically labeled CH$_3$OH and HCN in the original ice mixture. A similar amino acid mechanism was suggested in a different calculation, with the one difference that COOH was produced through CO$_2$ + H\cite{Holtom05}.

Experimental support of the presence of some amino acid formation during ice photolysis comes from a different experiment that employed {\it in situ} mass spectrometry in conjunction with reactive ion scattering (RIS) and low-energy sputtering (LES)\cite{Lee09}.  In this experiment glycine was detected after irradiating an interstellar ice-analog film composed of water, methylamine, and CO$_2$. This experiment demonstrates that {\it some} amino acids can be produced in VUV-irradiated ice, if the starting ice contains species of moderate complexity. Analysis of the products suggest that glycine formed in the ice through photocleavage of C-H and N-H bonds in methyl amine, followed by reactions of the dissociated H atom with CO$_2$ to form HOCO, and further reactions between HOCO and the dehydrogenated methyl amine to form glycine, i.e. consistent with previous theoretical calculations. 

More complex, ring based amino acids form in UV-irradiated ices that contain PAHs, for example naphthalene in a H$_2$O+NH$_3$ ice mixture exposed to extreme or vacuum UV radiation\cite{Chen08}.

\subsection{Formation of Sugars, Polymers and Other Prebiotic Molecules}

In addition to amino acids, sugars, small polymers, and other large complex organics are also of prebiotic interest. Such molecules are frequently observed to form during UV photo processing of interstellar ice analogs, either in conjunction with amino acids or on their own (if nitrogen is excluded from the ice mixture). The  ice photochemistry residues discussed in the previous section, for example also contained 2,5-diaminopyrrole, 2,5-diaminofuran, and 1,2,3-triaminopropane (Fig. \ref{fig:aa}).  If a simpler ice mixture composed of NH$_3$ and CH$_3$OH is irradiated, urea, glycolic acid, glycerol, hydroxyacetamide, glycerolic acid, and glycerol amide are all present in the photochemistry residue\cite{Nuevo10}. The glycerol (derivatives) is of special interest since it is a sugar, and the glycerol backbone is central to all lipids known as triglycerides. A proposed formation pathway for glycerol in this ice is a reaction between two CH$_2$OH radicals and H$_2$CO, all produced from photodissociation of CH$_3$OH.

Another interesting family of ice photoproducts is related to hexamethylenetetramine (HMT, C$_6$H$_{12}$N$_4$). This family is the most abundant photoproduct left in the residue when exposing H$_2$O:CH$_3$OH:NH$_3$:CO:CO$_2$ = 2:1:1:1:1 (similar to experiments aimed at amino acid production) at 12~K during deposition in a HV chamber\cite{MunozCaro03}. In difference to amino acids, HMT production can be monitored with infrared spectroscopy during ice irradiation, and none was detected until the ice sublimated and only the residue remained. In addition to HMT, five HMT derivatives were detected in the residue: methyl-HMT, hydroxy-HMT, methanyl-HMT, amin-aldehyd-HMT, and methanyl-aldehyd-HMT, demonstrating that once HMT forms a diverse set of addition reactions can take place. HMT production under these conditions is robust, but the yield of different HMT products vary by factors of a few up to a factor of 10 in `identical experiments', demonstrating some sensitivity to the exact experimental conditions, or alternatively a real chemical heterogeneity in the residue. Some of this variation may be due to reactions in the gas-phase during deposition, which is difficult to characterize; in the described experiments the sample and molecular beam are both irradiated during the experiment. More recent experiments where the ice is only irradiated after the ice has already been deposited does not produce the same kind of HMT-rich residue as was observed in the co-deposited experiments\cite{MunozCaro14}. It is unclear, however, how much residue that would be expected in the latter kind of experiment when taking into account the limited UV penetration depth into the ice. Based on isotopic labeling the main formation pathway of HMT and its derivatives were proposed to be through photo processing of CH$_3$OH producing H$_2$CO, and CH$_2$NH$_2$ from reactions with NH$_3$. Three CH$_2$NH$_2$ radicals then react to form a ring followed by reactions with three H$_2$CO molecules to form HMT. This is then followed by derivatization.

Even larger molecular structures than HMT have been found in similar experiments that begin with mixtures of saturated and unsaturated hydrocarbons with or without a source of nitrogen\cite{Dartois04,Dartois05}. These experiments initially follow a similar protocol as outlined above, i.e. the ice is irradiated during deposition at 10 K. Spectral features that can be identified with large polymers or amorphous carbon are not present at 10 K, but appear upon heating to 40 K. The observed polymer thus forms in the ice at cryogenic temperatures. The experiments thus demonstrate that very large organic molecules can form through ice photochemistry if enough carbon, UV photons and heat are supplied. 

If the initial ice mixture contains PAHs in addition to the small molecules, UV irradiation produces some ring-based photoproducts. UV irradiation of different PAHs in ices result in production of aromatic alcohols, quinones, ethers, and nucleobases (and ring-bases amino acids as reported above)\cite{Bernstein99,Bernstein02b,Nuevo09,Nuevo12,Materese13,Ashbourn07,Elsila06}. VUV photolysis of various coronene-ice mixtures result in additions of amino, methyl, methoxy, cyano/isocyano, and acid functional groups to the initial PAH molecule\cite{Bernstein02b}. 
Irradiation of pyrimidine in H$_2$O ice mixtures that also contained NH$_3$, CH$_3$OH and CH$_4$ results in the formation of the nucleobase cytosine\cite{Nuevo14}. The cytosine production appears to be quite sensitive to the details of the ice mixture, however\cite{Nuevo09,Nuevo12,Materese13}.
VUV irradiation of anthracene in H$_2$O results in oxidization to form anthracene ketones including 1,4-anthraquinone, and 9,10-anthraquinone. Quinones are fundamental to biochemistry, and their formation in several irradiated PAH:ice mixtures ice is therefore of interest\cite{Ashbourn07}. For example the polycyclic aromatic nitrogen heterocycle (PANH) quinoline is readily oxidized when embedded in water ice mixed with methane and methanol and exposed to VUV irradiation\cite{Elsila06}.

\subsection{Formation Pathways in Complex Ice Mixtures}

So far this section has focused on detections and yields of prebiotically interesting molecules in ice photochemistry experiments. A mechanistic understanding is key, however, to apply these results to astrophysical environments. This kind of understanding has been challenging to develop. As the complexity of the initial ice mixture increases, it becomes increasingly difficult to connect products with specific reactants/radicals, especially if the photodissociation cross sections and branching ratios of the original ice constituents are not well known. There are, however, some tools that facilitates this process, of which isotopic substitution is perhaps the most powerful. In isotopic substitution experiments the initial ice reactants are isotopically labeled by e.g. substituting the carbon in one molecule with carbon-13. The isotopic composition of the photoproducts is then compared with the isotopic composition of the parent molecules to deduce the reaction pathways. This has been most notably applied to deduce the formation pathway of glycine and other amino acids during ice photolysis.

The question on whether glycine forms from HCN hydrogenation or not in ice analogues can be generalized to whether amino acids in UV-irradiated ices form mainly through a Strecker-type synthesis (a reaction in water between hydrogen cyanide (HCN), ammonia
(NH$_3$), and an aldehyde (RCHO)) or a radical-radical mechanism in the ice. This has been addressed by isotopically labeled ice photochemistry experiments where glycine (NH$_2$CH$_2$COOH) and the larger amino acids alanine (NH$_2$CH(CH$_3$)COOH) and serine\\ (NH$_2$CH(CH$_2$OH)COOH) are formed\cite{Elsila07}. Perhaps surprisingly, the amino acids appear to form through multiple pathways, and the major reaction pathways did not match either of the proposed reaction mechanisms. Instead the authors suggested a modified radical-radical mechanism\cite{Elsila07}. 

The starting ice in these experiments was H$_2$O:CH$_3$OH:NH$_3$:HCN. The sample preparation and analysis followed previous experiments\cite{Bernstein02}. First, the relative importance of NH$_3$ and HCN was tested by excluding one of the two nitrogen containing reactants from some of the experiments. All three amino acids were produced in all irradiated ices, except for serine, which was only produced in HCN-containing ices. Glycine was reduced by 20\% when removing NH$_3$ and 90\% when removing HCN. Alanine was also reduced when removing NH$_3$, but somewhat enhanced in the absence of HCN. Serine was unaffected by the removal of NH$_3$ and not detected in the absence of HCN. This result alone demonstrates that there is not a single pathway that all amino acid (precursors) form within UV irradiated ices. With the help of isotopically labeled C and N, the authors demonstrated that in the case of serine two of the carbons originate from HCN, and one from CH$_3$OH: the acid and central carbon come from HCN and the side chain carbon from CH$_3$OH. The role of HCN was confirmed by the fact that almost all nitrogen in serine comes from HCN (rather than NH$_3$). 

Based on a similar analysis glycine incorporates either one or two carbons from HCN (i.e. a more mixed formation pathway). The acid C always comes from HCN, while the center carbon comes from either HCN or CH$_3$OH at almost equal probability. Glycine also gets up to 10\% of its nitrogen atoms from NH$_3$, consistent with the presence of some glycine in experiments where HCN is absent. In 60\% of the cases both the C and N came from HCN, suggesting that the original C-N bond had been preserved consistent with theoretical work\cite{Woon02}. An important conclusion of this paper is that while HCN aids in amino acid formation in photolyzed ices, NH$_3$ and CH$_3$OH can be used instead as C and N sources. Amino acid production in UV radiated ices can thus be expected under a range of interstellar ice conditions.

\subsection{Effects of Circularly Polarized UV Radiation}

So far all the discussed ice photochemistry has been induced by unpolarized light and as a result enantiomers have formed at equal abundances, i.e. racemic mixtures. On Earth, most chiral molecules only exist in one enantiomeric form. This is well understood in the presence of biology since enantiomerically pure complex chemical systems are more stable. It is less well understood where the initial enantiomeric excess came from, and whether it arose on Earth (before or after the onset of biology) or was delivered from Space. 

Extraterrestrial origins of enantiomeric excesses frequently invoke circularly polarized light (CPL)\cite{Bailey98}.  Circularly polarized (UV) radiation is present in the ISM and can be quite strong around massive young stars due to dichroic
scattering on aligned grains by a magnetic field in reflection nebulae close to regions containing massive
stars\cite{Lucas05}. CPL is expected to interact with chiral molecules asymmetrically, and exposure of interstellar ices to CPL could therefore result in a UV photochemistry that favors either destruction or production of left or right handed molecules. The resulting enantiomeric excess in the ice (residue) could then seed the Solar System and young earth with an eantiomeric excess of key prebiotic molecules.

To address whether this process could be active in interstellar ices, there have been several experiments aimed at detecting enantiomeric excesses in ice residues following ice exposure to circularly polarized UV radiation. Early experiments found that UV-CPL exposure to interstellar ice analogs did not result in definitive detection of enantiomeric excesses\cite{Nuevo06}. These experiments employed the LURE SU5 synchrotron beamline to expose 80~K ice to circularly polarized ultraviolet light (UV CPL) at 167 nm (7.45 eV). The resulting eantiomeric excess in the room temperature ice residue was at most 1\%. The main source of error in this study was the area integration of the GC peaks due to substantial overlap between enantiomeric peaks in the GC-MS spectra.  

\begin{figure}[htbp]
\begin{center}
\includegraphics[width=6in]{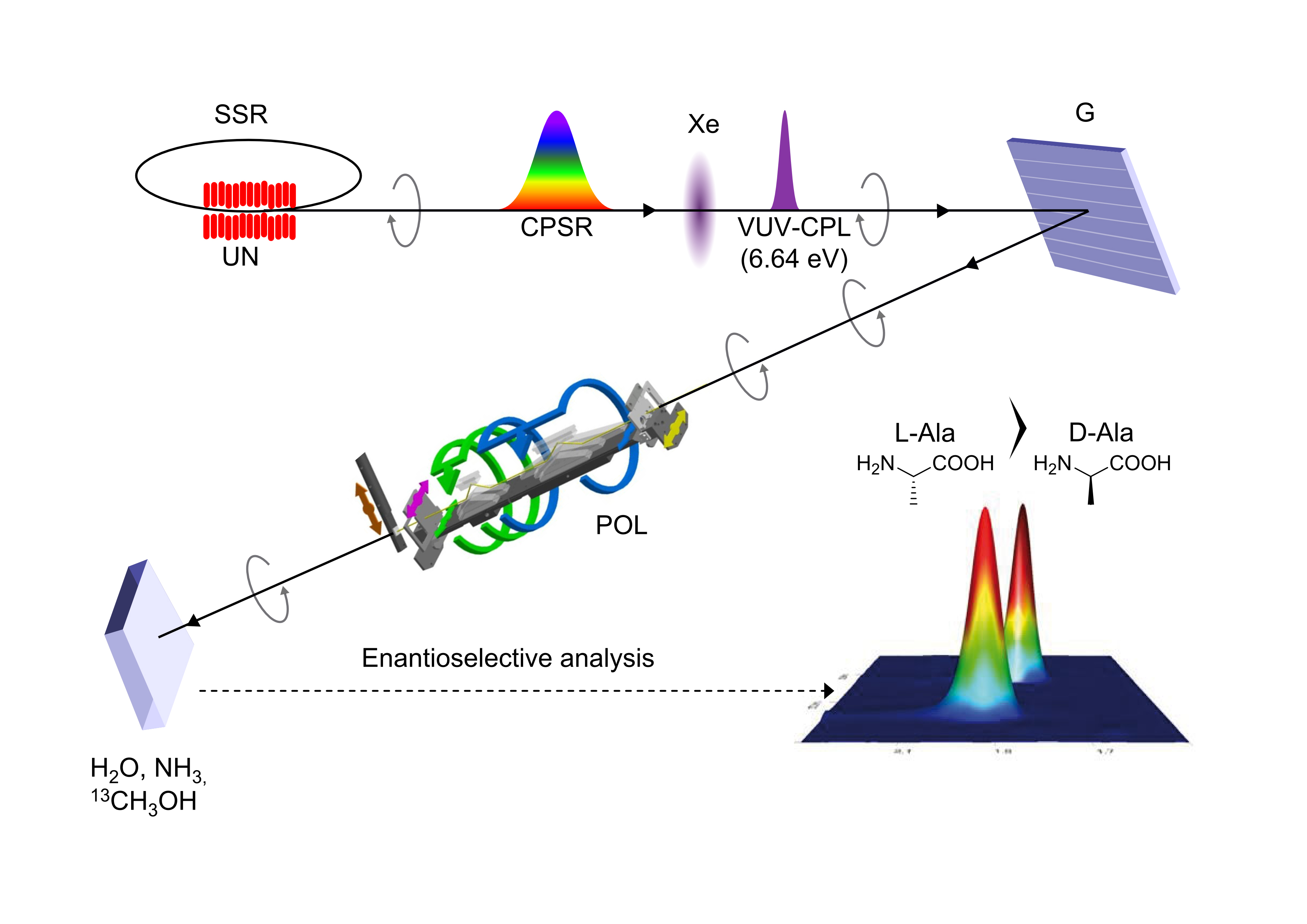}
\caption{Schematic of the experimental setup for the asymmetric photochemical synthesis of amino acids from achiral interstellar precursor molecules. VUV circularly polarized synchrotron radiation (CPSR) is produced by the HU640 undulator (UN) on the SOLEIL storage ring (SSR) and collected by the DESIRS beamline\cite{Nahon13}. The CPSR pulse irradiates the condensed H$_2$O, NH$_3$, and $^{13}$CH$_3$OH mixture at T = 80 K. Reproduced with permission from Ref. 103. Copyright 2011 AAS.}
\label{fig:cpl_scheme}
\end{center}
\end{figure}

Figure \ref{fig:cpl_scheme} presents the process and instruments that were used to make the first conclusive measurement of an enantioneric excess in a chiral molecule following UV-CPL irradiation of an initially achiral interstellar ice analog (H$_2$O:CH$_3$OH:NH$_3$ 2:1:1 at 80 K)\cite{deMarcellus11}. Two important developments enabled this detection: the use of high-flux of UV-CPL radiation, generated by a synchrotron facility, and the application two-dimensional gas chromatography/time-of-flight mass spectrometry (GC$\times$GC-TOFMS), which allows for better enantiomer peak separation. The chosen ice composition and temperature were loosely modeled on observed ISM ices; higher CH$_3$OH and NH$_3$ concentration and ice temperature compared to typical ISM conditions were used to maximize chemical reactivity in the ice. Except for using a polarized UV source, the experimental procedure was very similar to previous experiments aimed at forming amino acids through ice photolysis, i.e. thick ices were exposed VUV radiation as they are being deposited, followed by ice sublimation, and extraction and analysis of the remaining, refractory residue.  The hydrolyzed ice photochemistry residue was analyzed using enantioselective multidimensional gas chromatography. The result was an enantiomeric excess of up to 1.34\% (in isotopically labeled alanine). The sign and absolute value of the enantiomeric enhancement depended on  the helicity and number of UV-CPL photons per deposited molecule, consistent with expectations. Based on infrared monitoring of simpler species during irradiation, photofragments/radicals diffuse and combine during irradiation and the expectation is that the first stereogenic centers form at this point. Following ice sublimation, the residue was further irradiated for about 10 hr at room temperature, to potentially favor enantioselective photoreactions based on a photon-molecule asymmetry transfer. It is important to note that based on these experiments alone, it cannot be excluded that some or all of the chiral products formed at room temperature rather than in the ice. The achieved enantiomeric excess is comparable with some L excesses measured in meteorites. 

\begin{figure}[htbp]
\begin{center}
\includegraphics[width=6in]{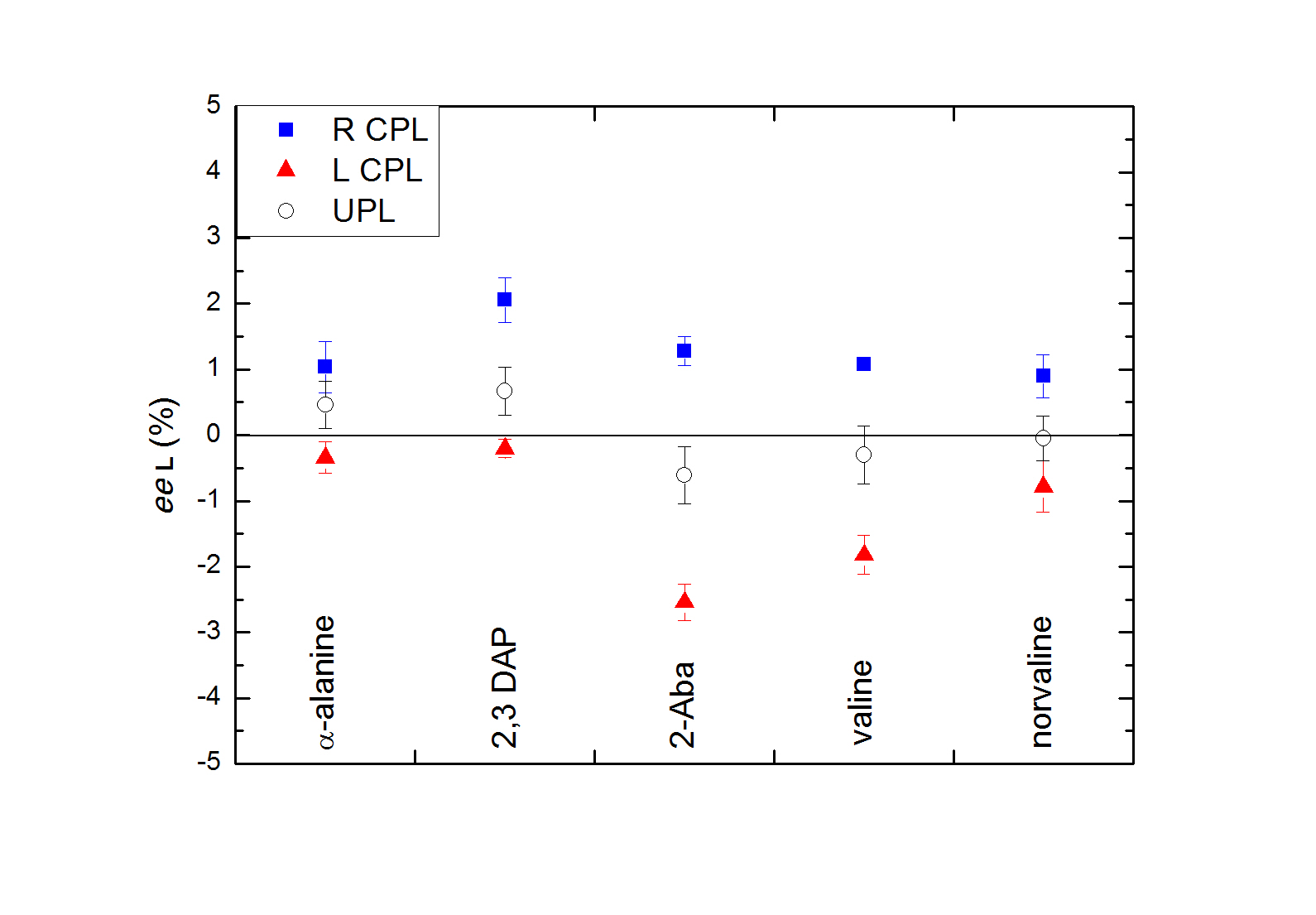}
\caption{Enantiomeric excess determined by enantioselective GC-TOFMS, measured in five different amino acids labeled with $^{13}$C. Experimental values are obtained from three residues of initially achiral circumstellar ice analogs irradiated by CPL at
10.2 eV. Blue squares were irradiated by R CPL, red triangles by L CPL, and white circles by unpolarized light (UPL). The enantiomeric excesses
are of the same sign in all five amino acids for a given helicity of CPL. Reproduced with permission from Ref. 104. Copyright 2011 AAS.}
\label{fig:cpl_aa}
\end{center}
\end{figure}

Follow-up experiments by the same group have provided a more comprehensive view of enantio-selective photochemistry in ices\cite{Modica14}. Figure \ref{fig:cpl_aa} shows the enantiomeric excess in five amino acids exposed to 10.2 eV VUV radiation with L, R or no polarization; the expected L and R excesses are reproduced for multiple species. Another important result is that irradiation at room temperature of the residue is not needed to produce these enantiomeric excesses. The initial ice mixture was also exposed to VUV-CPL at a different VUV energy, which also produced enantiomeric excesses. The formation of enantiomeric excesses in ice photochemistry residues thus appears to be a robust process. Based on these experiments UV-CPL ice exposure is a possible pathway to delivery of enantiomeric excesses to the early Earth, though the kinetics and reaction mechanisms require further exploration to evaluate how plausible this scenario is.

\section{Ice Photolysis vs Ice Radiolysis}

This section aims to compare the chemical response of ices to UV irradiation (ice photolysis), to electron bombardment, proton or ion bombardment (radiolysis), and X-ray photolysis. In astrophysical environments ices are exposed to all these forms of energy. Hence, understanding how ices responds to energy deposition of different kinds is important to predict the overall chemical evolution of interstellar ices. Further, analyzing the different responses of the ice to e.g. UV photons and electrons can provide important clues on the mechanisms that underpin the observed UV photochemistry.

\subsection{UV vs X-rays}

There are very few astrochemistry ice experiments that directly compare the effects of VUV and higher energy photons on ice chemistry.  One exception is a study on soft X-ray and VUV dissociation of CH$_3$OH ice into H$_2$CO\cite{Ciaravella10}, where a 8~K CH$_3$OH ice was irradiated with 0.3 keV soft X-rays under UHV conditions. The X-ray flux was low and the total dose for the experiment was only 10$^{-6}$ photons molecule$^{-1}$. This is many orders of magnitude less than in typical VUV photolysis experiments. The outcome of the X-ray irradiation was compared with 15--60 s of VUV photolysis of an identical ice. Even this very short VUV irradiation results in 70--900$\times$ higher VUV photon dose compared to the X-ray experiments. H$_2$CO is synthesized during both VUV and X-ray irradiation of the CH$_3$OH ice with a similar yield per irradiation dose. The mechanisms are potentially different, however. X-rays can dehydrogenate through ionization of bound H atoms directly or through secondary electrons. The first dehydrogenation step is then followed by a second one, while VUV photolysis can produce H$_2$CO in a single step. The proposed different mechanisms have not been confirmed by a kinetic analysis.

Broadband soft X-rays (up to 2 keV) have also been used to process ice mixtures\\ (H$_2$O:CO$_2$:NH$_3$:SO$_2$ 10:1:1:1) and condensed prebiotically important molecules using the spherical grating monochromator beamline at the Brazilian Synchrotron Light Source (LNLS) in Campinas, Brazil\cite{Pilling11,Pilling15}. Neither study has a direct VUV ice photolysis counterpart, but the derived photodissociation cross sections can still be compared with those found in VUV experiments. The dissociation cross section of parental species was on the order of $(2-7) \times 10^{-18}$ cm$^2$, comparable to many VUV dissociation cross sections. 

\subsection{UV vs. Electrons}

In an astrochemical context, high-energy cosmic rays interactions with molecules and grains in molecular clouds generate secondary electrons of a range of energies. Of especial importance is the track of keV electrons generated inside of ices as a cosmic ray passes through\cite{Bennett05}. To simulate this process there are many astrochemistry ice experiments that employ high energy electrons (on the order of keV) as an energy source\cite{Bennett07a,Bennett07b,Bennett11, Jones11, Kaiser13,Mason14,Jones14,Jones14b,Materese15}. Some of these experiments have explored identical ices to those used in VUV photolysis experiments and it should thus be possible to constrain the relative impact of radiolysis and photolysis on ice chemistry. Comparisons are complicated, however, by the fact that the energy content of electrons of $\sim$1 keV is several orders higher than that of VUV photons. Any observed differences between photolysis and radiolysis ice experiments may therefore {\it a priori} be due to either a difference in energy deposition per particle/photon or intrinsic excitation differences between electrons and photons. The comparisons are further complicated by different penetration depths of electrons of different energies, i.e. the penetration depth for 2~keV electrons is $\sim$100~nm, while it may be as low as 20 monolayers for 100 eV electrons\cite{Arumainayagam10,Mason14}. VUV penetration depths are similar to that of keV electrons.

There are a handful of astrochemistry studies that employ low-energy electrons of more comparable energy to VUV photons\cite{Boamah14,Mason14}. Low-energy electron-induced reactions in condensed matter in astrochemistry and other fields have recently been reviewed\cite{Arumainayagam10}. These experiments can be compared with VUV ice photolysis more straightforwardly, as long as the ices are thin enough that the electrons can penetrate through the ice, i.e. on the order of 10s of monolayers. A few experiments also compare the effects of high energy and low energy electrons\cite{Mason14}. In experiments where CH$_3$OH ice is processed using both low and higher energy electrons, the products and yields show little energy dependence; 100 eV, 1 keV or 5 keV bombardment produce similar yields. The similar product composition suggests that regardless of the initial electron energy, the bulk of ice destruction and formation is caused by secondary low-energy electrons, following thermalization of the incident electrons in the ice. Even in high-energy electron bombardment, the chemistry may thus be driven by electrons of comparable energies to VUV photons, though this requires more experiments to confirm. The similar yields may appear puzzling considering the different penetration depths of the different kinds of electrons, but can probably be explained by stronger interactions between low-energy electrons and ice.

Similarly to VUV irradiation, electron irradiation of simple ice mixtures is an efficient pathway to complex molecules. Molecules as complex as dipeptides have been observed following electron-irradiation of ice mixtures of CO$_2$, NH$_3$ and hydrocarbons\cite{Kaiser13}. When comparing VUV and high energy electron irradiation of the same pure ices or simple ice mixtures, the identified products are remarkably similar. That is, electron-irradiated 10~K pure CH$_3$OH ice and a CO:CH$_3$OH ice mixture result in the production of the the same photoproducts as detected in ice photochemistry experiments\cite{Bennett07a,Bennett07b,Oberg09d}, despite slightly different temperatures (10 vs. 20~K), and real differences in dose.  In the electron bombardment experiment $\sim$1.4 eV/molecule was injected in the ice, while in the VUV experiment $\sim$10x more energy is deposited into the ice. Some differences in the product composition was observed, however, and the extracted CH$_3$OH dissociation branching ratios were substantially different between the VUV and electron experiments. A kinetic analysis of the electron-irradiation experiments resulted in a derived CH$_2$OH:OCH$_3$:H$_2$CO:CH$_4$ dissociation  branching ratio of 7:10:0.5:0.6. In the VUV experiments  there was no direct evidence of unimolecular CH$_4$ and H$_2$CO formation (though the latter is probably present) and the CH$_2$OH/OCH$_3$ branching ratio was estimated to $>$5. It is unclear whether this reported difference would persist if the experiments were analyzed consistently, but if it does, it could provide a tool to distinguishing between VUV and energetic electron driven ice chemistry.

Pure CH$_3$OH ice has also been radiolyzed using low energy ($<$20~eV) electrons\cite{Boamah14}. 85~K ice was irradiated by electrons with energies of 7--20~eV and then heated up in TPD experiments during which products were identified through a combination of sublimation temperatures and m/z patterns as measured by a QMS. No quantitative analysis was attempted and only qualitative comparisons with VUV experiments are therefore possible. Compared to VUV photolysis, low-energy electrons seem to produce the same products with the one addition of methoxymethanol. Methoxymethanol was also identified following irradiation of condensed methanol with 55 eV electrons, but not following irradiation with VUV, 1 keV electrons or 5 keV electrons by other groups. The authors suggest that this could be used to distinguish electron radiolysis from VUV photolysis of ices. It is unclear whether this difference is real, however, since this species has not been actively searched for in CH$_3$OH ice photolysis experiments and no constraint on its (lack of) production during ice photolysis thus exists. Considering that CH$_3$O and CH$_2$OH are both known to form in photolysis experiments, there is no reason why methoxymethanol should not form as well, though its relative abundance might be different if the radical production branching ratios are different for photolysis and radiolysis. In general, these radiolysis experiments suggest that the low-energy electron irradiation chemistry can proceed analogously to VUV photolysis, i.e. a mechanism involving formation of methoxy (CH$_3$O) and hydroxymethyl (CH$_2$OH) radicals via electron impact excitation followed by radical-radical coupling. This is preferred over ionizing reactions, since all primary radiolysis products are observed following electron radiation at an incident electron energy of 7 eV, below the ionization energy for methanol. 

Another ice chemistry that has been studied both during VUV and energetic electron irradiation is formamide formation in NH$_3$ and CO containing ices\cite{Jones11}. Based on the electron irradiation experiments the main formation pathway of this molecule, regardless of energy source, involves the cleavage of a N-H in NH$_3$ to form NH$_2$ and H. The H reacts with CO to form HCO and then radical combination takes place to form NH$_2$HCO. The H+CO reaction has a barrier. Despite this barrier several studies have shown that the H+CO reaction is fast with thermalized H, which is explained by competition between diffusion and reaction of thermalized H \cite{Watanabe03,Fuchs09}. In these experiments an ice surface was bombarded with H atoms, which may {\it a priori} have been incompletely thermalized. Experiments using non-energetic ground state H-atoms within a neon-matrix as a starting point for H addition to CO reach the same conclusion, however, i.e. that reactions occur as long as diffusion-mediated encounters between H and CO take place\cite{Pirim11}.  In the case of formamide formation, the H is produced energetic in the ice, which may help to overcome the entrance barrier in the H+CO reaction, but it may also be that reactions between thermalized H and CO dominate. VUV and electron irradiation of CO$_2$:NH$_3$ ices also appear to produce a similar chemistry though there has been no quantitative comparison\cite{Jones14,Jones14b}.

VUV radiation and 1.2 keV electron do result in different ice resides following irradiation and heating of low-temperature N$_2$:CH$_4$:CO (100:1:1) ice\cite{Materese15}. The residues were studied by FTIR spectroscopy, X-ray absorption near-edge structure (XANES) spectroscopy, and analyzed by gas chromatography coupled to a mass spectrometer (GC-MS). The XANES analysis showed important differences in the residue structure and N/C and O/C ratios between the two sets of experiments. In the VUV irradiated ice the N/C and O/C ratios were $\sim$0.5 and 0.3, respectively, while in the electron irradiated ice they were $\sim$0.9 and 0.2, reflecting the greater efficiency at which 1.2 keV electrons can dissociate N$_2$ compared to VUV photons. Both experiments resulted in nitrile and amide production, but only the VUV experiment produced  aromatic compounds and carboxyl groups. 

In addition to these studies aimed at complex molecule formation, there are comparable VUV and electron studies on the chemistry of pure CO and pure N$_2$ ices\cite{Gerakines96,Jamieson06,Jamieson07,Wu12}. Pure CO photolysis and radiolysis both result in the production of a range of carbon oxides and carbon chains, including C$_3$, C$_2$O, C$_3$O, CO$_2$, C$_3$O$_2$. In UV photolysis experiments, the chemistry is initiated by a reaction between excited CO and ground state CO to form CO$_2$ and C. The same pathway has been assumed to drive the chemistry when CO is exposed to energetic electrons, and this assumption is consistent with detailed kinetic modeling\cite{Jamieson06}. In the case of N$_2$ photochemistry EUV photons and electrons seem to also initiate a similar chemistry, beginning with the excitation of a N$_2$ molecule to a dissociative state where it may then fragment. Specifically there was no evidence for ion formation in experiments on N$_2$ radiolysis with energetic electrons\cite{Jamieson07}. In summary, electrons and photons appear to have a very similar effect on ice chemistry, regardless of whether the initial ice mixture is simple or complex.

\subsection{UV vs. Protons and Ions}

Ion irradiation experiments with protons, He$^+$ or heavier ions are meant to simulate cosmic ray bombardment of interstellar ices. Such ion studies go back almost as far as VUV ice experiments in astrochemistry. Similar to electron and VUV irradiation, ion bombardment of simple, astrophysical relevant ices have been shown to produce amino acids and other complex molecules\cite{Hudson08}. Here we focus on studies that either comprise both VUV and ion bombardment experiments, or can be directly compared to VUV experiments discussed above.

In a series of such studies, identical ices were exposed to VUV radiation and ions and the product compositions compared\cite{Gerakines00,Gerakines01,Gerakines04}. VUV exposure and proton bombardment of CO$_2$:H$_2$O ice mixtures both result in the production of carbonic acid (H$_2$CO$_3$) as well as CO and CO$_3$. Differences in observed product amount are primarily attributed to differences in penetration depth, i.e. the yields normalized to energy absorption are indistinguishable for the two energy sources. The formation mechanism of the acid may be different in the two experiments, however, since acid production in the ion bombarded ice is inhibited when sulfur hexafluoride, which consumes electrons and thus inhibits ion formation, is added to the ice. Such formation rate reduction is not observed when sulfur hexafluoride is added to the ice prior to VUV processing. 

\begin{figure}[htbp]
\begin{center}
\includegraphics[width=5in]{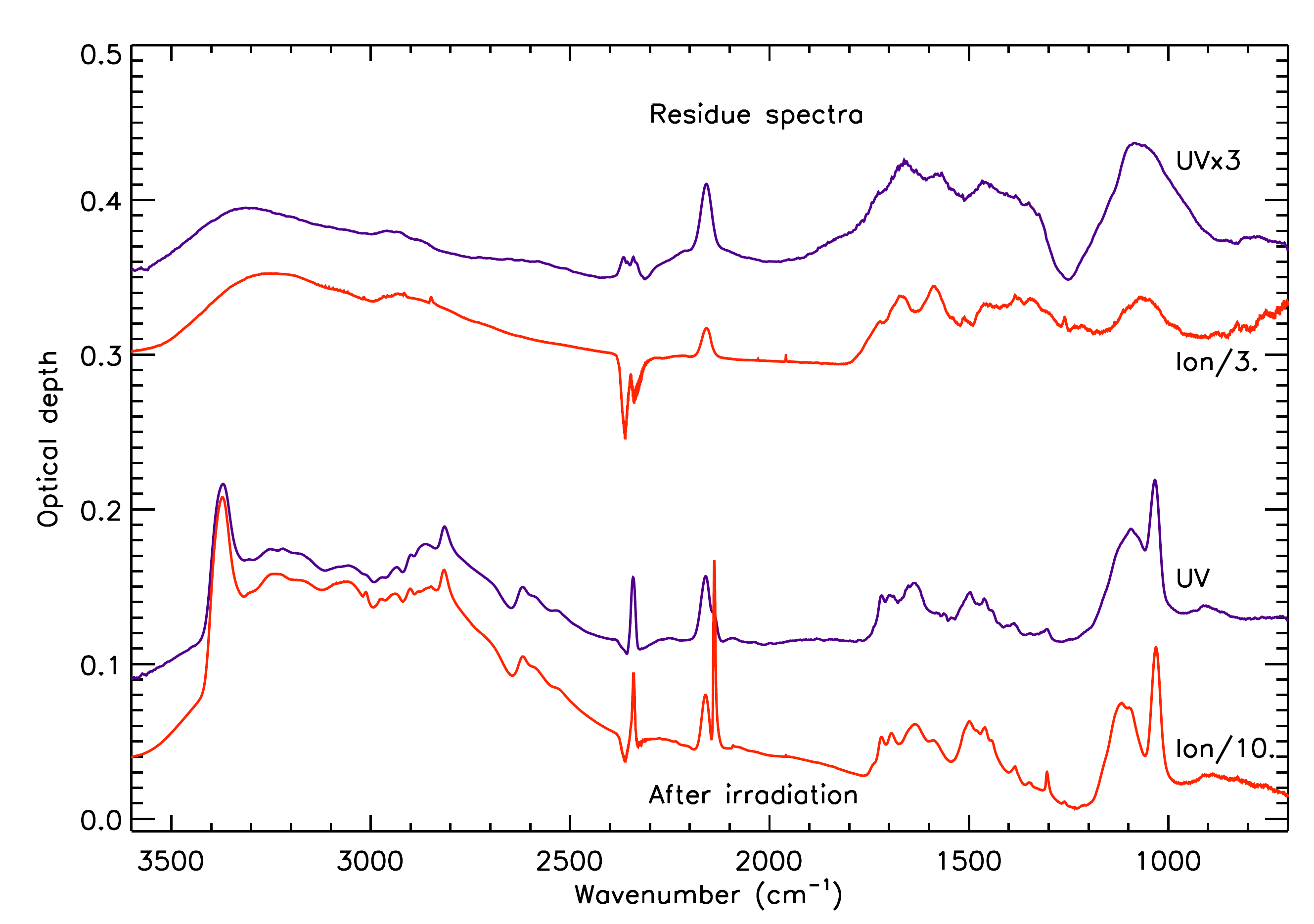}
\caption{Comparison of the chemistry induced in a CH$_3$OH:NH$_3$ = 1:1 ice mixture after UV or 620 MeV ion irradiation. The corresponding residue spectra at room temperature are shown as well. Reproduced with permission from Ref. 179. Copyright 2014 EDP Sciences.}
\label{fig:pd_ion}
\end{center}
\end{figure}

One recent comparison between VUV and ion irradiation ice chemistry is shown in Fig. \ref{fig:pd_ion}. CH$_3$OH:NH$_3$ ice was irradiated with either VUV photons or heavy ions from the GANIL accelerator and  the chemical evolution of the ice was monitored, including the composition of the post-warm-up residue, {\it in situ}\cite{MunozCaro14} . The deposited energy doses were similar for ion beams and UV photons to allow a direct comparison. A variety of organic species were detected, most of them common to both radiation sources, during irradiation, warm-up, and in the residues. The only clear difference between the two experiments was a higher abundance of CO in the ion irradiation experiments. This kine of ice mixtures, which can be dissociated by both VUV and ions, thus seem to produce similar products following VUV and ion irradiation\cite{MunozCaro14}.

H$_2$O:CO$_2$:CH$_3$OH (1:1:1) and H$_2$O:CO$_2$:CH$_4$ (1:1:1) ices exposed to VUV or protons also result in similar production compositions and reaction yields. Two substantial differences were noted, however:  (1) the formate ion is observed only for proton-irradiated CH$_3$OH-containing ices, and (2) the rate of CO  formation is measured to be 2--4 times higher in VUV-photolyzed ices. Pure HCN and HCN:NH$_3$:H$_2$O 18~K ice mixtures have also been exposed to VUV photons and energetic protons. Infrared spectra acquired after processing of pure HCN appears similar for both energy sources. Qualitatively, radiolysis and photolysis also produce similar product spectra when the starting ices are HCN:NH$_3$, HCN:H$_2$O, and HCN:NH$_3$:H$_2$O ice mixtures. When yields could be quantified, these were generally similar as well. 

There are several experiments on ion bombardment of CH$_3$OH ices\cite{Palumbo99,Baratta02,Modica10,Islam14}, including one that quantitatively compared the outcome of VUV and He$^+$ irradiation (with 30~keV ions)\cite{Baratta02} using infrared spectroscopy. All comparisons were made normalized to the energy deposition or radiation dose into the ice. At early times, VUV and ion irradiation result in similar ice destruction and product formation rates. At longer times ion irradiation is more efficient at destroying the original ice. This difference is explained by evolving ice optical properties with radiation exposure due to build-up of a refractory residue. The residue is opaque at UV wavelengths, but not to energetic ions. Another observed difference was that CO and CO$_2$ formation in the CH$_3$OH ice was more efficient in the VUV irradiated ice compared to the ion irradiated one,  consistent with other experiments that have found that a main effect of ion irradiation is the carbonization of the sample, i.e. the preferential loss of hydrogen, oxygen, and nitrogen. The mechanism behind this difference has not been fully explained. 

CH$_4$:H$_2$O ices have also been ion irradiated and the outcome is very similar to VUV-irradiated ices in terms of product composition and how this composition depends on the initial mixing ratio of H$_2$O and CH$_4$. There has been no quantitative study comparing the relative efficiencies of ions and VUV in driving this chemistry, however\cite{Moore98,Oberg10b}. A case where the outcome of VUV and ion (proton) irradiation is very different is the CO$_2$ production rate in CO ice. This is readily  explained by the lack of efficient dissociative CO excitation channels around Lyman-$\alpha$, however, rather than any fundamental difference between photon and particle processing\cite{Loeffler05}. 

Similarly when a CH$_3$OH:N$_2$ 16~K ice is exposed to ion and Lyman${\alpha}$ photons, either separately or simultaneously, the most obvious difference in outcome can be explained by lack of N$_2$ dissociation in the VUV experiments\cite{Islam14}. The samples were analyzed {\it in situ} using transmission infrared spectroscopy. After processing a CH$_3$OH:N$_2$ sample, the intensity of the methanol bands was observed to decrease at the same rate in all cases. CO$_2$, CO, H$_2$CO, CH$_4$, N$_2$O, HNCO, and OCN$^-$ all formed in the ice mixture during ion bombardment. After UV photolysis, the same CO$_2$, CO, H$_2$CO, and CH$_4$ formed, but no N-bearing species. This is consistent with expectations of that Lyman-$\alpha$ radiation cannot dissociate N$_2$ while 200 keV protons (and the accompanying cascade of secondary electrons) can. Similar differences between VUV and ion bombardment irradiation have been observed in other studies using N$_2$ as a starting material\cite{Hudson02,Moore03}, which can generally be explained by the turn on or off of N$_2$ dissociation channels.

In addition to these experiments that contain direct comparisons or are similar enough to compare to other VUV experiments, there are several ion bombardment experiments that could readily be repeated with VUV irradiation, including explorations of the interactions between H$_2$O:NH$_3$ and H$_2$O:NH$_3$:CO ices and  heavy, highly charged, and energetic ions (46 MeV $^{58}$Ni$^{13+}$)\cite{Pilling10}.  The average dissociation cross-sections of water, ammonia, and carbon monoxide in this study was $\sim$2, 1.4, and 1.9 $\times10^{-13}$, respectively, i.e. $\sim$5 orders of magnitude higher than for 10~eV VUV photons. When normalized to the energy per incident particle or photon, the cross sections are of the same order of magnitude for photons and ions, except for CO, where the cross section is lower for VUV photons\cite{Pilling10,Pilling10b}.

\section{Summary and Concluding Remarks}

Interstellar ice photochemistry is an efficient pathway to chemical complexity in space. It is a source of prebiotic amino acids and sugars, and may be the original source of enantiomeric excess on the nascent Earth. How much and under which conditions such complex organics can form during star and planet formation is still largely unknown, however, due to largely unconstrained ice photochemistry kinetics and mechanisms. The quantitative time-resolved studies that do exist on ice photochemistry have demonstrated that all steps of the physiochemical process, from initial photodissociation to radical reactions to form new molecules, are difficult to predict from theory or analogy with gas-phase photochemical reactions. The ice environment regulates the effective photodissociation cross section, the radical production branching ratio, radical diffusion rates and therefore radical combination rates, as well as the relative importance of reaction barriers. Experiments have provided some constraints on these processes for specific ice constituents and ice matrices, but much work remains to obtain a comprehensive and quantitative understanding of interstellar ice photochemistry and the role it plays in seeding nascent planets with prebiotic molecules and precursors. Below we summarize the findings in this review as well as provide some final recommendations.

\begin{enumerate}
\item Measured VUV photodissociation rates of pure and mixed ices are low compared to equivalent measurements for gas-phase species, which can be explained by a combination of ice opacity (most measurements are done on thick ices), and the cage effect, which results in immediate recombinations of some of the photo produced radicals. Because of the importance of the cage effect, gas phase photodissociation cross sections cannot be applied to ice chemistry models.
\item Photodissociation branching ratios into different radicals have been estimated for CH$_4$ and CH$_3$OH ice and the results are substantially different from gas-phase measurements. Branching ratios cannot then be taken from gas-phase measurements when modeling ice photochemistry.
\item Diffusion and/or reorientation of radicals often, but not always, regulate photochemistry kinetics. In ices that are too cold for thermal diffusion to be active, complex molecule formation can be explained by a combination of non-thermal diffusion and reactions between radicals formed at neighboring sites. It is important to note that both processes may be less important in astrophysical environments where radical production rates are lower. Measured product yields and kinetics in laboratory ices cannot be extrapolated directly to interstellar ices without taking into account the different radical production time scales in the two settings.
\item Photochemistry kinetics in ice mixtures often differ substantially from the photochemistry kinetics observed in pure ices due to a combination of new radical formation pathways, larger separation of radicals (if the matrix is inert), and different diffusion environments. These aspects must be taken into account when using data from pure ice experiments to model interstellar ice chemistry.
\item Many prebiotically interesting molecules, including amino acids, form in ice photochemistry experiments with ice compositions inspired by observed interstellar ices. It is important to realize, however, that the experiments are often designed to maximize yields, and that most of the individual amino acids seem to form during residual hydrolysis rather than through pure ice chemistry. A more detailed understanding of ice chemistry kinetics is required to predict the typical amino acid concentration in interstellar ices, and thus the abundance delivered to comets and further to nascent planets.
\item Electrons, Ions and X-rays can induce an ice chemistry similar to than of VUV photons. Initial radical formation branching ratios may be different for different sources of energy, but this has yet to be firmly established. The subsequent initial non-thermal diffusion step may also depend somewhat on the details of the dissociation process. The diffusion barriers themselves, and thus the thermal diffusion rate, should by contrast not depend on the details of the radical production step. Based on experiments the similarities of photolysis and radiolysis experiments are more important than the differences. That is, product compositions in photolysis and radiolysis ice experiments are often remarkably similar. This suggests that the overall chemical evolution of interstellar ices will mainly depend on the amount of energy deposited into the ice and not on in what kind it is delivered. 
\end{enumerate}

Ice photochemistry is clearly a plausible pathway to chemical complexity in space. It is not the only one however. To evaluate its importance compared to other sources of chemical complexity it is key to develop experimental constraints on the mechanisms and kinetics that drive ice photochemistry and related chemistries in a range of plausible ice morphologies. Existing experiments have demonstrated the difficulty in extrapolating such kinetics from gas to ice, and between different ice systems. As more experiments are carried out, more patterns may emerge however. Energy-resolved experiments seems to be an especially promising avenue for progress, since they enable direct comparison between the initial excitation step and the chemistry. Perhaps the most important missing information is how different radicals diffuse through realistic ices, and how thermal diffusion compares with non-thermal diffusion at different ice temperatures. Constraining these processes is absolutely crucial to extrapolate laboratory results on any kind of ice chemistry to interstellar settings where radical production rates are slow and time scales long.

\begin{acknowledgement}

The author thanks Edith Fayolle and Jennifer Bergner, and four anonymous referees for valuable feedback on the manuscript. The author also acknowledges funding from the Simons Collaboration on the Origins of Life (SCOL), award number 321183, an Alfred P. Sloan fellowship, and a Packard fellowship.

\end{acknowledgement}

\providecommand*\mcitethebibliography{\thebibliography}
\csname @ifundefined\endcsname{endmcitethebibliography}
  {\let\endmcitethebibliography\endthebibliography}{}

\end{document}